\documentclass[fleqn,usenatbib]{mnras}

\usepackage[T1]{fontenc}

\DeclareRobustCommand{\VAN}[3]{#2}
\let\VANthebibliography\thebibliography
\def\thebibliography{\DeclareRobustCommand{\VAN}[3]{##3}\VANthebibliography}

\usepackage{graphicx}	% Including figure files
\usepackage{amsmath}	% Advanced maths commands
\usepackage{amssymb}	% Extra maths symbols

\usepackage{booktabs}
\usepackage[capitalize]{cleveref}
\usepackage{dcolumn}
\usepackage{mathtools}
\usepackage{microtype}
\usepackage{upgreek}
\usepackage{siunitx}
\usepackage{tabularx}
\usepackage{wasysym}
\usepackage{xcolor}
\usepackage{xparse}

\newcolumntype{Y}{>{\centering\arraybackslash}X}

\NewDocumentCommand \ii{}{\mathrm{i}}
\NewDocumentCommand \e{}{\mathrm{e}}
\RenewDocumentCommand \d{}{\mathrm{d}}
\RenewDocumentCommand\pi{}{\uppi}
\NewDocumentCommand \N{}{\mathcal{N}}

\NewDocumentCommand \D{}{\mathcal{D}}

\NewDocumentCommand \vect{m}{\boldsymbol{#1}}
\NewDocumentCommand \matr{m}{\boldsymbol{\mathrm{#1}}}

\DeclareMathOperator{\dirac}{\updelta_\text{Dirac}}
\DeclareMathOperator{\kronecker}{\updelta}

\RenewDocumentCommand \upi{m}{\int \D #1 \,}
\NewDocumentCommand \usi{m}{\int \d #1 \,}

\NewDocumentCommand \td{mm}{\frac{\d #1}{\d #2}}
\NewDocumentCommand \pd{mm}{\frac{\partial #1}{\partial #2}}

\NewDocumentCommand \obs{}{{\text{obs}}}
\NewDocumentCommand \rec{}{{\text{rec}}}
\NewDocumentCommand \ext{}{{\text{ext}}}
\NewDocumentCommand \lin{}{{\text{lin}}}
\NewDocumentCommand \gal{}{{\text{g}}}
\NewDocumentCommand \LG{}{{\text{LG}}}
\NewDocumentCommand \com{}{\text{com}}
\NewDocumentCommand \lum{}{\text{lum}}

\NewDocumentCommand \sig{}{\sigma_8}
\NewDocumentCommand \siglin{}{\sig^\lin}
\NewDocumentCommand \siggal{}{\sig^\gal}
\NewDocumentCommand \fsig{}{f \sig}
\NewDocumentCommand \fsiglin{}{f \siglin}

\NewDocumentCommand \zref{}{\bar{z}}

\NewDocumentCommand \norm{m}{\hat{#1}}
\NewDocumentCommand \tenssmooth{m}{\tilde{#1}}

\NewDocumentCommand \mpc{s}{%
    \IfBooleanTF #1%
        {\mathrm{Mpc}}%
        {\; \mathrm{Mpc}}%
}
\NewDocumentCommand \hmpc{s}{%
    \IfBooleanTF #1%
        {h^{-1} \, \mathrm{Mpc}}%
        {\; h^{-1} \, \mathrm{Mpc}}%
}
\NewDocumentCommand \hgpc{s}{%
    \IfBooleanTF #1%
        {h^{-1} \, \mathrm{Gpc}}%
        {\; h^{-1} \, \mathrm{Gpc}}%
}
\NewDocumentCommand \kms{s}{%
    \IfBooleanTF #1%
        {\mathrm{km} \, \mathrm{s}^{-1}}%
        {\; \mathrm{km} \, \mathrm{s}^{-1}}%
}

\NewDocumentCommand \rmax{}{r_\mathrm{max}}
\NewDocumentCommand \reff{}{r_\mathrm{eff}}
\NewDocumentCommand \rsmooth{s}{r_\text{s}^\rec}

\NewDocumentCommand \ssmooth{s}{%
    \IfBooleanTF #1%
        {r_\text{s}}%
        {r_\text{s}^\text{min}}%
}

\NewDocumentCommand \czmin{s}{cz_\text{min}^{v\text{-}v}}

\NewDocumentCommand \bext{s}{%
    \IfBooleanTF #1%
        {B^\ext}%
        {\vect{B}^\ext}%
}

\NewDocumentCommand \bint{s}{%
    \IfBooleanTF #1%
        {B^\text{int}}%
        {\vect{B}^\text{int}}%
}

\NewDocumentCommand \drec{s}{%
    \IfBooleanTF #1%
        {\delta^\rec}%
        {\norm{\delta}^\rec}%
}
\NewDocumentCommand \vrec{s}{%
    \IfBooleanTF #1%
        {v^\rec}%
        {\vect{v}^\rec}%
}
\NewDocumentCommand \brec{s}{%
    \IfBooleanTF #1%
        {B^\rec}%
        {\vect{B}^\rec}%
}

\NewDocumentCommand \vobs{s}{%
    \IfBooleanTF #1%
        {v^\obs}%
        {\vect{v}^\obs}%
}
\NewDocumentCommand \vLG{s}{%
    \IfBooleanTF #1%
        {V^\LG}%
        {\vect{V}^\LG}%
}
\NewDocumentCommand \mvLG{s}{%
    \IfBooleanTF #1%
        {V^\LG_\text{mock}}%
        {\vect{V}^\LG_\text{mock}}%
}

\NewDocumentCommand \hobs{}{h^\obs}

\NewDocumentCommand \sspace{}{\mbox{$s$-space}}
\NewDocumentCommand \rspace{}{\mbox{$r$-space}}

\NewDocumentCommand \vvcomp{s}{\mbox{$v$-$v$} comparison}

\title[Constrained realizations from 2MRS]{Constrained realizations of 2MRS density and peculiar velocity fields: growth rate and local flow}

\author[R. Lilow and A. Nusser]{
Robert Lilow$^{1}$\thanks{E-mail: \href{mailto:rlilow@campus.technion.ac.il}{rlilow@campus.technion.ac.il}}
and Adi Nusser$^{1}$\thanks{E-mail: \href{mailto:adi@physics.technion.ac.il}{adi@physics.technion.ac.il}}
\\
$^{1}$Department of Physics, Technion, Haifa 3200003, Israel
}

\date{Accepted XXX. Received YYY; in original form ZZZ}

\pubyear{2021}

\usepackage{newtxtext,newtxmath}
\begin{document}
\label{firstpage}
\pagerange{\pageref{firstpage}--\pageref{lastpage}}
\maketitle

% Abstract of the paper
\begin{abstract}
We generate constrained realizations (CRs) of the density and peculiar velocity fields within $200 \hmpc$ from the final release of the Two-Micron All-Sky Redshift Survey (2MRS) -- the densest all-sky redshift survey to date. The CRs are generated by combining a Wiener filter estimator in spherical Fourier-Bessel space with random realizations of log-normally distributed density fields and Poisson-sampled galaxy positions. The algorithm is tested and calibrated on a set of semi-analytic mock catalogs mimicking the environment of the Local Group (LG), to rigorously account for the statistical and systematic errors of the reconstruction method. By comparing our peculiar velocity CRs with the observed velocities from the Cosmicflows-3 catalog, we constrain the normalized linear growth rate to $\fsiglin = 0.367 \pm 0.060$, which is consistent at the $1.1 \sigma$ level with the latest Planck results as well as other direct probes. Simultaneously, we estimate a bulk flow contribution from sources beyond the 2MRS reconstruction volume of $\bext* = 199 \pm 68 \kms$ towards $l = 299 \pm 18^\circ$, $b = 8 \pm 19^\circ$. The total reconstructed velocity field at the position of the LG, smoothed with a $1 \hmpc$ Gaussian, is $685 \pm 75 \kms$ towards $l = 270.6 \pm 6.6^\circ$, $b = 35.5 \pm 7.2^\circ$, in good agreement with the observed CMB dipole. The total reconstructed bulk flow within different radii is compatible with other measurements. Within a $50 \hmpc$ Gaussian window we find a bulk flow of $274 \pm 50 \kms$ towards $l = 287 \pm 9^\circ$, $b = 11 \pm 10^\circ$. The code used to generate the CRs and obtain these results, dubbed CORAS, is made publicly available.
\end{abstract}

% Select between one and six entries from the list of approved keywords.
% Don't make up new ones.
\begin{keywords}
cosmology: observations -- galaxies: statistics -- dark matter -- large-scale structure of Universe
\end{keywords}

%%%%%%%%%%%%%%%%%%%%%%%%%%%%%%%%%%%%%%%%%%%%%%%%%%

%%%%%%%%%%%%%%%%% BODY OF PAPER %%%%%%%%%%%%%%%%%%

%%%%%%%%%%%%%%%%%%%%%%%%%%%%%%%%%%%%%%%%
\section{Introduction}
\label{sec:intro}
The observed minuscule temperature fluctuations in the primordial cosmic microwave background (CMB) firmly support the standard cosmological paradigm of gravitational instability as the driver for the formation of the late clumpy universe. In this paradigm, the cosmological background governs the evolution of the large-scale structure of the underlying mass density field. Galaxies form in virialized objects (halos) of the gravitationally dominant dark matter (DM) and are thus a direct, albeit biased \citep{kaiser_spatial_1984,bardeen_statistics_1986}, tracer of the mass density field. Therefore, analysis of the large-scale structure of the observed galaxy distribution has become a standard probe of the parameters of the cosmological background.

Our focus here is the nearby low-redshift (less than $200 \hmpc$) distribution of galaxies and the corresponding large-scale flow pattern of galaxies. Galaxies at low redshift are beyond the peak of their merger and star formation activity. This potentially yields a tighter link between the galaxy and DM distribution than at high redshifts, and simplifies the extraction of cosmological information. Another advantage offered by low-redshift data is the availability of accurate measurements of peculiar velocities in addition to the distribution of galaxies. While the galaxy distribution is a biased tracer of the underlying mass density, the equivalence principle implies that the galaxies share the same velocity field as the underlying mass. Utilizing these distinct data allows us to probe additional aspects of the cosmological models, e.\,g.~deviations from the standard model \citep{hellwing_clear_2014}, and to better control systematic and statistical uncertainties.

One approach to analysing the data is via auto- and cross-correlation functions. Correlation analysis of the galaxy distribution in redshift surveys can be contrasted with predictions from cosmological models to probe cosmological parameters and the (biasing) relation between the DM and the galaxy distribution \citep[e.\,g.][]{blake_wigglez_2011,percival_2df_2004,samushia_interpreting_2012,beutler_6df_2012,howlett_clustering_2015,achitouv_consistency_2017,blake_power_2018}. Complementary information can also be obtained from correlations in the observed velocities \citep[e.\,g.][]{johnson_6df_2014,huterer_testing_2017,dupuy_estimation_2019}, and by cross-correlating the two \citep[e.\,g.][]{nusser_velocitydensity_2017,adams_improving_2017,qin_redshift-space_2019,adams_joint_2020}.

Another approach involves a direct comparison of the measured peculiar velocities with those reconstructed from the galaxy distribution in redshift surveys \citep[e.\,g.][]{kaiser_large-scale_1991,hudson_optical_1994,nusser_prediction_1994,pike_cosmological_2005, davis_local_2011,turnbull_cosmic_2012,ma_comparison_2012,carrick_cosmological_2015,said_joint_2020,boruah_cosmic_2020}. The reconstruction relies on a relation implied by gravitational instability theory between the velocity and density. In a nutshell, the velocity is simply proportional to the gravitational force field per unit mass (acceleration) multiplied by the Hubble time with a proportionality constant depending on the matter content of the universe. This approach is almost free of cosmic variance. Theoretically, only the motion of a single object (e.\,g.~the local group of galaxies) is sufficient to infer relevant cosmological parameters. 

A multitude of methods for the reconstruction of the density and peculiar velocity fields have been proposed \citep[e.\,g.][]{bertschinger_recovering_1989,yahil_redshift_1991,nusser_prediction_1994,fisher_wiener_1995,bistolas_nonlinear_1998,zaroubi_wiener_1999,kitaura_recovering_2010,jasche_bayesian_2010,courtois_three-dimensional_2011,kitaura_initial_2013,jasche_bayesian_2013,wang_reconstructing_2013,carrick_cosmological_2015,lavaux_bayesian_2016,jasche_physical_2019,graziani_peculiar_2019,kitaura_cosmic_2020,zhu_reconstruction_2020}. However, an accurate and precise recovery of the velocity field from galaxy redshift surveys is impossible. $a)$ The density field is sampled by a finite number of discrete tracers (galaxies). Therefore, ``shot noise" prevents an accurate derivation of the density field itself. $b)$ Any dynamical relation adopted for the velocity reconstruction is associated with systematic and random uncertainties and cannot capture all dynamical effects even if the full density field is given, especially in regions with large density contrasts. $c)$ The redshift survey provides redshifts and not distances. Incoherent motions cause an artificial smearing (e.\,g.~fingers-of-god) of structures along the line of sights over a few Mpc scales. This smearing presents another barrier in the ability to recover the velocity on small scale. Finally, $d)$ redshift surveys probe finite volumes, while the velocity field in the survey is affected by the gravitational tug of the large scale structure outside the survey volume. 

Therefore, it is necessary to apply a smoothing procedure on the galaxy distribution to suppress the effects of nonlinear evolution, incoherent motions and shot noise. The reconstructed velocity therefore miss small scale modulations that are present in the observed velocities in distance indicator catalogs. This is highly relevant in the comparison of reconstructed versus observed galaxy velocities, where the mismatch in the probed scales can lead to systematic uncertainties in the inferred parameters \citep{davis_comparison_1996,berlind_biased_2000,nusser_biasing_2020}. In this respect, the smoothing procedure by means of orthogonal modes in the space of the data, as done in \cite{davis_local_2011}, has the great advantage that both the reconstructed and observed velocities are matched on the same scales. The procedure yields unbiased estimates of the cosmological parameters but does not fully exploit the data on small scales.

An alternative approach is the Wiener filtering methodology, which under certain assumptions is designed to exploit the full information content of the data. The Wiener filter (WF) is the optimal linear filter in the sense that it minimizes the variance of the underlying signal around the reconstructed one \citep{zaroubi_wiener_1995}. It leaves regions (and scales) where the data has a high signal-to-noise-ratio mostly unaltered, but suppresses the estimated signal in noise-dominated regions. Therefore, the WF estimates of the density and velocity fluctuations are biased towards zero. Accordingly, a direct inference of cosmological information (e.\,g.~the power spectrum/correlation) from the WF fields is biased. Also, the comparison of these fields with independent observations must employ proper statistical tools that incorporate this bias. Furthermore, the effective smoothing length of the WF is data-dependent, complicating the interpretation of cosmological information obtained by comparison of different date sets. Both of these complications can be partially alleviated using the technique of constrained realizations (CRs) \citep{bertschinger_path_1987,hoffman_constrained_1991,zaroubi_wiener_1999}. This technique aims at sampling the statistical distribution of all possible underlying fields compatible with the observed data on the full range of scales. Marginalizing over this distribution yields unbiased estimates of the inferred cosmological parameters. It furthermore provides a faithful estimation of the observational errors in both these parameter estimates and the reconstructed fields themselves.

In this paper, we present the CORAS framework for the fast generation of COnstrained Realizations from All-sky Surveys.\footnote{\label{foot:coras_link}CORAS is publicly available at \href{https://github.com/rlilow/CORAS}{https://github.com/rlilow/CORAS}.} Following the method devised in \cite{fisher_wiener_1995}, it is based on applying the WF to a spherical Fourier-Bessel (SFB) expansion of the density field, which is suitable for a flux-limited (almost) all-sky redshift survey, as both redshift-space distortions (RSDs) and galaxy selection preserve the statistical isotropy of structure relative to the observer. To derive the peculiar velocity field associated with the galaxy distribution in redshift space, we adopt the linear theory relation between density and velocity. It has been shown in \cite{keselman_performance_2017} that for scales $\gtrsim 5 \hmpc$ such a linear reconstruction yields generally better results than nonlinear methods when applied to realistic mock redshift catalogues. The authors of \cite{keselman_performance_2017} suspect this to be caused by the higher sensitivity of nonlinear reconstructions to the information on small scales, which are more strongly contaminated by nonlinear RSDs (e.\,g.~fingers-of-god) and selection effects. Altogether, this yields a numerically very efficient, conservative reconstruction algorithm.

The SFB-WF reconstruction method has previously been successfully applied in \cite{fisher_wiener_1995,webster_wiener_1997} to the IRAS 1.2-Jy survey, in \cite{schmoldt_density_1999} to the IRAS PSCz survey and in \cite{erdogdu_reconstructed_2006} to an earlier release of the Two-Micron All-Sky Redshift Survey (2MRS) with a lower flux limit of $K_s \leq 11.25$ \citep{huchra_2mass_2005}.

In CORAS we further extend this method by implementing an algorithm for the fast generation of CRs, allowing a faithful error assessment accounting for the correlations in the density and velocity fields. For this, we adapt the Hoffman-Ribak method \citep{hoffman_constrained_1991} for generating CRs of Gaussian fields to capture the non-Gaussianity of small-scale structures as well as the shot noise. This is achieved by generating random galaxy catalogs Poisson-sampled from log-normal density realizations, closely following the approach of \cite{agrawal_generating_2017}.

In this work, we apply CORAS to the final release of 2MRS with a flux limit of $K_s \leq 11.75$, containing nearly 45,000 galaxies, 90 \% of which within an effective depth of $z \approx 0.05$ \citep{huchra_2mass_2012,macri_2mass_2019}. Compared to the earlier 2MRS release, the number of galaxies approximately doubled, making it the densest all-sky redshift survey to date, thus significantly improving the reconstruction accuracy and precision compared to previous applications of the SFB-WF method. We reconstruct the matter density contrast and peculiar velocity fields in a spherical volume of comoving radius $\rmax = 200 \hmpc$. In linear theory, this depends on the ratio of growth function and galaxy bias, $\beta = f/b$. By multiplying $\beta$ with the galaxy density fluctuation amplitude on a scale of $8 \hmpc$, $\siggal$, we obtain the normalized growth rate $\fsig = \beta \siggal$, which only combines cosmological parameters \citep{pike_cosmological_2005}. The galaxy fluctuation amplitude $\siggal$ is estimated from 2MRS itself. To constrain $\fsig$, we use the publicly available Cosmicflows-3 (CF3) catalog of galaxy distances \citep{tully_cosmicflows-3_2016} and minimize the scatter between reconstructed and observed peculiar velocities with a maximum-likelihood (ML) estimator. We simultaneously constrain the bulk flow contribution from sources beyond the 2MRS reconstruction volume as well as the velocity of the Local Group (LG) with respect to the Cosmic Microwave Background (CMB).

While the ML estimator and the standard deviation between different CRs provide an excellent estimate of the errors due to observational distance uncertainties and shot noise, there are also errors introduced by the approximations assumed in our modelling itself. Since these intrinsic reconstruction errors and biases are difficult to model theoretically, we perform an extensive test of CORAS on a set of mock 2MRS and CF3 galaxy catalogs. We generate these from the publicly available \textsc{sage} semi-analytic galaxy catalog of the \textsc{MultiDark} simulation run MDPL2 \citep{klypin_multidark_2016,riebe_multidark_2013},\footnote{The galaxy catalog has been downloaded from the CosmoSim database (\href{https://www.cosmosim.org}{https://www.cosmosim.org}).} selecting subvolumes of the simulation box that closely represent the LG environment. The results of this test are then used to calibrate our constraints on $\fsig$ and to obtain realistic estimates of the total reconstruction errors of CORAS.

The paper is structured as follows. In \cref{sec:method} we present the methodology behind CORAS, starting with a short reminder of linear theory and RSDs. Afterwards, SFB decomposition, WF estimator and CR generation are discussed, before closing this section by describing the velocity-velocity comparison method. We then describe 2MRS and CF3 as well as the data preparation steps in \cref{sec:data}. The generation of mock galaxy catalogs and the test of CORAS on those are described in \cref{sec:test_on_mocks}. The parameter constraints resulting from the velocity-velocity comparison are discussed in \cref{sec:parameter_constraints}, and the reconstructed radial velocities are directly compared to the observed ones in \cref{sec:comparison_to_observed_velocity}. Afterwards, we analyse the reconstructed fields on a grid in \cref{sec:reconstructed_fields_on_grid}, where we also discuss the uncertainty due shot noise as well as the reconstructed bulk flow. We conclude by summarizing and discussing our results in \cref{sec:conclusion}. Technical details on various aspects are described in \cref{sec:sigma_galaxy_estimator,sec:SFB_details,sec:selection_function_estimator,sec:generating_random_signal_and_data_realizations,sec:test_on_mocks,sec:tensor_smoothing}.

%%%%%%%%%%%%%%%%%%%%%%%%%%%%%%%%%%%%%%%%
\section{Methodology}
\label{sec:method}
We begin by discussing the different theoretical ingredients that CORAS relies on to generate CRs from all-sky surveys, including all the assumptions and approximations made. We denote by $\vect{r}$ and $\vect{v} = \dot{\vect{r}}$ the comoving coordinate and comoving peculiar velocity of a mass tracer (a galaxy), respectively. The mean density of the universe is $\bar \rho$ and the density contrast is $\delta = \rho / \bar{\rho} - 1$. The total mass density parameter is $\Omega_\mathrm{m}=\bar \rho/\rho_\mathrm{c}$, where $\rho_\mathrm{c}$ is the critical density of the universe. The Hubble function is $H(t)=\dot a/a$ with $a$ being the scale factor related to the redshift by $z=1/a-1$. Furthermore, as usual, $f$ denotes the logarithmic derivative of the linear growth factor $D_+$ with respect to $a$, $f= \d \ln D_+ / \d \ln a$ \citep{peebles_large-scale_1980}, which for a $\Lambda$CDM cosmology can approximately be expressed in terms of the dimensionless matter density parameter, $f \approx \Omega_\mathrm{m}^{0.55}$ \citep{linder_cosmic_2005}.

%%%%%%%%%%%%%%%%%%%%
\subsection{Linear growth and bias}
\label{sec:method:linear_theory}
We assume the matter density contrast $\delta = \rho / \bar{\rho} - 1$ and the comoving peculiar velocity $\vect{v}$ (relative to the CMB) to be related via linear theory,
\begin{equation}
    \vect{\nabla}_r \cdot \vect{v} = - f H \, \delta \,.
	\label{eq:method:linear_relation_between_density_contrast_and_peculiar_velocity}
\end{equation}
Since we can not observe the matter density contrast directly, we need to relate it to the galaxy density contrast $\delta^\gal$. Assuming a linear galaxy bias $b$ such that $\delta^\gal = b \, \delta$, we obtain $\vect{\nabla}_r \cdot \vect{v} = - \beta H \, \delta^\gal$, with the biased growth rate $\beta = f / b$. Due to the luminosity dependence of the galaxy bias and the radially increasing minimal observable luminosity in a flux-limited survey, the mean bias of the observed galaxies and thus also $\beta$ strongly depend on distance. To estimate this dependence, we express the bias as the ratio of the nonlinear galaxy and matter density fluctuation amplitudes, $b(r) = \siggal(r) / \sig$, and estimate $\siggal(r)$ directly from the survey, as described in \cref{sec:sigma_galaxy_estimator}.\footnote{When referring to the linear matter density fluctuation amplitude, we write $\siglin$ instead.} We can then correct for the radial bias dependence by defining the normalized density contrast
\begin{equation}
	\norm{\delta}(\vect{r}) \coloneqq \frac{\delta^\gal(\vect{r})}{\siggal(r)} = \frac{\delta(\vect{r})}{\sig} \,,
	\label{eq:method:normalized_density_contrast}
\end{equation}
whose relation to $\delta$ only depends on the cosmological parameter $\sig$ and not on any survey characteristics anymore. In terms of $\norm{\delta}$, \cref{eq:method:linear_relation_between_density_contrast_and_peculiar_velocity} reads
\begin{equation}
	\vect{\nabla}_r \cdot \vect{v} = - \fsig H \, \norm{\delta} \,.
	\label{eq:method:linear_relation_between_normalized_density_contrast_and_peculiar_velocity}
\end{equation}

Due to its redshift dependence, the parameter combination $\fsig H$ still has an implicit distance dependence that has to be accounted for when solving \cref{eq:method:linear_relation_between_normalized_density_contrast_and_peculiar_velocity} for the velocity. The redshift-dependence of $H$ is absorbed into the comoving coordinates by working in units $\hmpc$. For $\fsig$ we simply neglect this dependence and use its value at the mean redshift $\zref$ of the survey. This is a good approximation for a shallow survey such as 2MRS, where $\zref \approx 0.025$ and deviations of $\fsig$ around its mean value are less than $2 \%$. For reconstructions of deeper future all-sky surveys, however, the full redshift dependence needs to be taken into account.

%%%%%%%%%%%%%%%%%%%%
\subsection{Redshift space distortions and discrete data}
\label{sec:method:redshift_space_distortions}
Galaxy redshift surveys like 2MRS provide redshifts (rather than real distances) and angular positions of a set of $N^\gal$ galaxies. In a flux-limited survey we only observe a fraction $\phi(r)$ of galaxies at a given distance, where $\phi(r)$ is the radial selection function of the survey. The latter is obtained directly from the observations using the $F/T$-estimator described in \cref{sec:selection_function_estimator}.

Let $\vect{s}$ be the comoving redshift coordinate defined as
\begin{equation}
	\vect{s} = d_\com(z^\obs) \, \frac{\vect{r}}{r} \approx \vect{r} + \frac{v_r}{H} \, \frac{\vect{r}}{r} \,.
	\label{eq:method:comoving_redshift_coordinate}
\end{equation}
where $r = d_\com(z)$ and $v_r$ are, respectively, the true comoving distance and radial peculiar velocity of a galaxy. For the distance-redshift relation we assume a flat $\Lambda$CDM cosmology with $\Omega_\mathrm{m}$ taken from the Planck-18 results \citep{aghanim_planck_2020-1}.\footnote{We use the Planck-18 results for the TT,TE,EE+lowE+lensing combination: $\Omega_\mathrm{m} = 0.3153$, $\Omega_\mathrm{b} = 0.04930$, $h = 0.6736$, $n_\mathrm{s} = 0.9649$, $\siglin = 0.8111$.} However, for the low-redshift survey 2MRS the dependence of our results on $\Omega_\mathrm{m}$ is only marginal.

A simple estimator of the normalized density contrast field in \sspace\ is given by a weighted sum of Dirac delta distributions at the \sspace\ positions $\vect{s}_j$ of the $N^\gal$ observed galaxies,
\begin{equation}
    \begin{split}
    	\norm{\delta}^\mathrm{D}(\vect{s}) &= \frac{1}{\siggal(s)} \, \biggl[ \frac{1}{\bar{n}^\gal} \sum_{j=1}^{N^\gal} \, \frac{1}{\phi(s_j)} \, \dirac(\vect{s} - \vect{s}_j) - 1\biggr] \\
    	&\eqqcolon \frac{1}{\bar{n}^\gal} \sum_{j=1}^{N^\gal} \, \norm{w}(s_j) \, \dirac(\vect{s} - \vect{s}_j) - \phi(s) \, \norm{w}(s) \, .
    \end{split}
	\label{eq:method:estimator_of_density_field}
\end{equation}
The superscript ``$\mathrm{D}$'' denotes the observed data, $\bar{n}^\gal$ is the mean galaxy number density, and we defined the weighting function
\begin{equation}
	\norm{w}(s) \coloneqq \frac{1}{\phi(s) \, \siggal(s)} \,.
	\label{eq:method:weighting_function}
\end{equation}
The inverse selection function accounts for the expected total number of galaxies per an observed one.\footnote{The selection function should actually be evaluated at the galaxies' \rspace\ distances, but $\phi(r)$ and $\phi(s)$ agree at 0\textsuperscript{th} order in $v_r$.} The mean galaxy number density is estimated via
\begin{equation}
	\bar{n}^\gal = \frac{1}{V} \sum_{j=1}^{N^\gal} \, \frac{1}{\phi(s_j)} \,,
	\label{eq:method:mean_galaxy_density_estimator}
\end{equation}
where $V$ is the spherical reconstruction volume of radius $\rmax$.

Galaxy redshifts are commonly provided in heliocentric reference frame, but they can be transformed to any other frame moving at any arbitrary velocity. In studies of the local large-scale structures, two reference frames are commonly considered: the frame defined by the CMB on the one hand and that comoving with the LG, at $\vLG* = 620 \pm 15 \kms$ towards $l = 271.9 \pm 2.0^\circ$ and $b = 29.6 \pm 1.4^\circ$ with respect to the CMB \citep{aghanim_planck_2020}, on the other hand.

We live in a moderate-density cosmic neighborhood with a highly coherent flow out to a few Mpc \citep{sandage_redshift-distance_1986,tully_our_2008}. For nearby galaxies, the redshifts in the LG frame (hereafter LG redshifts) are therefore close to the cosmological redshifts and provide a good proxy for the actual distances. Accordingly, the LG frame \sspace\ distribution of those galaxies is close to their \rspace\ distribution. If we instead use redshifts in the CMB frame (hereafter CMB redshifts) as a distance proxy, the large coherent motion of nearby galaxies relative to the CMB will introduce a dipole modulation in their \sspace\ distribution: galaxies in the direction of $\vLG$ will appear more distant and those in the opposite direction less distant than they actually are. Some of the latter will acquire negative redshifts and need to be excluded. Those typically lie within distances of $\vLG* \cos \theta / H$, where $\theta$ is the angle between the angular galaxy position and the direction of $\vLG$.

For distant galaxies, the opposite is true: CMB redshifts are on average closer to the cosmological redshifts, thus providing a better distance proxy. Therefore, the CMB frame \sspace\ distribution is close to the \rspace\ distribution of those galaxies, whereas their LG frame \sspace\ distribution exhibits a dipole modulation: Galaxies in the direction of $\vLG$ appear too close and those in the opposite direction too distant. Thus, on one side some galaxies beyond $\rmax$ would be included, while on the other side some galaxies within $\rmax$ are excluded.

There is also a drawback to working with LG redshifts which is related to testing and calibrating the results. Generating mock catalogs matching the LG properties can be challenging even in the largest simulations available. In particular, LG candidates which satisfy the local coherence and amplitude of the velocity field are rare, see \cref{sec:test_on_mocks:mock_generation}. In our analysis we will thus consider both choices of redshift reference frames.

%%%%%%%%%%%%%%%%%%%%
\subsection{Spherical Fourier-Bessel decomposition}
\label{sec:method:SFB_decomposition}
The underlying cosmological fluctuations are statistically homogeneous and isotropic, yielding diagonal two-point correlators in Fourier space. {In realistic redshift surveys, the homogeneity is broken by two effects: First, the radial peculiar velocities of galaxies introduce redshift-space distortions. Second, in flux-limited surveys, the fraction of observable galaxies decreases with distance, causing a coupling between different Fourier modes.} However, in the absence of additional angular selection, statistical isotropy is preserved. This isotropy is best exploited by working in spherical Fourier-Bessel (SFB) space, where the different angular modes remain independent and only the radial modes are coupled. In the following, we focus on the main aspects of the SFB space and refer to \cref{sec:SFB_details} and \cite{fisher_wiener_1995} for details. 

Given $\norm{\delta}(\vect{r})$ in terms of the spherical coordinates $\vect{r} = (r,\vartheta,\varphi)$, its SFB expansion within a spherical volume of radius $\rmax$ is defined in terms of spherical Bessel functions $j_l$ and spherical harmonics $Y_{lm}$,
\begin{equation}
	\norm{\delta}(\vect{r}) \approx \sum_{l=0}^{l_\text{max}} \, \sum_{m=-l}^l \sum_{n=1}^{n_\text{max}(l)} C_{ln} \, \norm{\delta}^r_{lmn} \, j_l(k_{ln} r) \, Y_{lm}(\vartheta, \varphi) \, , 
	\label{eq:method:SFB_expansion_of_the_normalized_density_contrast}
\end{equation}
where $\norm{\delta}^r_{lmn}$ are the SFB coefficients, $k_{ln}$ are the radial Fourier modes, and $C_{ln}$ are normalization coefficients. Only in the limit of $l_\text{max}$ and $n_\text{max}(l)$ tending to infinity, the SFB base functions form a complete set and \cref{eq:method:SFB_expansion_of_the_normalized_density_contrast} becomes exact. In practice we need to truncate the sum at some finite number of modes, thus limiting the radial and angular resolution of the decomposed function. The choice of $\rmax$, $l_\text{max}$ and $n_\text{max}(l)$ is discussed in \cref{sec:data:2MRS}.

The discreteness of the radial spectrum is a consequence of considering only a finite survey volume, and different possible choices for the boundary conditions are discussed in \cite{fisher_wiener_1995,erdogdu_reconstructed_2006}. They argue that the most natural and conservative boundary conditions correspond to a vanishing density contrast at $r \geq \rmax$. In combination with the continuity of the gravitational force field at $r = \rmax$, this yields the \cref{eq:app:radial_SFB_modes,eq:app:SFB_normalization_coefficient_for_potential_boundary_conditions} for $k_{ln}$ and $C_{ln}$, respectively. The SFB base functions then form an orthogonal set, allowing an inversion of \cref{eq:method:SFB_expansion_of_the_normalized_density_contrast} to obtain the coefficients $\norm{\delta}^r_{lmn}$ via
\begin{equation}
	\norm{\delta}^r_{lmn} = \int\limits_0^{\mathclap{\rmax}} \d r \, r^2 \int \d \Omega \; j_l(k_{ln} r) \, Y^*_{lm}(\vartheta, \varphi) \, \norm{\delta}(\vect{r}) \,.
	\label{eq:method:SFB_coefficients_of_density}
\end{equation}
Furthermore, since the linear velocity field can be expressed as a potential gradient and the SFB base functions are eigenfunctions of the Laplace operator, we can invert \cref{eq:method:linear_relation_between_normalized_density_contrast_and_peculiar_velocity} to express the components of the velocity field directly in terms of $\norm{\delta}^r_{lmn}$, see \cref{eq:app:radial_component_of_peculiar_velocity_field_from_density_contrast_SFB_coefficients,eq:app:theta_component_of_peculiar_velocity_field_from_density_contrast_SFB_coefficients,eq:app:phi_component_of_peculiar_velocity_field_from_density_contrast_SFB_coefficients}. We can thus immediately evaluate both $\delta(\vect{r})$ and $v(\vect{r})$ once the coefficients $\norm{\delta}^r_{lmn}$ are reconstructed from the observed galaxy distribution.

For a galaxy redshift survey, the \rspace\ coefficients $\norm{\delta}^r_{lmn}$ are obtained in two steps. First, we insert the estimator of $\norm{\delta}(\vect{s})$, \cref{eq:method:estimator_of_density_field}, into the \sspace\ equivalent of \cref{eq:method:SFB_coefficients_of_density},
\begin{equation}
	\norm{\delta}^{s,\mathrm{D}}_{lmn} = \frac{1}{\bar{n}^\gal} \sum_{j=1}^{N^\gal} \, \norm{w}(s_j) \, j_l(k_{ln} s_j) \, Y^*_{lm}(\vartheta_j, \varphi_j) - \kronecker_{l0} \, \norm{M}^s_n \,,
	\label{eq:method:normalized_density_contrast_data}
\end{equation}
where $\norm{M}^s_n$ is  the \sspace\ monopole contribution defined in \cref{eq:app:SFB_density_contrast_monopole_contribution_redshift_space}. As shown in detail in \cite{fisher_wiener_1995}, the SFB coefficients in $s$- and \rspace\ can then be related by inserting the continuity equation $\bigl(1+\norm{\delta}(\vect{r})\bigr) \, \d^3 r = \bigl(1+\norm{\delta}(\vect{s})\bigr) \, \d^3 s$ into the \sspace\ equivalent of \cref{eq:method:SFB_coefficients_of_density}, and Taylor-expanding all remaining functions of redshift distance around the respective real distance. To first order in the peculiar velocities, this yields
\begin{equation}
	\norm{\delta}^{s,\mathrm{D}}_{lmn} = \sum_{n'}^{n_\text{max}(l)} (\matr{Z}_l)_{nn'} \, \Bigl[\norm{\delta}^{r,\mathrm{D}}_{lmn} + \kronecker_{l0} \, \norm{M}^r_n \Bigr] \,,
	\label{eq:method:relation_between_redshift_and_real_space_density_contrast_SFB_coefficients_via_coupling_matrix}
\end{equation}
where the matrix $\matr{Z}_l$, defined in \cref{eq:app:coupling_matrix_between_redshift_and_real_space_density_contrast_SFB_coefficients}, describes the coupling between the radial modes of the $s$- and \rspace\ density contrasts. The \rspace\ monopole contribution $\norm{M}^r_n$, defined in \cref{eq:app:SFB_density_contrast_monopole_contribution_real_space}, is introduced to correct for any spurious mean density contrast introduced by the RSD correction.

In overdense regions the mapping from $s$- to \rspace\ can become multi-valued. The linearised RSD correction does not take this into account and is thus inaccurate in those regions. Alongside other nonlinear small-scale effects, this can be mitigated by smoothing the reconstructed fields on a scale of a few $\hmpc*$, as discussed in \cref{sec:method:wiener_filter}.

%%%%%%%%%%%%%%%%%%%%
\subsection{Wiener filtering}
\label{sec:method:wiener_filter}
If the the observed data were in fact only the (linearly) redshift-space distorted true underlying density signal, it would be sufficient to invert the coupling relation in \cref{eq:method:relation_between_redshift_and_real_space_density_contrast_SFB_coefficients_via_coupling_matrix}. In reality, however, the data are noisy and a simple inversion of RSDs is generally unstable and greatly amplifies the noise \citep{zaroubi_wiener_1995,fisher_wiener_1995}. To mitigate this effect, we will apply the Wiener filter (WF) $\matr{W}$ to the data,
\begin{equation}
	\norm{\delta}^\mathrm{W} \coloneqq \matr{W} \, \norm{\delta}^\mathrm{D} = \bigl\langle \norm{\delta}^\mathrm{S} \norm{\delta}^\mathrm{D} \bigr\rangle \, \bigl\langle \norm{\delta}^\mathrm{D} \norm{\delta}^\mathrm{D} \bigr\rangle^{-1} \, \norm{\delta}^\mathrm{D} \,,
	\label{eq:method:general_definition_of_the_wiener_filter}
\end{equation}
where $\bigl\langle {\norm{\delta}}^\mathrm{S} \norm{\delta}^\mathrm{D} \bigr\rangle$ and $\bigl\langle \norm{\delta}^\mathrm{D} \norm{\delta}^\mathrm{D} \bigr\rangle$ denote the normalized density signal-data cross-correlation and data-data auto-correlation, respectively. Essentially, this filter suppresses the noise-dominated parts of the data while leaving signal-dominated data mostly unaffected. The WF minimizes the variance of the residual field $\Delta \norm{\delta} \coloneqq \norm{\delta}^\mathrm{S} - \norm{\delta}^W$, \citep{zaroubi_wiener_1995}
\begin{equation}
	\begin{split}
		\bigl\langle \Delta \norm{\delta} \, \Delta \norm{\delta} \bigr\rangle
		&= \bigl\langle \norm{\delta}^\mathrm{S} \norm{\delta}^\mathrm{S} \bigr\rangle - \matr{W} \, \bigl\langle \norm{\delta}^\mathrm{D} \norm{\delta}^\mathrm{S} \bigr\rangle\,.
		\label{eq:method:variance_of_signal_around_wiener_filter}
	\end{split}
\end{equation}

If the probability distribution function (PDF) of the signal given the data,
\begin{equation}
    P\bigl(\norm{\delta}^\mathrm{S} \big| \norm{\delta}^\mathrm{D} \bigr) = \frac{P\bigl(\norm{\delta}^\mathrm{D} \big| \norm{\delta}^\mathrm{S} \bigr) \, P\bigl(\norm{\delta}^\mathrm{S}\bigr)}{P\bigl(\norm{\delta}^\mathrm{D}\bigr)} \,,
\end{equation}
were Gaussian, the WF estimate would equal both the conditional mean and the most probable realization of the signal given the data. In our situation this is not the case. While the density fluctuations in the early universe were exceptionally close to Gaussian \citep{aghanim_planck_2020-1}, the subsequent nonlinear gravitational evolution leads to the development of non-Gaussian features. It is found that the evolved underlying density field approximately follows a log-normal distribution \citep{coles_lognormal_1991,kofman_evolution_1994}. The observed data are the positions of galaxies, which we model as a set of point-like tracers sampled from the underlying log-normal density field by a Poisson process. Nevertheless, the WF remains a useful estimator for the reconstructed signal, as it keeps its variance-minimizing and noise-suppressing properties for arbitrary distributions of signal and data.

In real space, the correlation functions entering the WF in \cref{eq:method:general_definition_of_the_wiener_filter} are given by \citep{bertschinger_large-scale_1992}
\begin{align}
	\bigl\langle \norm{\delta}^\mathrm{S}(\vect{r}_1) \, \norm{\delta}^\mathrm{D}(\vect{r}_2) \bigr\rangle &= \xi_{\norm{\delta}}(|\vect{r}_1 - \vect{r}_2|) \,,
	\label{eq:method:signal_data_covariance} \\
	\bigl\langle \norm{\delta}^\mathrm{D}(\vect{r}_1) \, \norm{\delta}^\mathrm{D}(\vect{r}_2) \bigr\rangle &= \xi_{\norm{\delta}}(|\vect{r}_1 - \vect{r}_2|) + \frac{\norm{w}(r_1)}{\bar{n}^\gal} \, \dirac(\vect{r}_1 - \vect{r}_2) \,,
	\label{eq:method:data_data_covariance}
\end{align}
where $\xi_{\norm{\delta}} = \xi_\delta / \sig^2$ denotes the correlation function of the normalized density contrast, $\norm{\delta}$. The second term in \cref{eq:method:data_data_covariance} is the contribution from shot noise, which increases towards larger distances where a smaller fraction of galaxies is observed.

As shown in detail in \cite{fisher_wiener_1995}, one can perform the SFB transforms of the correlation functions in \cref{eq:method:signal_data_covariance,eq:method:data_data_covariance}, and combine the results with the definition of the WF, \cref{eq:method:general_definition_of_the_wiener_filter}, as well as the linear RSD correction, \cref{eq:method:relation_between_redshift_and_real_space_density_contrast_SFB_coefficients_via_coupling_matrix}, to obtain the overall expression for the SFB modes of our normalized density contrast estimator,
\begin{equation}
    \begin{split}
	\norm{\delta}^{r,\mathrm{W}}_{lmn} &= \sum_{n'}^{n_\text{max}(l)} \, \Bigl(\matr{S}_l \, \bigl(\matr{S}_l + \matr{N}_l\bigr)^{-1}\Bigr)_{nn'} \, \norm{\delta}^{r,\mathrm{D}}_{lmn'} \\
	&= \sum_{n'}^{n_\text{max}(l)} \, \Bigl(\matr{S}_l \, \bigl(\matr{S}_l + \matr{N}_l\bigr)^{-1} \, \matr{Z}_l^{-1}\Bigr)_{nn'} \, \norm{\delta}^{s,\mathrm{D}}_{lmn'} \,.
    \end{split}
	\label{eq:method:minimum_variance_estimate_in_SFB_space}
\end{equation}
The components of the signal correlation matrix $\matr{S}_l$ are given in \cref{eq:app:signal_matrix}, those of the noise correlation matrix $\matr{N}_l$ in \cref{eq:app:noise_matrix}. $\matr{S}_l$ depends on the normalized density contrast power spectrum $P_{\norm{\delta}} = P_\delta / \sig^2$, for which we use $P_\delta$ provided by Cosmic Emu \citep{heitmann_miratitan_2016}, assuming a flat $\Lambda$CDM cosmology and the cosmological parameters given in the Planck-18 results \citep{aghanim_planck_2020-1}. We compute the value of the nonlinear $\sig$ from this power spectrum, finding $\sig = 0.8963$.

The WF density estimator in \cref{eq:method:general_definition_of_the_wiener_filter} does not guarantee positivity of the density field since it contains no information on the full density PDF beyond the correlation function. It can thus yield negative densities in very underdense regions. This contamination can be reduced by convolving the reconstructed density field obtained after applying the WF with a Gaussian smoothing kernel with a width $\rsmooth$ of a few $\hmpc*$. This is easily achieved by multiplying the SFB coefficients of the normalized density contrast with the SFB-transformed Gaussian kernel,
\begin{equation}
    \norm{\delta}^r_{lmn} \rightarrow \norm{\delta}^r_{lmn} \, 
    \exp\biggl[-\frac{(k_{ln} \, \rsmooth)^2}{2}\biggr] \,,
    \label{eq:method:SFB_space_smoothing}
\end{equation}
before evaluating the \rspace\ density or velocity fields via \cref{eq:method:SFB_expansion_of_the_normalized_density_contrast,eq:app:radial_component_of_peculiar_velocity_field_from_density_contrast_SFB_coefficients,eq:app:theta_component_of_peculiar_velocity_field_from_density_contrast_SFB_coefficients,eq:app:phi_component_of_peculiar_velocity_field_from_density_contrast_SFB_coefficients}.

For applications where strict positivity of $1+\norm{\delta}$ is required an additional simple remapping of density contrasts below some small threshold value
$\norm{\delta}_\textsc{t}$ can be performed, e.\,g.
\begin{equation}
	\norm{\delta} \rightarrow (1 + \norm{\delta}_\textsc{t}) \, \e^{\norm{\delta} - \norm{\delta}_\textsc{t}} - 1 \quad \text{if $\norm{\delta} < \norm{\delta}_\textsc{t}$.}
	\label{eq:method:remapping_of_underdensities}
\end{equation}
As we are mainly interested in the reconstructed velocity field in this work, though, no such post-processing is necessary.

%%%%%%%%%%%%%%%%%%%%
\subsection{Constrained realizations}
\label{sec:method:constrained_realizations}
When applied to a flux-limited survey, the WF density and velocity fields tend to zero at large distances where the noise dominates the underlying signal. For many applications, it is useful to create realizations of these fields which compensate for the WF-suppressed power in a way that is compatible with the observed data. This can be achieved in the framework of constrained realizations (CRs) \citep{bertschinger_path_1987}, which deals with the task of generating random fields of a given statistical distribution that satisfy a set of constraints. Hoffman and Ribak developed an optimal algorithm for the case of a zero-mean Gaussian random field with constraints that are linear in the field \citep{hoffman_constrained_1991}. It relies on the fact that the variance of a Gaussian field around the constrained mean field is independent of the actual values of the constraints. This allows us to generate CRs by simply adding random residual field realizations to the constrained mean of the field.

In our reconstruction problem neither the signal nor the data are Gaussian random fields. However, if we replace the constrained mean by the WF estimate, we nevertheless satisfy the essential property that the variance of the residual density contrast, \cref{eq:method:variance_of_signal_around_wiener_filter}, does not depend on the actual value of the data $\norm{\delta}^\mathrm{D}$. This suggests that we define the density contrast CR $\norm{\delta}^\mathrm{C}$ in complete analogy to the Hoffman-Ribak method: First, we generate random signal and data realizations, $\norm{\delta}^\mathrm{R,S}$ and $\norm{\delta}^\mathrm{R,D}$, and compute the corresponding WF estimate $\norm{\delta}^\mathrm{R,W} = \matr{W} \, \norm{\delta}^\mathrm{R,D}$. Afterwards, the random residual field realization $\Delta\norm{\delta}^\mathrm{R} = \norm{\delta}^\mathrm{R,S} - \norm{\delta}^\mathrm{R,W}$ is added to the WF estimate of the actual data,
\begin{equation}
	\norm{\delta}^\mathrm{C} \coloneqq \Delta\norm{\delta}^\mathrm{R} + \norm{\delta}^\mathrm{W}
	\,.
	\label{eq:method:constrained_realizations_around_wiener_filter}
\end{equation}
By construction, $\norm{\delta}^\mathrm{C}$ obeys
\begin{equation}
	\bigl\langle \bigl(\norm{\delta}^\mathrm{C} - \norm{\delta}^\mathrm{W}\bigr)^p\bigr\rangle_\mathrm{R} = \bigl\langle\bigl(\Delta\norm{\delta}^\mathrm{R}\bigr)^p\bigr\rangle_\mathrm{R} = \bigl\langle(\Delta\norm{\delta})^p\bigr\rangle = \bigl\langle\bigl(\norm{\delta}^\mathrm{S} - \norm{\delta}^\mathrm{W}\bigr)^p\bigr\rangle
	\label{eq:method:moments_of_constrained_realizations_around_wiener_filter}
\end{equation}
for arbitrary powers $p$, where $\langle\dotsc\rangle_\mathrm{R}$ denotes the average over the random realizations $\norm{\delta}^\mathrm{R,S}$ and $\norm{\delta}^\mathrm{R,D}$, which are drawn from the same PDFs we assume for $\norm{\delta}^\mathrm{S}$ and $\norm{\delta}^\mathrm{D}$, respectively. According to \cref{eq:method:moments_of_constrained_realizations_around_wiener_filter}, the distribution of both $\norm{\delta}^\mathrm{C}$ and $\norm{\delta}^\mathrm{S}$ around $\norm{\delta}^\mathrm{W}$ is precisely the same, independent of the specific choice of signal and data PDFs. This is exactly the property we desired. In fact, $\norm{\delta}^\mathrm{C}$ precisely fills up the power missing in the plain WF estimate, as can be shown by a combined average over random and actual signal and data realizations,
\begin{equation}
	\begin{split}
		\Bigl\langle\bigl\langle \norm{\delta}^\mathrm{C} \, \norm{\delta}^\mathrm{C}\bigr\rangle_\mathrm{R}\Bigr\rangle
		&= \bigl\langle \norm{\delta}^\mathrm{S} \, \norm{\delta}^\mathrm{S}\bigr\rangle \,.
		\label{eq:method:2-point_correlation_of_constrained_realization}
	\end{split}
\end{equation}

\Cref{eq:method:2-point_correlation_of_constrained_realization} also shows that the CRs capture the full 2-point correlations of the actual signal. The same will not hold for higher-order $p$-point correlators, though, since the CRs defined in \cref{eq:method:constrained_realizations_around_wiener_filter} do not sample the full conditional PDF $P\bigl[\norm{\delta}^\mathrm{S}|\norm{\delta}^\mathrm{D}\bigr]$. This also means that the constrained density field realizations can take negative values in very underdense regions -- just like the WF estimate. To alleviate this problem, we can thus apply the same smoothing and, if required, remapping to $\norm{\delta}^\mathrm{C}$ as described in \cref{eq:method:SFB_space_smoothing,eq:method:remapping_of_underdensities} for $\norm{\delta}^W$ in \cref{sec:method:wiener_filter}.

Details on the numerical implementation of generating random log-normal signal and Poisson-sampled data realizations as well as their combination into CRs of the density contrast and peculiar velocity fields according to \cref{eq:method:constrained_realizations_around_wiener_filter} can be found in \cref{sec:generating_random_signal_and_data_realizations}.

%%%%%%%%%%%%%%%%%%%%
\subsection{Velocity-velocity comparison and parameter estimation}
\label{sec:method:parameter_inference}
According to \cref{eq:method:linear_relation_between_normalized_density_contrast_and_peculiar_velocity}, the normalized growth rate $\fsig$ can be inferred by matching the radial peculiar velocity reconstructed from the smoothed density with independently observed velocities (hereafter \vvcomp). In practice, we compare the observed and reconstructed distance moduli since the distance modulus is the actual observable with normally distributed errors in a galaxy distance catalog such as CF3 \citep{tully_cosmicflows-3_2016}. To relate distance modulus $\mu$ and radial peculiar velocity $v_r$, we evaluate $\mu$ at the cosmological redshift $z = z^\obs - v_r/c$ and expand it to linear order in $v_r$,
\begin{equation}
	\mu(z) = 25 + 5 \log_{10}\biggl(\frac{d_\lum(z)}{\mpc*}\biggr) \approx \mu\bigl(z^\obs\bigr) - \eta(z^\obs) \, v_r(\vect{s}) \,.
	\label{eq:method:expected_distance_modulus_linearized_in_peculiar_velocity}
\end{equation}
Here, $d_\lum$ denotes the luminosity distance and $\eta$ is defined as
\begin{equation}
	\eta(z) \coloneqq - \pd{\mu}{v_r}(z) \, \biggr|_{v_r=0} = \frac{5}{\ln(10)} \frac{1}{c \, d_\lum(z)} \, \pd{d_\lum}{z}(z) \,.
	\label{eq:method:distance_modulus_derivative_with_respect}
\end{equation}
To minimize the spatial Malmquist bias effect \citep{aaronson_velocity_1982}, the radial peculiar velocity field $v_r$ in \cref{eq:method:expected_distance_modulus_linearized_in_peculiar_velocity} is not evaluated at the measured distance inferred from $\mu$, which is subject to large errors, but rather at the observed redshift distance, $\vect{s} = (s, \vartheta^\obs, \varphi^\obs)$ with $s = d_\com(z^\obs)$.

The reconstruction of the peculiar velocity field requires knowledge of the mass distribution also outside the survey volume. The velocity dipole component presents a special case. Its value on a shell at radius $r$ is completely independent of the mass distribution beyond $r$ when the peculiar velocity is expressed relative to the reconstructed LG motion, $\vrec_\LG(\vect{r}) = \vrec(\vect{r}) - \vrec(0)$ \citep{nusser_prediction_1994}, where $\vrec$ is defined relative to the CMB frame. From this point of view, it is advantageous to perform the \vvcomp\ using $\vrec_\LG$ rather than $\vrec$. However, the disadvantage of working with $\vrec_\LG(r)$ is that it approaches $-\vLG$ at large $r$. Since this reflex dipole is actually significant, it may dominate the statistical comparison of the velocities at large distance.

In order to avoid this situation, we make the comparison for galaxies beyond certain redshift cuts $\czmin$. Instead of working with $\vrec_\LG$, we express all peculiar velocities relative to the CMB and allow for a constant (bulk) velocity offset $\bext$ between the reconstructed and observed velocities of the galaxies satisfying the cut,
\begin{equation}
	\vect{v}^\text{rec} \rightarrow \vect{v}^\text{rec} + \bext \,.
	\label{eq:method:adding_external_dipole_velocity_contribution}
\end{equation}
This mitigates the effect of the external field and at the same time yields a comparison at the level of velocity fluctuations on top of a constant flow. We stress that $\bext$ only enters at the level of the peculiar velocities. It does not affect the redshifts at which the galaxies are placed. The choice of redshift frame is discussed in \cref{sec:method:redshift_space_distortions}.

Furthermore, we allow the dimensionless Hubble parameter $\hobs$ used in the computation of the observed galaxy distances to vary freely. This allows for a ``breathing mode'' which reflects uncertainties in the determination of the absolute distance scale in the galaxy distance catalog and in determining the actual mean number density of galaxies within the reconstruction volume. In our analysis, $\hobs$ is treated as a nuisance parameter. Together with $\fsig$ there are thus five free parameters to be fixed, which we collect into the parameter vector $\vect{\Theta} = (\fsig, \bext, \hobs)$.

Let $\mu^\obs$ be the vector of the $N^\obs$ observed distance moduli and $\mu^\mathrm{rec}$ that of the respective distance moduli predicted from our reconstructed velocity, $\mu_j^\text{rec} = \mu(z_j^\obs) - \eta(z_j^\obs) \, v_r^\text{rec}(\vect{s}_j)$. Then $\fsig$ is estimated by maximizing the log-likelihood $\mathcal{L}_\mu$ of the difference vector $\Delta \mu = \mu^\obs - \mu^\mathrm{rec}$,
\begin{equation}
	 \mathcal{L}_\mu = -\frac{1}{2} \, \Bigl[\Delta \mu^\top C_{\Delta\mu}^{-1} \Delta \mu + \ln{\det{C_{\Delta\mu}}}\Bigr] \,,
	\label{eq:method:log-likelihood}
\end{equation}
where $C_{\Delta\mu}$ is the covariance matrix of $\Delta \mu$. In principle, we could determine the most probable parameter values $\vect{\Theta}^\mathrm{mp}$ using the WF estimate of $v_r^\rec$ and including its 1-point variance in $C_{\Delta\mu}$. However, this approach would completely ignore the spatial correlations between the peculiar velocities and their scatter. To properly take these into account, we will instead resort to the velocity CRs $\vect{v}^\mathrm{C}$. For each CR, we compute $\vect{\Theta}^\mathrm{mp}(\vect{v}^\mathrm{C})$ by maximizing $ \mathcal{L}_\mu$, assuming that $\vect{v}^\mathrm{C}$ describes the true velocity field. $C_{\Delta\mu}$ is thus just given by the covariance of the observed distances, which is diagonal for CF3,
\begin{equation}
	C_{\Delta\mu} = C_{\mu^\obs} = \mathrm{diag}\Bigl[\bigl(\sigma_{\mu,j}^\obs\bigr)^2\Bigr]_{j = 1}^{N^\obs} \,.
\end{equation}
The actual scatter in the reconstructed velocity field is then accounted for by considering the distribution of $\vect{\Theta}$ over a sufficiently large set of CRs.

The PDF of the parameters is thus described by
\begin{equation}
	P(\vect{\Theta}) = \upi{\vect{v}^\mathrm{C}} \, P\bigl(\vect{\Theta} | \vect{v}^\mathrm{C}\bigr) \, P\bigl(\vect{v}^\mathrm{C}\bigr) \approx \frac{1}{N^\mathrm{C}} \, \sum_{\alpha=1}^{N^\mathrm{C}} \, P\bigl(\vect{\Theta} | \vect{v}^\mathrm{C}_\alpha\bigr) \,,
	\label{eq:method:parameter_likelihood}
\end{equation}
where marginalizing over the distribution $P\bigl(\vect{v}^\mathrm{C}\bigr)$ of velocity field CRs is approximated by averaging over a set of $N^\mathrm{C}$ CRs. $P\bigl(\vect{\Theta} | \vect{v}^\mathrm{C}\bigr)$ is the conditional PDF of the parameters given a specific velocity CR, which we assume to be a Gaussian around the most probable parameter vector $\vect{\Theta}^\mathrm{mp}$ obtained by maximizing the log-likelihood $ \mathcal{L}_\mu$ in \cref{eq:method:log-likelihood}, i.\,e.
\begin{equation}
    \begin{split}
	- 2 \ln{P\bigl(\vect{\Theta} | \vect{v}^\mathrm{C}\bigr)} &= \bigl[\vect{\Theta} - \vect{\Theta}^\mathrm{mp}\bigl(\vect{v}^\mathrm{C}\bigr)\bigr]^\top \, C_\Theta^{-1}\bigl(\vect{v}^\mathrm{C}\bigr) \, \bigl[\vect{\Theta} - \vect{\Theta}^\mathrm{mp}\bigl(\vect{v}^\mathrm{C}\bigr)\bigr] \\
	&\phantom{=} + \ln{\det{C_\Theta\bigl(\vect{v}^\mathrm{C}\bigr)}}.
    \end{split}
\end{equation}
The parameter covariance matrix $C_\Theta(\vect{v}^\mathrm{C})$ is estimated as the inverse of the Fisher matrix,
\begin{equation}
    \begin{split}
	\bigl[C_\Theta^{-1}(\vect{v}^\mathrm{C})\bigr]_{ab} &= \biggl\langle \frac{\partial^2  \mathcal{L}_\mu}{\partial \Theta_a \partial \Theta_b}\biggr\rangle^\obs \, \biggr|_{\vect{\Theta} = \vect{\Theta}^\mathrm{mp}(\vect{v}^\mathrm{C})} \\
	&= \sum_{j=1}^{N^\obs} \, \frac{1}{\bigl(\sigma_{\mu,j}^\obs\bigr)^2} \, \pd{\mu_j^\text{rec}}{\Theta_a} \, \pd{\mu_j^\text{rec}}{\Theta_b} \, \biggr|_{\vect{\Theta} = \vect{\Theta}^\mathrm{mp}(\vect{v}^\mathrm{C})} \,,
    \end{split}
\end{equation}
where $\langle\dotsm\rangle^\obs$ denotes the average over all $N^\obs$ observed group distances.\footnote{The derivative with respect to $\fsig$ is performed numerically via central differencing, since $\fsig$ enters via the linear RSD correction in a non-trivial way. The other derivatives are performed analytically.}

The maximum-likelihood (ML) estimate of $\vect{\Theta}$ is given by the mean of the distribution in \cref{eq:method:parameter_likelihood},
\begin{equation}
	\begin{split}
		\Theta_a^\text{ML} &= \usi{\vect{\Theta}} \, \Theta_a \, P\bigl(\vect{\Theta}\bigr) \\
		&\approx \frac{1}{N^\mathrm{C}} \, \sum_{\alpha=1}^{N^\mathrm{C}} \, \Theta_a^\mathrm{mp}\bigl(\vect{v}^\mathrm{C}_\alpha\bigr) \,.
	\end{split}
	\label{eq:method:optimal_parameter_as_sum_over_realizations}
\end{equation}
The associated $1 \sigma$ errors are obtained from the variance
\begin{align}
	\sigma_{\Theta,a}^2 &= \usi{\vect{\Theta}} \, \Theta_a^2 \, P\bigl(\vect{\Theta}\bigr) - \bigl(\Theta_a^\text{ML}\bigr)^2
	\label{eq:method:parameter_variance_as_sum_over_realizations} \\
	&\approx \frac{1}{N^\mathrm{C}} \, \sum_{\alpha=1}^{N^\mathrm{C}} \, \Bigl(\Theta_a^\mathrm{mp}\bigl(\vect{v}^\mathrm{C}_\alpha\bigr)\Bigr)^2 - \bigl(\Theta_a^\text{ML}\bigr)^2 + \frac{1}{N^\mathrm{C}} \, \sum_{\alpha=1}^{N^\mathrm{C}} \, \Bigl(\sigma_{\Theta,a}^\obs\bigl(\vect{v}^\mathrm{C}_\alpha\bigr)\Bigr)^2 \,,
	\nonumber
\end{align}
with $\bigl(\sigma_{\Theta,a}^\obs\bigl(\vect{v}^\mathrm{C}\bigr)\bigr)^2 = \bigl[C_\Theta\bigl(\vect{v}^\mathrm{C}\bigr)\bigr]_{aa}$. The first two terms in \cref{eq:method:parameter_variance_as_sum_over_realizations} describe the variance of the most probable parameter value $\Theta_a^\mathrm{mp}$ due to the scatter between different CRs. Hence, they quantify the uncertainty caused by shot noise in the galaxy redshift catalog. The last term, on the other hand, is the average variance of parameter $\Theta_a$ per CR, and thus quantifies the uncertainty caused by errors in the observed distance moduli.

%%%%%%%%%%%%%%%%%%%%%%%%%%%%%%%%%%%%%%%%
\section{Redshift survey and peculiar velocity catalog}
\label{sec:data}

%%%%%%%%%%%%%%%%%%%%
\subsection{2MRS}
\label{sec:data:2MRS}
We use CORAS to generate CRs of the density and velocity fields from the latest release of the Two-Micron All-Sky Redshift Survey (2MRS) \citep{huchra_2mass_2012,macri_2mass_2019}. 2MRS is a flux-limited catalog that covers 91\% of the sky and provides redshifts for 44,572 galaxies with a $K_s$-band magnitude of $K_s \leq 11.75$. We fix the outer reconstruction boundary radius to $\rmax = 200 \hmpc$, containing 98\% of the 2MRS galaxies.

In order to mitigate the effect of the fingers-of-god in the reconstruction, we use the corresponding galaxy group catalog of \cite{tully_galaxy_2015}, placing galaxies at the mean redshift of their associated group. Other group catalogs are available \citep[e.\,g.][]{lim_galaxy_2017,lambert_2mass_2020}, but we prefer \cite{tully_galaxy_2015} because it is also used in the grouping of CF3 galaxies, thus ensuring an optimal match between redshift positions in both 2MRS and CF3. Galaxies with negative redshifts and those outside of the reconstruction volume, $s>\rmax$, are excised. We further impose a partial volume limit within a distance $30 \hmpc$ by keeping only galaxies sufficiently bright to be observable if placed at that distance. This reduces potential biases due to selection effects within this local sub-volume that are not properly captured by the $F/T$ selection function estimator described in \cref{sec:selection_function_estimator}.

The statistical isotropy of the observed galaxy distribution is slightly violated by the zone of avoidance (ZOA) near the galactic plane. To reinstate the isotropy, we populate the ZOA with copies of neighbouring galaxies, following the approach of \cite{yahil_redshift_1991}. For this, we separate the survey volume into bins in radius, latitude and longitude, and draw the number of galaxies to be copied into one of the ZOA bins from a Poisson distribution whose mean is given by the mean number of galaxies in the adjacent latitudinal bins. Those galaxies are then assigned the redshift and longitude of one of the galaxies chosen at random from the adjacent bins as well as a random latitude drawn from a uniform distribution.

The choice of $l_\text{max}$ and $n_\text{max}(l)$ in the SFB decomposition of the density and velocity fields is a trade-off between computational speed and spatial resolution. We take $l_\text{max} = 60$ and $n_\text{max}(l)$ such that $k_{nl} \, \rmax \leq 120$, corresponding to a transverse resolution $\pi \, r/60 \approx 10 \, r/\rmax \hmpc$ and a radial resolution $\pi \, \rmax / 120 \approx 5 \hmpc$. With this choice, there are on average less than 0.55 observed galaxies per resolved volume element at any radius. The information gain offered by probing smaller scales is negligible. The WF field on these noise dominated scales would simply approach zero.

%%%%%%%%%%%%%%%%%%%%
\subsection{Cosmicflows-3}
\label{sec:data:CF3}
To constrain the normalized growth rate $\fsig$ and the
bulk flow contribution from sources beyond the 2MRS reconstruction volume, $\bext$, we compare the reconstructed radial peculiar velocities to the directly observed velocities in the galaxy distance catalog Cosmicflows-3 (CF3) \citep{tully_cosmicflows-3_2016}. CF3 is a large compilation of galaxy distances derived using various methods, e.\,g.~the Tully-Fisher relation, the Fundamental Plane and type Ia supernovae luminosities. The catalog contains $\sim 17,700$ entries in total, assigned to over $\sim 11,500$ groups. Whenever possible, these groups have been matched with those in the 2MRS group catalog \citep{tully_galaxy_2015}. There are $\sim 2700$ groups containing at least two members with redshifts and angular positions in 2MRS.

The observed CF3 velocities are obtained from the group-averaged redshifts and distance moduli. This reduces the contribution from incoherent small-scale flows in the \vvcomp. Additionally, it reduces the random uncertainty in the distance modulus of each group typically by $1/\sqrt{N}$, where $N$ is the number of galaxies in the group. We include groups that contain only a single member galaxy. Excising those was found to have an insignificant effect on the estimated parameters, while increasing the errors.

We only consider CF3 groups with redshift velocities $cz \leq 16,000 \kms$, which is the maximal redshift of the major CF3 component from 6dFGS \citep{springob_6df_2014,tully_cosmicflows-3_2016}. The small number of CF3 groups beyond this redshift where found to only have a negligible effect on the inferred value of $\fsig$. We further discard any remaining extreme CF3 outliers deviating by more than $5\sigma$ from the corresponding 2MRS-reconstructed distance moduli computed using $\rsmooth = 5 \hmpc$, Planck-18 cosmology \citep{aghanim_planck_2020-1} and $\hobs = 0.75$ \citep{tully_cosmicflows-3_2016}. The results are found to be robust against changing $\hobs$ to the Planck-18 value of the Hubble constant.

%%%%%%%%%%%%%%%%%%%%%%%%%%%%%%%%%%%%%%%%
\section{Test on semi-analytic mocks}
\label{sec:test_on_mocks}
The ML error estimate in \cref{eq:method:parameter_variance_as_sum_over_realizations} includes only the observational uncertainties due to shot noise in 2MRS and distance errors in CF3. It does not, however, account for errors resulting from the inability of our linear reconstruction method to recover the nonlinear galaxy velocities accurately and precisely. To test CORAS and quantify these errors, we generate realistic mock catalogs that reproduce the general survey characteristics of 2MRS and CF3 as closely as possible. For this, we use the $z=0$ snapshot of the \textsc{MultiDark} simulation run MDPL2 \citep{klypin_multidark_2016,riebe_multidark_2013}, which has a box side length of $1 \hgpc$, a mass resolution of $2.23 \cdot 10^9 \, M_{\astrosun}$ and provides a \textsc{RockStar} \citep{behroozi_rockstar_2013} halo catalog as well as different semi-analytic galaxy catalogs. Of those we choose to use the \textsc{sage} catalog, as it was found to represent the observed stellar mass function by far the best \citep{knebe_multidark-galaxies_2018}. This is important for us, since we will use the stellar mass as a proxy for absolute $K_s$ magnitude of the mock galaxies.

For all tests on the mocks, we adopt the cosmological parameters of the MDPL2 simulation,\footnote{$\Omega_\mathrm{m} = 0.307115$, $\Omega_\mathrm{b} = 0.048206$, $h = 0.6777$, $n_\mathrm{s} = 0.96$, $\siglin = 0.8228$}, which are compatible with the Planck-13 results \citep{ade_planck_2014-1}. In the WF, we use the Cosmic Emu power spectrum \citep{heitmann_miratitan_2016} derived from these parameters. The value of the nonlinear $\sig$ computed from this power spectrum is $\sig = 0.910$. We  point out that this differs by $\sim 4 \%$ from the value of 0.95 computed from the MDPL2 particle distribution in \cite{hollinger_assessing_2021}. We do not know the reason for this minor disagreement, but for consistency with the analysis of the actual data we will use the value computed from the Cosmic Emu power spectrum.

%%%%%%%%%%%%%%%%%%%%
\subsection{Mock catalog generation}
\label{sec:test_on_mocks:mock_generation}
We start by identifying spherical subvolumes of MDPL2 which closely represent the LG environment. We use the following set of criteria:
\begin{enumerate}
	\item Centered on an LG-like group of galaxies with minimal individual halo masses of $\SI{5e11} \, M_{\astrosun}$, a combined group halo mass of $1.5 - \SI{2.5e12} \, M_{\astrosun}$ and a maximal group radius of $4 \mpc$
	\item Center-of-mass LG velocity $\mvLG* = 500 - 700 \kms$
	\item No halos of mass above $\SI{e14}{M_{\astrosun}}$ within $13 \mpc$ distance
	\item Exactly one Virgo-like halo of mass $0.5 - \SI{1.5e15} \, M_{\astrosun}$ within a distance of $13 - 20 \mpc$
	\item LG infall velocity towards Virgo of $0 - 400 \kms$
	\item Angle between $\mvLG$ and Virgo direction of $35 - 50^\circ$
	\item Difference between $\mvLG$ and the velocity field at the origin, smoothed with a $5 \hmpc$ Gaussian, less than $100 \kms$
	\item Bulk flow of the smoothed velocity field in a $30 \hmpc$ sphere matching the observed value found in \cite{nusser_cosmological_2011} within $\pm 100 \kms$ ($\approx 3 \sigma$) in each galactic coordinate direction
	\item A minimum distance between different mock LGs of $200 \hmpc$ to avoid significant overlap between mock reconstruction volumes
\end{enumerate}
There are 17 subvolumes satisfying these criteria. For each of these, the velocity of the LG frame of reference is taken as the smoothed velocity field at the origin, rather than $\mvLG$. This helps to remove any large incoherent velocity component of the mock LG.

\begin{figure}
	\centering
	\includegraphics[width=\linewidth]{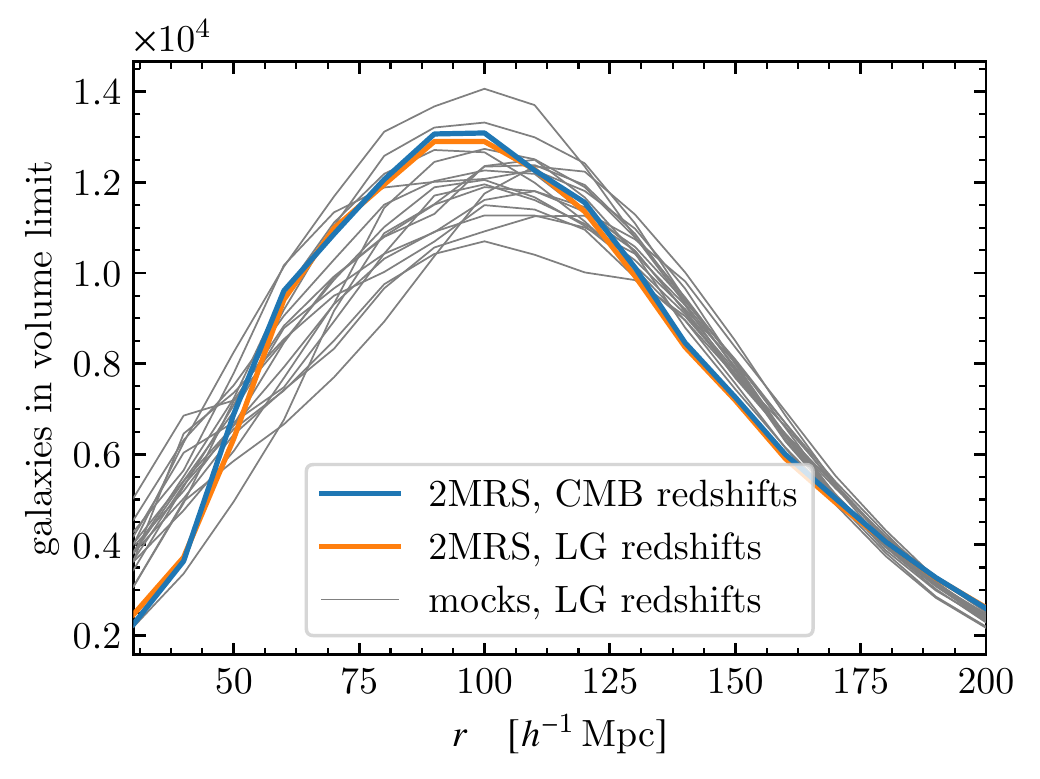}
	\caption{Radial dependence of the number of galaxies contained in volume-limited subsamples of 2MRS (thick lines) and the set of mock catalogs (thin grey lines). The two thick lines for 2MRS correspond to the CMB (blue) and LG (orange) frame redshifts, and are nearly identical. The same holds for the mocks, and only the LG redshift results are plotted for them.}
	\label{fig:galaxy_number_in_volume_limited_subsamples}
\end{figure}

To construct the mock 2MRS catalogs for each of the identified subvolumes, we assign each galaxy a cosmological redshift as well as observed CMB and LG frame redshifts based on their comoving distance and peculiar velocity. We estimate their absolute $K_s$-band magnitude based on their stellar mass $\mathcal{M}_*$,
\begin{equation}
	M_{K_s} = M_{K_s,\astrosun} - 2.5 \biggl(\log_{10} \mathcal{M}_* - \log_{10} \frac{\mathcal{M}}{L_{K_s}}\biggr) \,,
\end{equation}
where the absolute solar magnitude is $M_{K_s,\astrosun} = 3.27$ \citep{willmer_absolute_2018} and the logarithm of the stellar mass-to-light ratio was drawn from a normal distribution with mean and standard deviation $\log_{10}\mathcal{M} / L_{K_s} = -0.85 \pm 0.1$. The latter is compatible with the findings in \cite{bell_optical_2003} and approximately reproduces the luminosity distribution of 2MRS. This is illustrated by the number of galaxies above the luminosity threshold for different volume-limiting radii shown in \cref{fig:galaxy_number_in_volume_limited_subsamples} for both 2MRS and the mocks. The apparent $K_s$ magnitudes are then obtained via \cref{eq:selection_function_estimator:absolute_apparent_magnitude_relation}, where we use the same $k$-correction and luminosity evolution correction as for 2MRS. In addition, we rotate each MDPL2 subvolume such that the direction of Virgo and the plane it spans with the LG velocity vector match those in the actual Universe. This then defines a mock galactic plane, allowing us to simulate the ZOA. The mock 2MRS catalogs are finally compiled from all galaxies outside of the ZOA satisfying the 2MRS flux limit $K_s \leq 11.75$.

Mock 2MRS group catalogs are then created by combining all mock 2MRS galaxies sharing the same main \textsc{RockStar} halo into a group. In analogy to \cite{tully_galaxy_2015}, group redshifts are defined as unweighted averages over the group members, while the angular group positions are obtained by a luminosity weighted average. Note that this method of grouping is idealized, as it is based on the true \rspace\ positions of galaxies, while the actual 2MRS galaxies are grouped based on their \sspace\ positions. We expect this effect to be small, though.

\begin{figure*}
	\centering
	\includegraphics[width=\textwidth]{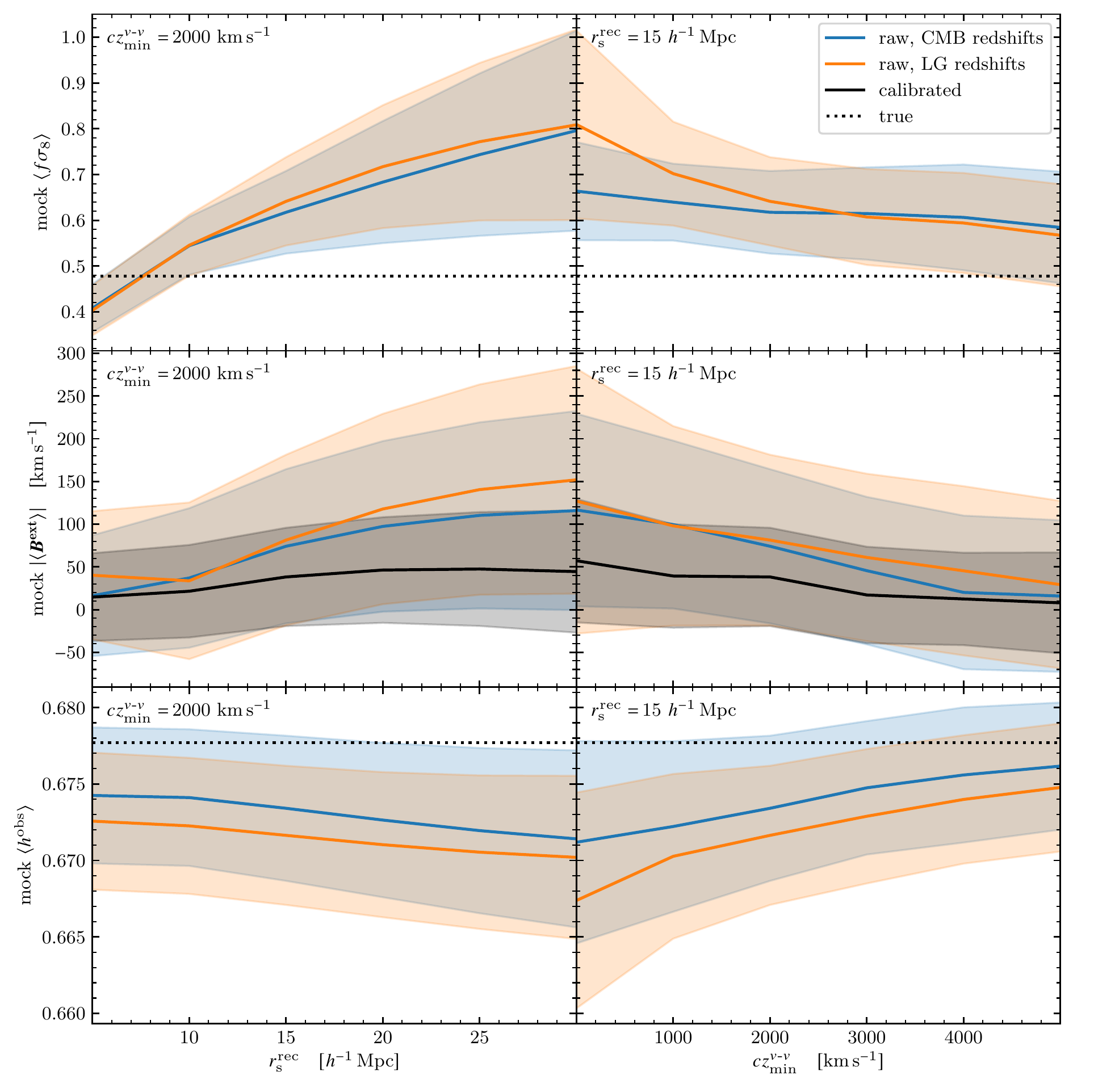}
	\caption{Average values of $\fsig$ (top), $\bext$ (middle) and $\hobs$ (bottom) inferred via \vvcomp\ from the set of mock catalogs as a function of the smoothing scale $\rsmooth$ (left) and lower redshift cutoff $\czmin$ (right). The blue and orange lines show the raw ML estimates obtained by placing galaxies at their CMB and LG frame redshifts, respectively. For $\bext$, the additional black line shows the calibrated value. The shaded areas mark the $1 \sigma$ scatter between individual mocks. For reference, the dotted lines in the top and bottom panels mark the respective true mock parameter values, $\fsig^\mathrm{true} = 0.4779$ and $\hobs{}^\mathrm{,true} = 0.6777$.}
	\label{fig:inferred_mock_parameters}
\end{figure*}

Compared to 2MRS, generating realistic mock CF3 catalogs is more difficult since CF3 is a compendium of all available galaxy distances and thus has no simple selection criteria. While one could choose to reproduce the observed redshift or distance distributions of CF3 as closely as possible, this might lead to unwanted selection biases if the respective distributions of mock galaxies are too different from those in the actual Universe. For example, such an approach could preferably select mock galaxies with unusually large positive or negative peculiar radial velocities at certain distances, to compensate for a difference in simulated and actual distributions of cosmological redshifts.

To avoid such biases, we use a selection criterion based on the richness of galaxy groups instead. We assume that distance measurements will be available for any sufficiently rich galaxy group, i.\,e.~any group containing some minimal number of observable galaxies. Specifically, we include any group of at least 3 mock 2MRS galaxies into our mock CF3 catalog. Additionally, we include all groups within a distance of $10 \hmpc$ to the mock LG. Each of the remaining groups is included with a probability chosen such that on average the total number of groups and the relative distribution of group richness in the actual CF3 catalog is reproduced. Observed distance moduli of groups are given by the unweighted averages over their members, with the distance modulus of each member galaxy being drawn from a normal distribution of width 0.43 around the true value, which is the mean uncertainty on single-galaxy distance moduli in CF3. The resulting mock distance-redshift distributions show good qualitative agreement with the distribution in the actual CF3 catalog.

Finally, we apply all data preparation steps described in \cref{sec:data:2MRS,sec:data:CF3} to the 17 mock 2MRS and CF3 catalogs.

%%%%%%%%%%%%%%%%%%%%
\subsection{Parameter estimation}
\label{sec:test_on_mocks:test_of_parameter_inference}
We apply the ML estimator described in \cref{sec:method:parameter_inference} to infer the parameters $\vect{\Theta} = (\fsig, \bext, \hobs)$ via the \vvcomp\ between the mock 2MRS and CF3 catalogs. We investigate the dependence of the results on the following choices:
\begin{itemize}
    \item placing the galaxies at their observed CMB or LG frame redshifts
    \item Gaussian smoothing of $\vrec$ on scales $5 \hmpc \leq \rsmooth \leq 30 \hmpc$
    \item imposing lower redshift cutoffs $0 \leq \czmin \leq 5000 \kms$ in the comparison
\end{itemize}
All results will be plotted as a function of $\rsmooth$ for fixed $\czmin = 2000 \kms$ and as a function of $\czmin$ for fixed $\rsmooth = 15 \hmpc$. The fixed value of $\czmin$ is chosen to exclude the region containing the highly coherent flow of nearby galaxies with the LG in the real Universe \citep{sandage_redshift-distance_1986,tully_our_2008}, which can create a mismatch between velocities reconstructed using either CMB or LG redshifts. The fixed value of $\rsmooth$ is chosen to smooth over small-scale flows violating our assumption of linear theory.

\Cref{fig:inferred_mock_parameters} shows the results averaged over the set of mocks, $\langle\fsig\rangle$, $|\langle\bext\rangle|$ and $\langle\hobs\rangle$, as well as their $1 \sigma$ scatter between individual mocks. For all parameters we observe good agreement between the CMB and LG redshift results within the $1 \sigma$ scatter. There is also only an insignificant dependence on $\czmin$, in particular for $\czmin \geq 2000 \kms$, where coherent local flows play no role.

For the normalized growth rate, $\langle\fsig\rangle$, we observe a significant increase with $\rsmooth$, though. This can be attributed to comparing smoothed reconstructed with raw observed velocities, which introduces an $\rsmooth$-dependent bias on the inferred value of $\fsig$ \citep{davis_comparison_1996,berlind_biased_2000,nusser_biasing_2020}. For reference, we marked the true mock value $\fsig^\mathrm{mock,true} = 0.4779$ with a dotted line. It crosses our estimate at $\rsmooth \approx 7.5 \pm 2 \hmpc$. Interestingly, this is larger than the crossing scale of $4-5 \hmpc$ found in \cite{berlind_biased_2000,carrick_cosmological_2015,hollinger_assessing_2021} by directly comparing simulated galaxy or halo velocities with the linear velocity prediction obtained from the Gaussian-smoothed density field of those tracers. While we work in \sspace\ to obtain the crossing scale, these authors use \rspace\ density fields, which might change this scale. Comparing with the noisy observed distance moduli rather than the true peculiar velocities directly could also affect the inferred crossing scale. Further investigations would be necessary to pinpoint the reason. In any case, we do not rely on this crossing scale, which indeed can be coincidental since linear theory is not expected to apply on these scales. Since we apply exactly the same analysis to both real and mock data, we directly use the full smoothing-scale dependence of $\fsig$ found in the mocks to calibrate the estimate of $\fsig$ inferred from the actual 2MRS and CF3 catalogs.

For the amplitude of the average bulk velocity contribution from sources beyond the reconstruction volume, $|\langle\bext\rangle|$, we also observe an increase with $\rsmooth$, albeit less significant. This $\rsmooth$-dependence of $\bext$ is a consequence of its implicit $\fsig$-dependence. As the $\fsig$ estimate grows with $\rsmooth$, so does the amplitude of the reconstructed survey-internal contribution to the bulk flow, $\bint$. To pertain a good fit to the observed total bulk flow, the external contribution $\bext$ changes accordingly,
\begin{equation}
    \bext(\fsig) + \bint(\fsig) \approx \bext(\fsig') + \bint(\fsig') \,.
    \label{eq:test_on_mocks:external_bulk_flow_correction}
\end{equation}

Since the mocks were only selected based on the flow close to the (mock) LG, their true $\bext$ is expected to be approximately randomly distributed between different mocks. In fact, we see that for $\rsmooth = 7.5 \hmpc$, where the $\fsig$ estimate happens to be closest to its true value, $\langle\bext\rangle$ is compatible with zero at the $1 \sigma$ level. The increase with larger $\rsmooth$ is the imprint of the general alignment of the \emph{internal} contribution $\bint$ with the (mock) LG velocity. We can correct for the $\fsig$-dependence by translating the raw ML estimates of $\bext$ via \cref{eq:test_on_mocks:external_bulk_flow_correction} to $\bext(\fsig^\text{true})$. We use the reconstructed internal bulk flow contributions $\bint$ within a sphere of radius $100 \hmpc$ for this, as this volume is large compared to all considered smoothing scales but also excludes the shot noise-dominated outer region of the reconstruction volume. In addition, we take the average of the calibrated results obtained for CMB and LG frame redshifts, treating the small deviation between those as a systematic error contribution. The resulting calibrated $|\langle\bext\rangle|$ is shown in the middle panel of \cref{fig:inferred_mock_parameters}. It is indeed nearly independent of both $\rsmooth$ and $\czmin$, and compatible with zero.

The average result for the dimensionless Hubble parameter, $\hobs$, shows no notable $\rsmooth$-dependence. It is slightly smaller than the actual value used in the MDPL2 simulation, which we marked with a dotted line, but still agrees within less than $\sim 1.5 \sigma$.

%%%%%%%%%%%%%%%%%%%%
\subsection{Reconstructed fields}
\label{sec:test_on_mocks:test_of_field_reconstruction}
We now compare the WF estimates of $\drec$ and $\vrec$ with the true simulated fields, both smoothed with a $5 \hmpc$ Gaussian. The reconstructed fields are obtained using LG redshifts, the true value $\fsig^\text{true}$ of the normalized growth rate employed in MDPL2 and the calibrated external bulk flow contribution $\bext(\fsig^\text{true})$ inferred for each mock at $\rsmooth = 15 \hmpc$ and $\czmin = 2000 \kms$.

\begin{figure*}
	\centering
	\includegraphics[width=\textwidth]{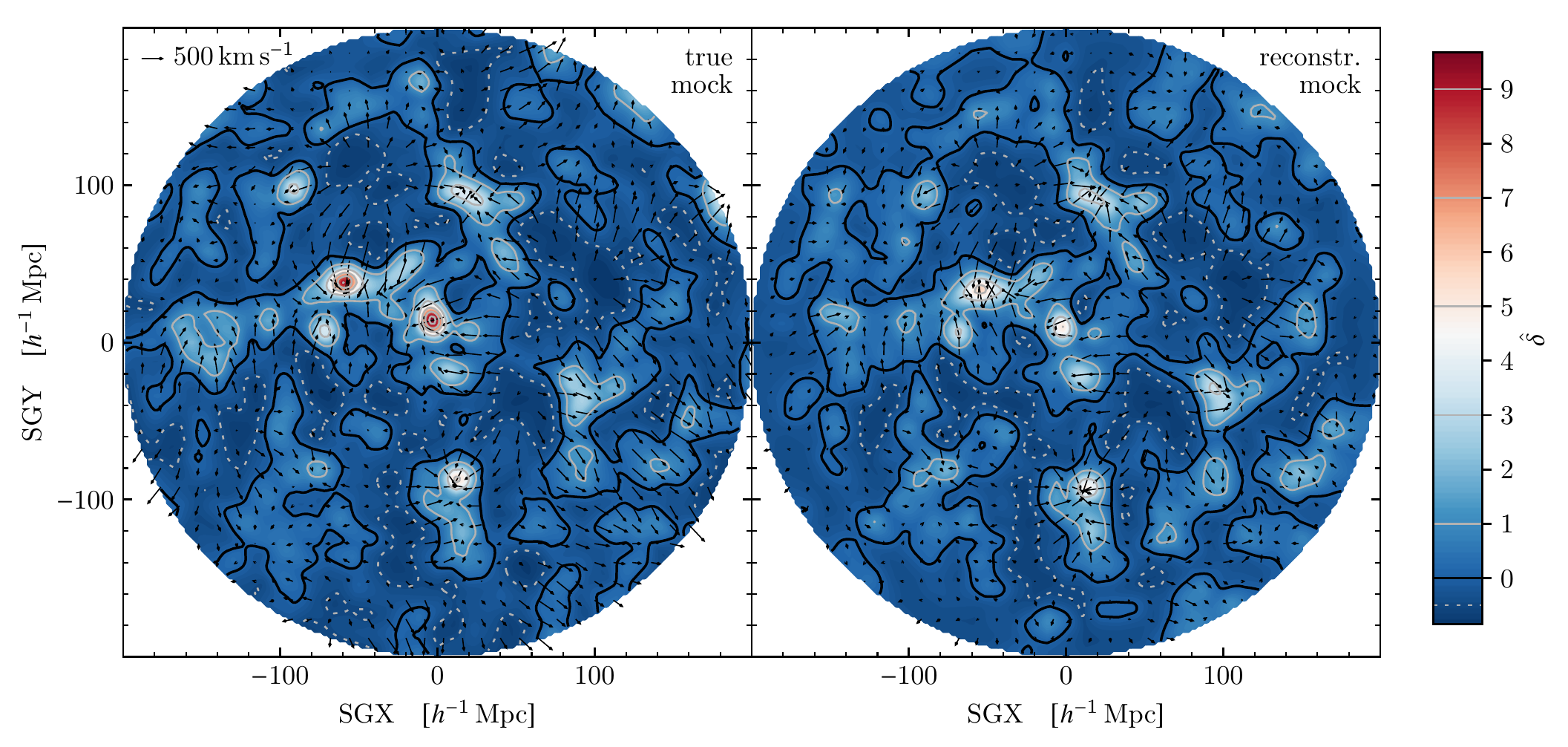}
	\caption{True (left) and reconstructed (right) normalized density contrast (heat map and contours) and peculiar velocity (arrows) in the supergalactic plane for one example of a mock. The contours show the values of $\hat{\delta}$ as marked in the colour bar. An arrow corresponding to $500 \kms$ is given as reference.}
	\label{fig:SGP_reconstructions}
\end{figure*}

In \cref{fig:SGP_reconstructions} we plot both fields in a slice through the mock supergalactic plane for one example of a mock. We find overall very good visual agreement for both the normalized density and the velocity field, with the largest differences being seen towards larger radii, as the uncertainty due to shot noise increases. All other mocks show a similar degree of visual agreement.

\begin{figure}
	\centering
	\includegraphics[width=\linewidth]{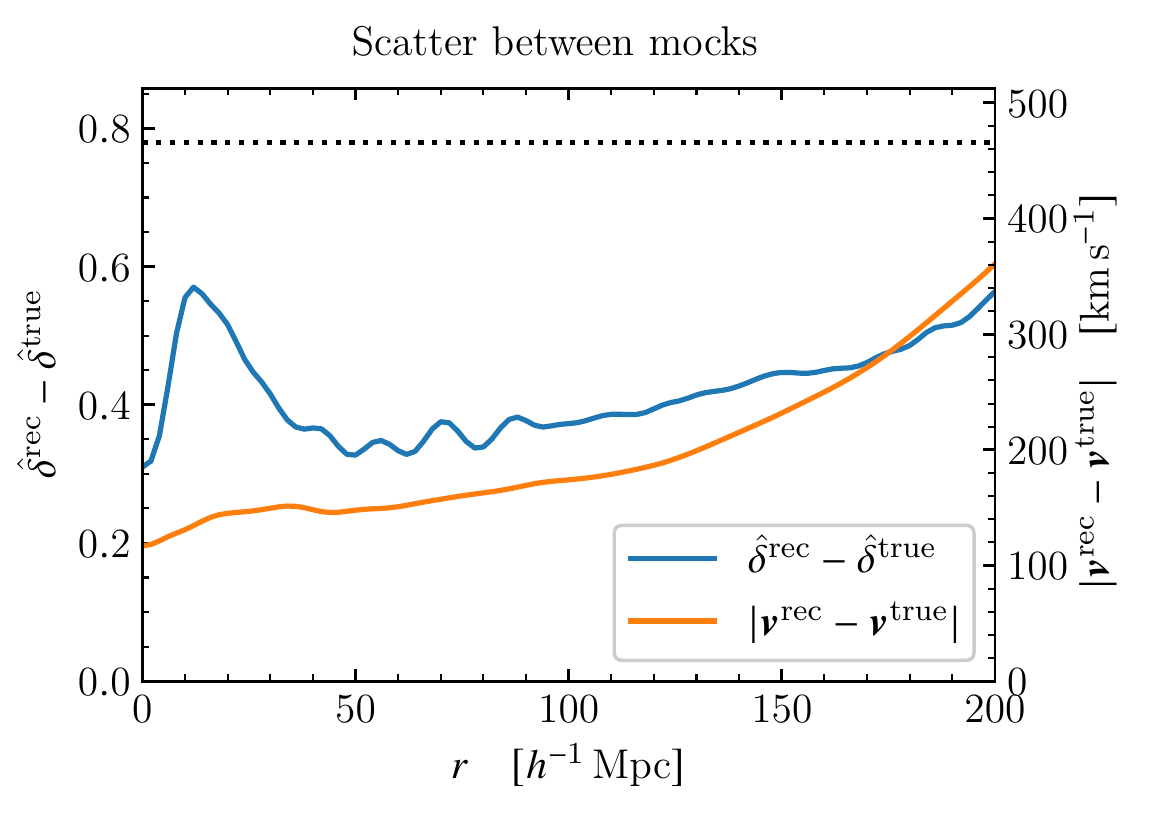}
	\caption{Residual standard deviations between reconstructed and true normalized density contrast (left axis, blue line) and peculiar velocity (right axis, orange line) over the set of mocks, shown as a function of radius. The dotted horizontal line marks the square roots of both fields' mock cosmic variances, $\sigma^\text{cos}_{\norm{\delta}} = 0.78$ and $\sigma_v^\text{cos} = 466 \kms$.}
	\label{fig:solid_angle_averaged_residual_fields}
\end{figure}

To quantify the intrinsic errors of the reconstruction method, we compute the residual variance between reconstructed and true fields over the set of mocks. We then average it over the full solid angle and take the square root. \Cref{fig:solid_angle_averaged_residual_fields} plots the results for both $\norm{\delta}$ and $v$. For reference, we also mark the square root of their mock cosmic variances, $\sigma_{\norm{\delta}}^\text{cos} = 0.78$ and $\sigma_v^\text{cos} = 466 \kms$. The error in $\norm{\delta}$ increases from $\sim 0.3$ at the origin to $\sim 0.55$ at $\rmax = 200 \hmpc$ ($\sim 0.4 \, \sigma^\text{cos}_{\norm{\delta}}$ to $0.7 \, \sigma^\text{cos}_{\norm{\delta}}$). It furthermore shows a peak around $r = 14 \hmpc$ caused by the mock Virgo overdensity being generally underestimated in the reconstruction. The error in $v$ does not display any notable features and increases from $\sim 100 \kms$ at the origin towards $\sim 350 \kms$ at the boundary ($\sim 0.2 \, \sigma^\text{cos}_v$ to $0.75 \, \sigma^\text{cos}_v$).

%%%%%%%%%%%%%%%%%%%%%%%%%%%%%%%%%%%%%%%%
\section{Parameter constraints from the velocity-velocity comparison}
\label{sec:parameter_constraints}
In the following, we apply the ML estimator described in \cref{sec:method:parameter_inference} to infer the parameters $\vect{\Theta} = (\fsig, \bext, \hobs)$ via the \vvcomp\ between the actual 2MRS and CF3 catalogs. The resulting ML estimates and their associated $1 \sigma$ ML errors, \cref{eq:method:optimal_parameter_as_sum_over_realizations,eq:method:parameter_variance_as_sum_over_realizations}, are plotted in \cref{fig:real_parameter_fsigma8_vext_and_h} for the same choices of redshift reference frame, $\rsmooth$ and $\czmin$ as for the mocks in \cref{sec:test_on_mocks:test_of_parameter_inference}.

%%%%%%%%%%%%%%%%%%%%
\subsection[Normalized growth rate $\fsig$]{Normalized growth rate $\boldsymbol{\fsig}$}
\label{sec:parameter_constraints:fsigma8}
\begin{figure*}
	\centering
	\includegraphics[width=\textwidth]{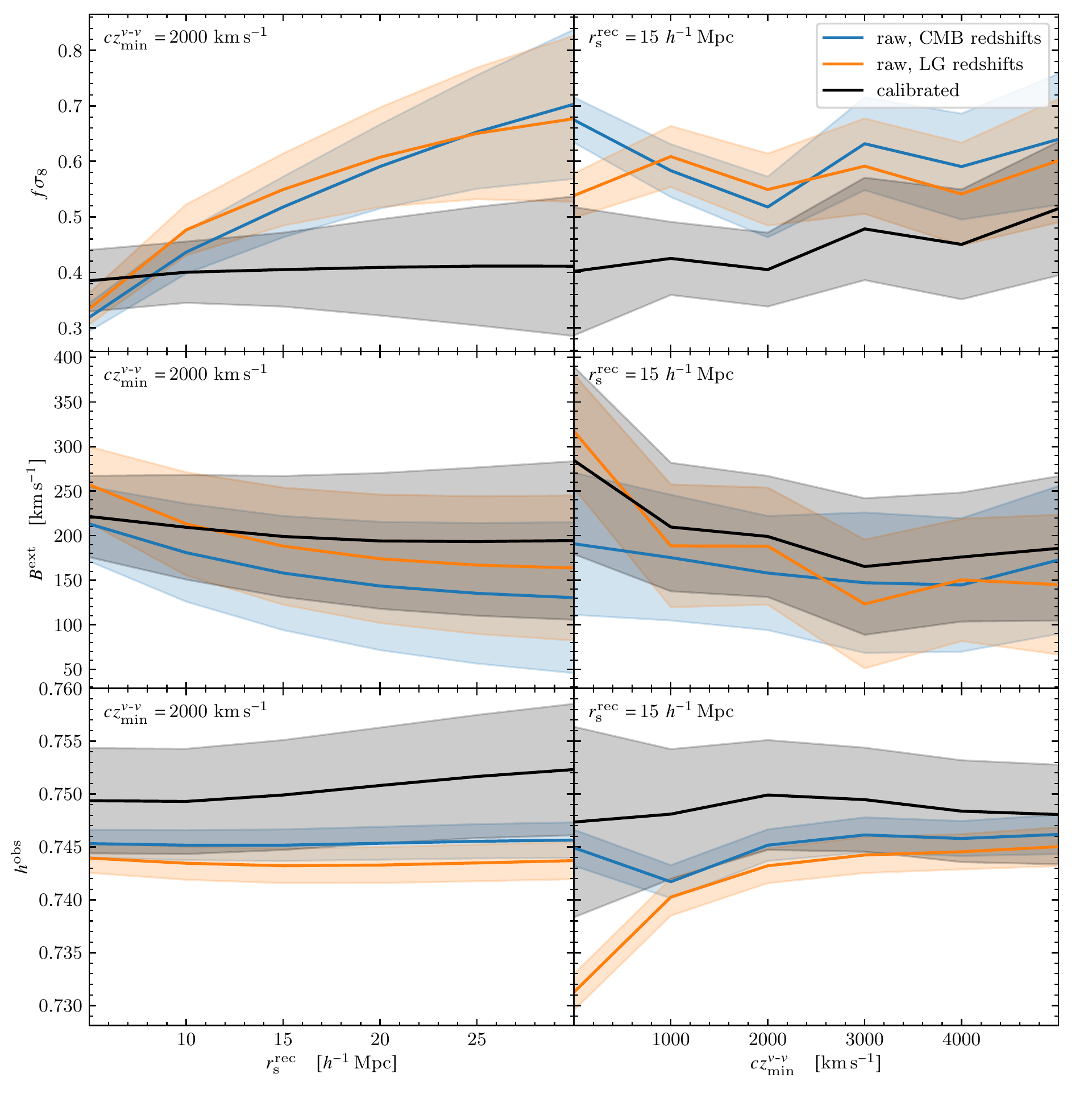}
	\caption{
	Values of $\fsig$ (top), $\bext$ (middle) and $\hobs$ (bottom) inferred via \vvcomp\ between 2MRS and CF3 as a function of the smoothing scale $\rsmooth$ (left) and lower redshift cutoff $\czmin$ (right). The blue and orange lines show the raw ML estimates obtained by placing galaxies at their CMB and LG frame redshifts, respectively. The black lines show the calibrated values. The shaded areas mark the $1 \sigma$ error.}
	\label{fig:real_parameter_fsigma8_vext_and_h}
\end{figure*}

The results for $\fsig$ are presented in the top panels of \cref{fig:real_parameter_fsigma8_vext_and_h}. The results for CMB and LG redshifts generally agree within $1 \sigma$, except for $\czmin < 1000 \kms$. This deviation can be attributed to the aforementioned highly coherent flow of nearby galaxies with the LG. While there is no clear dependence on $\czmin$, we observe the same increase with $\rsmooth$ as in the mocks.

To reduce the $\rsmooth$-dependent bias, we calibrate these results by multiplying them with the ratio $\fsig^\mathrm{true} / \langle \fsig\rangle$ found for the mocks. Here, $\fsig^\mathrm{true} = 0.4779$ is the true value of the (nonlinear) normalized growth rate employed in the MDPL2 simulation \citep{klypin_multidark_2016}. The results for CMB and LG frame redshifts show the same good level of agreement as the uncalibrated results. Therefore, we use their average as our overall calibrated estimate of $\fsig$, and treat the small deviation between the two reference frames as a systematic error contribution. The result is plotted as the black line in the figure and shows no notable dependence on $\rsmooth$. Only the estimated total $1 \sigma$ error grows with $\rsmooth$. The values for different $\czmin$ are also compatible with each other within their $1 \sigma$ errors.

For our final calibrated estimate of $\fsig$, we adopt $\czmin = 2000 \kms$, as this excludes the highly coherent flow near the LG \citep{sandage_redshift-distance_1986,tully_our_2008}, which is not accurately reproduced in the mocks. The excluded fraction of CF3 groups within $2000 \kms$ is $\sim 7.5 \%$. Increasing $\czmin$ any further would dismiss more groups than necessary and increase the overall error on $\fsig$.

Regarding the smoothing scale, the smallest overall error estimates are found from $\rsmooth = 5$ to $10 \hmpc$, matching the range of scales found to approximately match the true growth rate in the mocks. At the same time, however, our linear theory assumption does not correctly describe the nonlinear flows appearing on these small scales. In principle, our calibration should approximately capture the systematic errors arising due to the linear reconstruction method. Even though, it is prudent to avoid relying on the calibration on scales where nonlinear evolution and incoherent motions  play an important role. For $\rsmooth \geq 15 \hmpc$, linear theory is a better approximation and we expect the reconstruction errors to be captured more reliably. Hence, we choose to adopt the calibrated value for $\rsmooth = 15 \hmpc$ as our conservative final estimate of $\fsig$, given in \cref{tab:parameter_values}. For reference, we also list the results for $\rsmooth = 5$ and $10 \hmpc$ in the same table.

Assuming the Planck-18 cosmology \citep{aghanim_planck_2020-1}, we can furthermore translate the calibrated $\fsig$ to its linear value $\fsiglin$, by multiplying $\fsig$ with the ratio of the linear and nonlinear matter density fluctuation amplitudes, $\siglin / \sig = 0.905$, where $\sig$ has been computed from the adopted Cosmic Emu power spectrum. The result is also given in \cref{tab:parameter_values}.

\begin{table*}
	\centering
	\caption{Calibrated parameter values and errors inferred via \vvcomp\ between 2MRS and CF3, for a lower redshift cutoff $\czmin = 2000 \kms$ and different smoothing scales $\rsmooth$. The parameters are the nonlinear and linear calibrated normalized growth rate $\fsig$ and $\fsiglin$, bulk velocity contribution $\bext$ from sources beyond the 2MRS reconstruction volume, and dimensionless Hubble parameter $\hobs$ used to compute the observed CF3 distances. The velocities are given in galactic coordinates and relative to the CMB frame. The different sources of error are: 2MRS shot noise, CF3 distance errors, and systematic errors due to choice of redshift reference frame and, for $\fsig$, $\fsiglin$ and $\hobs$, also due to scatter in the calibration against the mocks. The results for $\rsmooth = 15 \hmpc$ are adopted as our conservative final parameter estimates.}
	
	\begin{tabularx}{\linewidth}{Y X Y Y Y Y Y}
		\toprule
		$\rsmooth \; [\hmpc*]$ & parameter & value & 2MRS error & CF3 error & systematic error & total error \\
		\midrule
		& $\fsig$ & 0.385 & $\pm 0.016$ & $\pm 0.027$ & $\pm 0.046$ & $\pm 0.056$ \\
		& $\fsiglin$ & 0.348 & $\pm 0.014$ & $\pm 0.025$ & $\pm 0.041$ & $\pm 0.050$ \\
		\cmidrule{2-7}
		& $\bext*_x \; [\kms*]$ & 100 & $\pm 36$ & $\pm 16$ & $\pm 8$ & $\pm 40$ \\
		& $\bext*_y \; [\kms*]$ & -188 & $\pm 40$ & $\pm 16$ & $\pm 20$ & $\pm 48$ \\
		5 & $\bext*_z \; [\kms*]$ & 59 & $\pm 38$ & $\pm 13$ & $\pm 3$ & $\pm 40$ \\
		\addlinespace
		& $\bext* \; [\kms*]$ & 221 & $\pm 39$ & $\pm 16$ & $\pm 17$ & $\pm 46$ \\
		& $l^\ext \; [{}^\circ]$ & 298 & $\pm 10$ & $\pm 4$ & $\pm 3$ & $\pm 11$ \\
		& $b^\ext \; [{}^\circ]$ & 15 & $\pm 11$ & $\pm 4$ & $\pm 1$ & $\pm 11$ \\
		\cmidrule{2-7}
		& $100 \, \hobs$ & 74.94 & $\pm 0.07$ & $\pm 0.12$ & $\pm 0.48$ & $\pm 0.50$ \\
		\midrule
		& $\fsig$ & 0.400 & $\pm 0.025$ & $\pm 0.028$ & $\pm 0.041$ & $\pm 0.055$ \\
		& $\fsiglin$ & 0.362 & $\pm 0.022$ & $\pm 0.025$ & $\pm 0.037$ & $\pm 0.050$ \\
		\cmidrule{2-7}
		& $\bext*_x \; [\kms*]$ & 97 & $\pm 50$ & $\pm 16$ & $\pm 8$ & $\pm 53$ \\
		& $\bext*_y \; [\kms*]$ & -180 & $\pm 55$ & $\pm 18$ & $\pm 17$ & $\pm 60$ \\
		10 & $\bext*_z \; [\kms*]$ & 44 & $\pm 53$ & $\pm 14$ & $\pm 1$ & $\pm 54$ \\
		\addlinespace
		& $\bext* \; [\kms*]$ & 209 & $\pm 54$ & $\pm 17$ & $\pm 15$ & $\pm 59$ \\
		& $l^\ext \; [{}^\circ]$ & 298 & $\pm 14$ & $\pm 5$ & $\pm 3$ & $\pm 15$ \\
		& $b^\ext \; [{}^\circ]$ & 12 & $\pm 15$ & $\pm 4$ & $\pm 1$ & $\pm 16$ \\
		\cmidrule{2-7}
		& $100 \, \hobs$ & 74.93 & $\pm 0.09$ & $\pm 0.12$ & $\pm 0.48$ & $\pm 0.50$ \\
		\midrule
		& $\fsig$ & 0.405 & $\pm 0.034$ & $\pm 0.029$ & $\pm 0.049$ & $\pm 0.067$ \\
		& $\fsiglin$ & 0.367 & $\pm 0.031$ & $\pm 0.027$ & $\pm 0.044$ & $\pm 0.060$ \\
		\cmidrule{2-7}
		& $\bext*_x \; [\kms*]$ & 95 & $\pm 58$ & $\pm 16$ & $\pm 6$ & $\pm 61$ \\
		& $\bext*_y \; [\kms*]$ & -172 & $\pm 64$ & $\pm 19$ & $\pm 20$ & $\pm 70$ \\
		15 & $\bext*_z \; [\kms*]$ & 28 & $\pm 62$ & $\pm 14$ & $\pm 2$ & $\pm 63$ \\
		\addlinespace
		& $\bext* \; [\kms*]$ & 199 & $\pm 63$ & $\pm 19$ & $\pm 18$ & $\pm 68$ \\
		& $l^\ext \; [{}^\circ]$ & 299 & $\pm 17$ & $\pm 5$ & $\pm 3$ & $\pm 18$ \\
		& $b^\ext \; [{}^\circ]$ & 8 & $\pm 18$ & $\pm 4$ & $\pm 1$ & $\pm 19$ \\
		\cmidrule{2-7}
		& $100 \, \hobs$ & 74.99 & $\pm 0.11$ & $\pm 0.12$ & $\pm 0.50$ & $\pm 0.52$ \\
		\bottomrule
	\end{tabularx}
	
	\label{tab:parameter_values}
\end{table*}

%%%%%%%%%%%%%%%%%%%%
\subsection{External bulk flow contribution}
\label{sec:parameter_constraints:velocities}
We now turn to the ML estimate of $\bext$, describing the
bulk flow contribution from sources beyond the 2MRS reconstruction volume. Its absolute value $\bext*$ is plotted in the middle panels of \cref{fig:real_parameter_fsigma8_vext_and_h}. The results obtained for CMB and LG redshifts agree well within the $1 \sigma$ ML errors, and are only weakly dependent on $\rsmooth$ and $\czmin \geq 1000 \kms$. The same is true for the direction of $\bext$. As described in \cref{sec:test_on_mocks:test_of_parameter_inference} for the mocks, we correct for the implicit $\fsig$ dependence of the raw $\bext$ estimate via \cref{eq:test_on_mocks:external_bulk_flow_correction}, using $\bint$ within a $100 \hmpc$ sphere, and translate $\bext$ for each $\rsmooth$ and $\czmin$ to the respective calibrated value of $\fsig$. The result is plotted as the black line and displays only a negligible remaining $\rsmooth$-dependence. As for $\fsig$, we adopt the calibrated value of $\bext$ for $\rsmooth = 15 \hmpc$ as our conservative final estimate. Together with the results for $\rsmooth = 5$ and $10 \hmpc$ it is given in \cref{tab:parameter_values}. The value is consistent with those found in \citep{carrick_cosmological_2015,boruah_cosmic_2020,stahl_peculiar-velocity_2021} for the external bulk flow contribution relative to the redshift compilation 2M++ \citep{lavaux_2m++_2011}, restricted to approximately the same reconstruction volume we consider.\footnote{2M++ combines 2MRS with the 6dF galaxy redshift survey DR3 \citep{jones_6df_2009} and the Sloan Digital Sky Survey (SDSS) DR7 \citep{abazajian_seventh_2009}. In \cite{carrick_cosmological_2015,boruah_cosmic_2020,stahl_peculiar-velocity_2021} a maximal reconstruction radius of $\rmax = 200 \hmpc$ is used for the regions covered by 6dF and SDSS, and $\rmax = 125 \hmpc$ for the rest.}

%%%%%%%%%%%%%%%%%%%%
\subsection[Dimensionless Hubble parameter $\hobs$ in Cosmicflows-3]{Dimensionless Hubble parameter $\boldsymbol{\hobs}$ in Cosmicflows-3}
\label{sec:parameter_constraints:hubble_parameter}
The ML estimate of the dimensionless Hubble parameter $\hobs$ used to compute the CF3 distances is shown in the bottom panels of \cref{fig:real_parameter_fsigma8_vext_and_h}. It displays no dependence on $\rsmooth$ and only a weak dependence on $\czmin \geq 1000 \kms$. Apart from $\czmin < 1000 \kms$, the region of highly coherent flow with the LG, the results for CMB and LG frame redshifts agree within $1 \sigma$. In \cref{sec:test_on_mocks:test_of_parameter_inference} we found for the mocks that the true value of $\hobs$ is systematically slightly underestimated. As for $\fsig$, we calibrate for this by multiplying the raw estimate of $\hobs$ with the ratio $\hobs{}^\mathrm{,true} / \langle \hobs\rangle$ found for the mocks, with $\hobs{}^\mathrm{,true} = 0.6777$ being the dimensionless Hubble parameter employed in the MDPL2 simulation \citep{klypin_multidark_2016}. The result, averaged over both considered redshift reference frames, is shown as the black line. The deviation between both frames is treated as an additional systematic error contribution. Again, the calibrated value for $\rsmooth = 15 \hmpc$ is adopted as our conservative final estimate and listed in \cref{tab:parameter_values} together with the results for $\rsmooth = 5$ and $10 \hmpc$. It is in excellent agreement with the value $\hobs = 0.75 \pm 0.02$ found by analysing the global in- and outflow in CF3, and consistent with the result $\hobs = 0.76 \pm 0.04$ obtained from the CF3 supernovae \citep{tully_cosmicflows-3_2016}. Although $\hobs$ is only a nuisance parameter in our analysis, this agreement presents a strong consistency check.

%%%%%%%%%%%%%%%%%%%%%%%%%%%%%%%%%%%%%%%%
\section{Quality of match between reconstructed and observed velocities}
\label{sec:comparison_to_observed_velocity}
The analysis of the \vvcomp\ has so far only focused on parameter inference. A weakness of this type of analysis is that it does not reveal any potential discrepancies between the velocities. We investigate here the quality of the agreement by means of a 
point-by-point comparison of galaxy group velocities and the correlation function of the velocity residual.

%%%%%%%%%%%%%%%%%%%%
\subsection{Point-by-point comparison}
\label{sec:comparison_to_observed_velocity:point_by_point_comparison}
Since $\vrec*_r$ is smoothed by default in the reconstruction procedure and $\vobs*_r$ is the actual observed velocity, we first need to bring the two quantities to a common smoothing scale. We thus apply an additional smoothing on a scale sufficiently larger than the one employed in $\vrec*_r$. In principle, we could expand $\vrec*_r$ and $\vobs*_r$ in orthogonal base functions as done in \cite{davis_local_2011}. However, for simplicity of presentation we apply here a Gaussian tensor-smoothing scheme similar to that employed in the POTENT method \citep{dekel_potential_1990}. It is described in \cref{sec:tensor_smoothing}. The smoothing scale is adapted to ensure a minimal number of CF3 groups to be contained in the smoothing window. In regions that are sufficiently densely sampled by CF3 it is set to a fixed minimal value $\ssmooth$. For all values of $\ssmooth$ considered, less than 2.3 \% of the CF3 group positions have an adaptive smoothing scale larger than the chosen $\ssmooth$.

Let $\tenssmooth{\vect{v}}$ denote the tensor-smoothed velocity field. As input for $\tenssmooth{v}^\rec_r$ we use the WF field smoothed with an $\rsmooth = 5 \hmpc$ Gaussian -- the smallest scale for which we have estimated the parameters $\vect{\Theta}$. To match the raw observed velocities as closely as possible, we use the uncalibrated ML estimates of $\fsig$ and $\bext$ found for this smoothing scale in the reconstruction. Accordingly, the input velocities for $\tenssmooth{v}^\obs_r$ are computed using the corresponding uncalibrated ML estimate of $\hobs$. The observed galaxies and groups are placed at their LG redshifts, which are locally closer to their actual distances. All further mathematical details are described in \cref{sec:tensor_smoothing}.

\begin{figure}
	\centering
	\includegraphics[width=\linewidth]{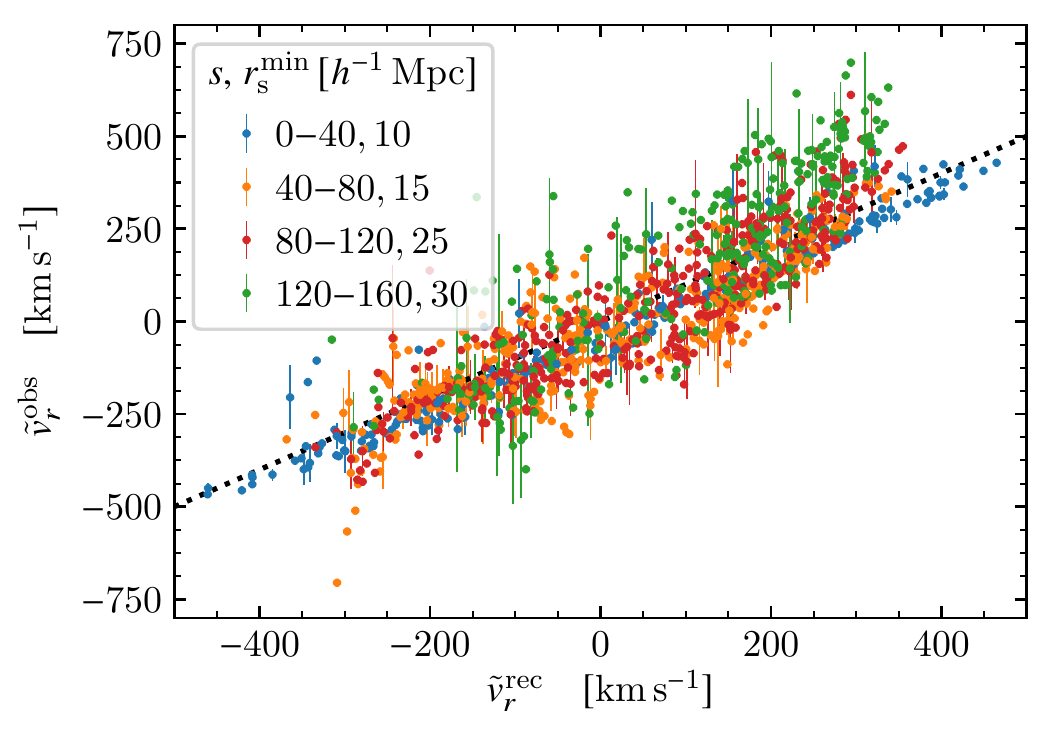}
	\caption{Comparison of tensor-smoothed observed and reconstructed radial velocities, $\tenssmooth{v}^\obs_r$ and $\tenssmooth{v}^\rec_r$, of 1500 randomly selected CF3 groups. The colours represent different bins in redshift distance $s$ as well as the adopted smoothing scales $\ssmooth$ in each bin, as indicated in the legend. The dotted black line marks the diagonal line $\tenssmooth{v}^\obs_r = \tenssmooth{v}^\rec_r$. For a random subset of 200 points, the error on $\tenssmooth{v}^\obs_r$ is marked.}
	\label{fig:radial_velocity_scatter}
\end{figure}

In \cref{fig:radial_velocity_scatter} we plot $\tenssmooth{v}^\obs_r$ against $\tenssmooth{v}^\rec_r$, both subject to the same tensor-smoothing and evaluated at the positions of 1500 randomly selected CF3 groups in four different redshift distance bins, each $40 \hmpc$ thick. To account for the increasing sparsity of CF3 groups with distance, we also increase the minimal tensor smoothing scale from $\ssmooth = 10 \hmpc$ to $\ssmooth = 30 \hmpc$ as indicated in the figure. For a random subset of 200 of the shown points, we also mark the errors on $\tenssmooth{v}^\obs_r$.\footnote{In comparison, the error on $\tenssmooth{v}^\rec_r$ is negligible.} We observe a strong correlation between $\tenssmooth{v}^\obs_r$ and $\tenssmooth{v}^\rec_r$. Relative to the errors on $\tenssmooth{v}^\obs_r$, the RMS of the residual velocity $\tenssmooth{v}^\obs_r - \tenssmooth{v}^\rec_r$ per bin is $1.8 \sigma$, $1.4 \sigma$, $1.1 \sigma$ and $1.0 \sigma$ from the inner- to the outermost bin. A fraction of the observed velocities in the outermost bin show a systematically larger outflow than in the reconstruction. However, given the observational errors, this discrepancy is statistically not significant. Such a slight discrepancy might, for example, be the result of a small mean underdensity in the reconstruction volume, which would not be accounted for in the reconstruction since the SFB boundary conditions impose a vanishing mean density contrast. Reconstructions of larger volumes based on upcoming galaxy surveys will allow us to investigate this further in future work. We also note that nearly all of the groups showing this outflow are from the 6dFGS Fundamental Plane data set \citep{springob_6df_2014}, which is the main CF3 contribution at those large distances. The 6dFGS peculiar velocities have previously been found to show some systematic deviations from model predictions \citep{springob_6df_2014}, and required additional morphology-dependent corrections in the zero-point calibration in CF3 \citep{tully_cosmicflows-3_2016}.

\begin{figure*}
	\centering
	\includegraphics[width=\textwidth]{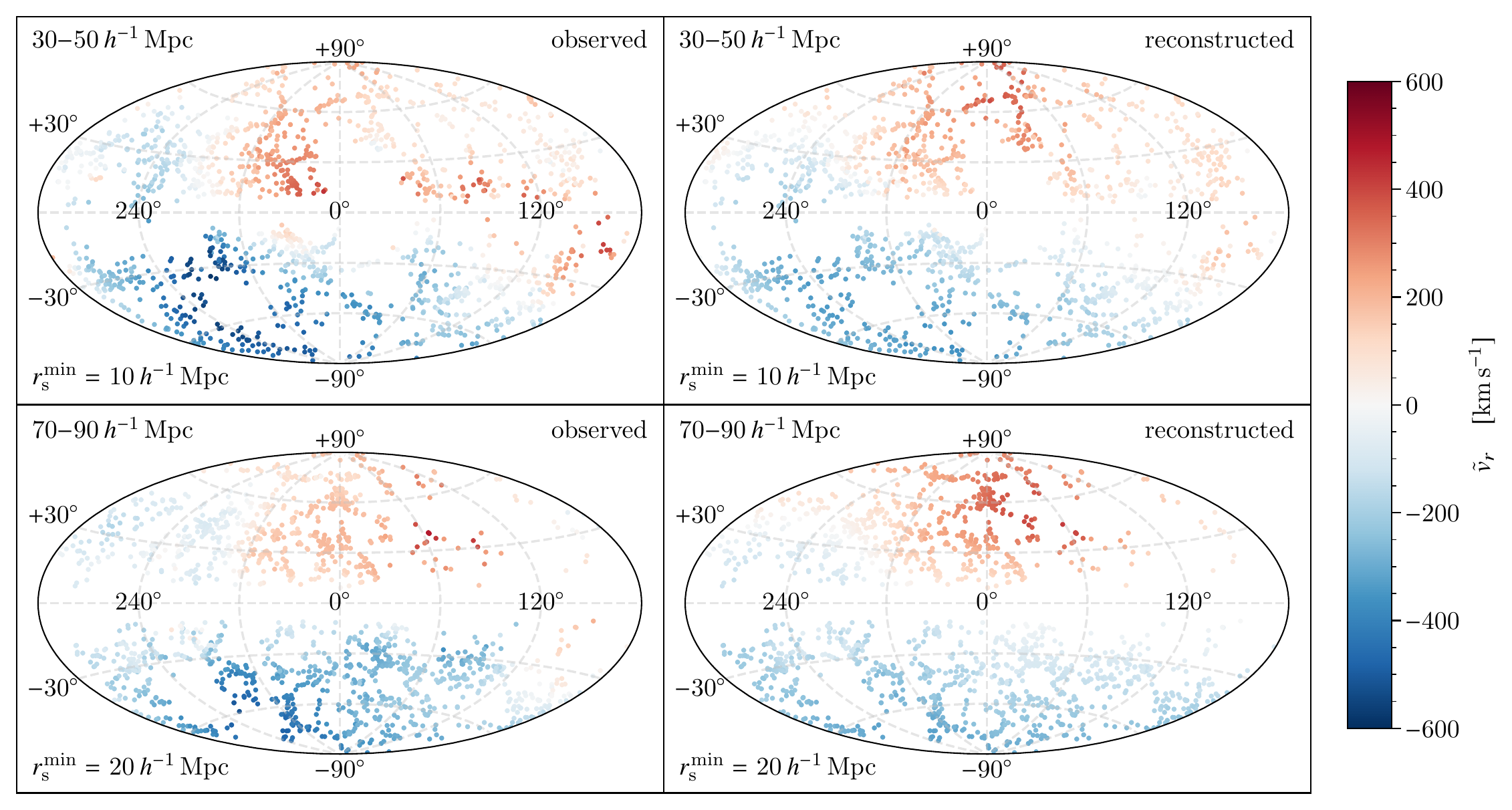}
	\caption{Tensor-smoothed observed (left) and reconstructed (right) radial velocities at the positions of the CF3 groups within $20 \hmpc$ thick spherical shells around $s = 40 \hmpc$ (top) and $s = 80 \hmpc$ (bottom), shown in galactic Aitoff projection. The adopted minimal smoothing scales $\ssmooth$ are denoted in the panels. Small-scale spatial fluctuations are highlighted by subtracting a $400 \kms$ bulk velocity in the direction of the observed LG motion $\vLG$.}
	\label{fig:radial_velocity_fields_aitoff_comparison}
\end{figure*}

\Cref{fig:radial_velocity_fields_aitoff_comparison} shows a comparison between $\tenssmooth{v}^\obs_r$ and $\tenssmooth{v}^\rec_r$ at the positions of the CF3 groups within two $20 \hmpc$ thick spherical shells in the Galactic Aitoff projection. We plot shells around $s = 40 \hmpc$ and $s = 80 \hmpc$, using the tensor smoothing scales $\ssmooth = 10 \hmpc$ and $\ssmooth = 20 \hmpc$, respectively. We find an overall good agreement between the radial flow patterns. The deviations between the observed and reconstructed velocity amplitudes in some directions are in line with the scatter seen in \cref{fig:radial_velocity_scatter}.

%%%%%%%%%%%%%%%%%%%%
\subsection{Correlation function}
\label{sec:comparison_to_observed_velocity:correlation_function}
To further quantify the agreement between observed and reconstructed velocities, we compute the pair correlation function of the radial velocities defined as \citep{gorski_cosmological_1989}
\begin{equation}
    \Psi(\Delta s) \coloneqq \frac{\sum_{(i,j)_s} \, v_{r,i} \, v_{r,j} \, \cos(\theta_{ij})}{\sum_{(i,j)_s} \, \cos^2(\theta_{ij})} \,.
    \label{eq:reconstructed_fields:correlation_function}
\end{equation}
The sum runs over all distinct pairs of observed CF3 groups separated by a distance $\Delta s$, and $\theta_{ij}$ denotes the angle between them. No tensor-smoothing is applied, as it does not affect $\Psi$ notably. We thus compute $\Psi$ directly from the reconstructed and observed input velocities as defined in \cref{sec:comparison_to_observed_velocity:point_by_point_comparison}. We only consider CF3 groups with $cz^\obs \leq 10,000 \kms$ to avoid noise contamination from the large observational velocity errors of more distant groups.

\begin{figure}
	\centering
	\includegraphics[width=\linewidth]{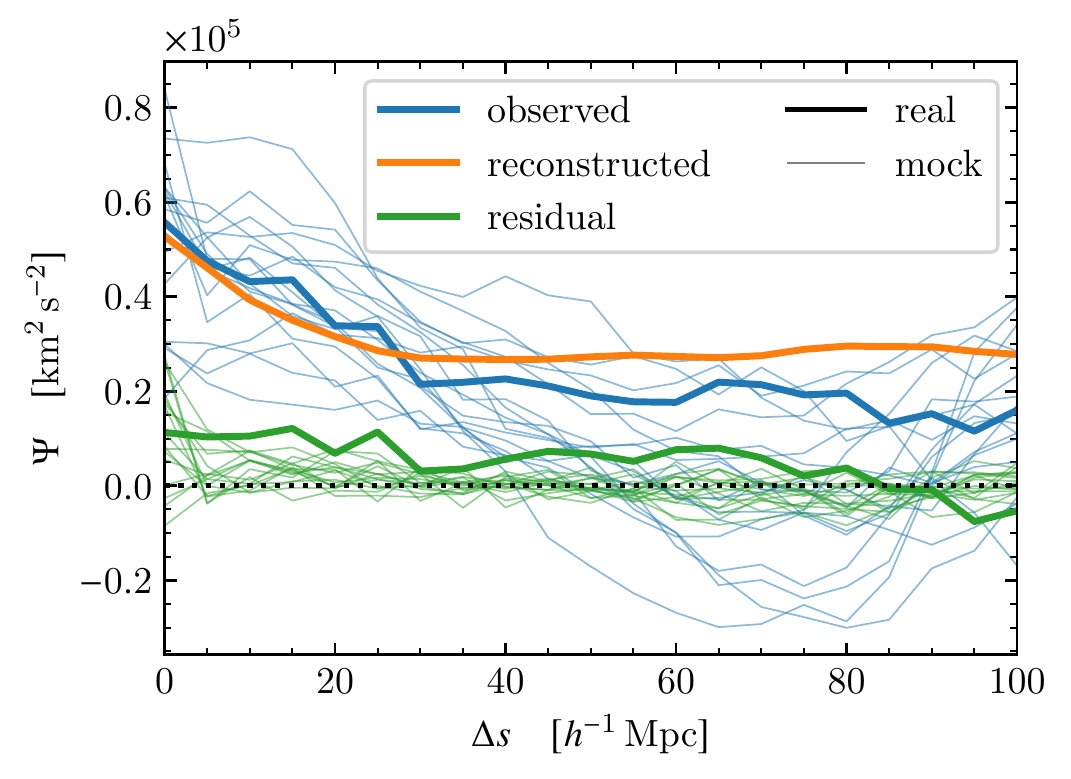}
	\caption{Correlation functions of $\vobs*_r$, $\vrec*_r$ and their residual as a function of the redshift distance separation between pairs of CF3 groups. The thick lines show the results for the actual data, the thin lines for the set of mocks. For clarity, the reconstructed mock correlations are omitted. Only groups with $cz^\obs \leq 10,000 \kms$ have been considered.}
	\label{fig:radial_velocity_correlation}
\end{figure}

\Cref{fig:radial_velocity_correlation} shows the result for $\Psi$ computed for the observed and reconstructed velocities as well as their residual. We see a good agreement between the correlation functions of observed and reconstructed radial velocities, in particular for separations $\Delta s \lesssim 30 \hmpc$. For larger separations, the reconstructed correlation is slightly larger than the observed one. The better agreement at smaller separations is expected because we are inferring  the parameters $\vect{\Theta}$ from a point-by-point \vvcomp, i.\,e.~at zero separation. Furthermore, the correlation in the residual velocity is strongly reduced. The small remaining residual correlation is approximately compatible with a residual bulk flow of $\sim 60$ to $80 \kms$, which is comparable to the uncertainty on our estimate of $\bext$. Overall, we thus find that our reconstruction is compatible with the observed flows over a wide range of scales.

For comparison, we also plot the observed and residual correlations found for the set of mocks as thin lines in \cref{fig:radial_velocity_correlation}. The distribution of observed mock correlations is found to be compatible with that in the actual CF3 catalog. For clarity, the reconstructed mock correlations are omitted. But the small residual mock correlations show that there is good agreement between reconstructed and observed mock correlations. The residual mock correlations are generally even closer to zero than for the actual data. A possible explanation is that the mock distance moduli are drawn from the same distribution, while CF3 is a compilation of different individual data sets. Minor systematic discrepancies between the zero-point calibrations of these data sets are likely to generate an additional residual correlation. Since this effect is small, we do not explore it further here.

%%%%%%%%%%%%%%%%%%%%%%%%%%%%%%%%%%%%%%%%
\section{Reconstructed fields on a grid}
\label{sec:reconstructed_fields_on_grid}
We now take a closer look at the WF estimates of both $\drec$ and $\vrec$, using observed LG redshifts and adopting the calibrated values of $\fsig$ and $\bext$ for $\rsmooth = 15 \hmpc$ listed in \cref{tab:parameter_values}. We compute these fields on a regular spherical grid with a radial resolution of $\Delta r = 2 \hmpc$ and an angular resolution of $\Delta \vartheta = \Delta \varphi = \pi / 50$. Unless stated otherwise, they are smoothed with an $\rsmooth = 5 \hmpc$ Gaussian.

%%%%%%%%%%%%%%%%%%%%
\subsection{Scatter between constrained realizations}
\label{sec:reconstructed_fields_on_grid:uncertainty}
\begin{figure}
	\centering
 	\includegraphics[width=\linewidth]{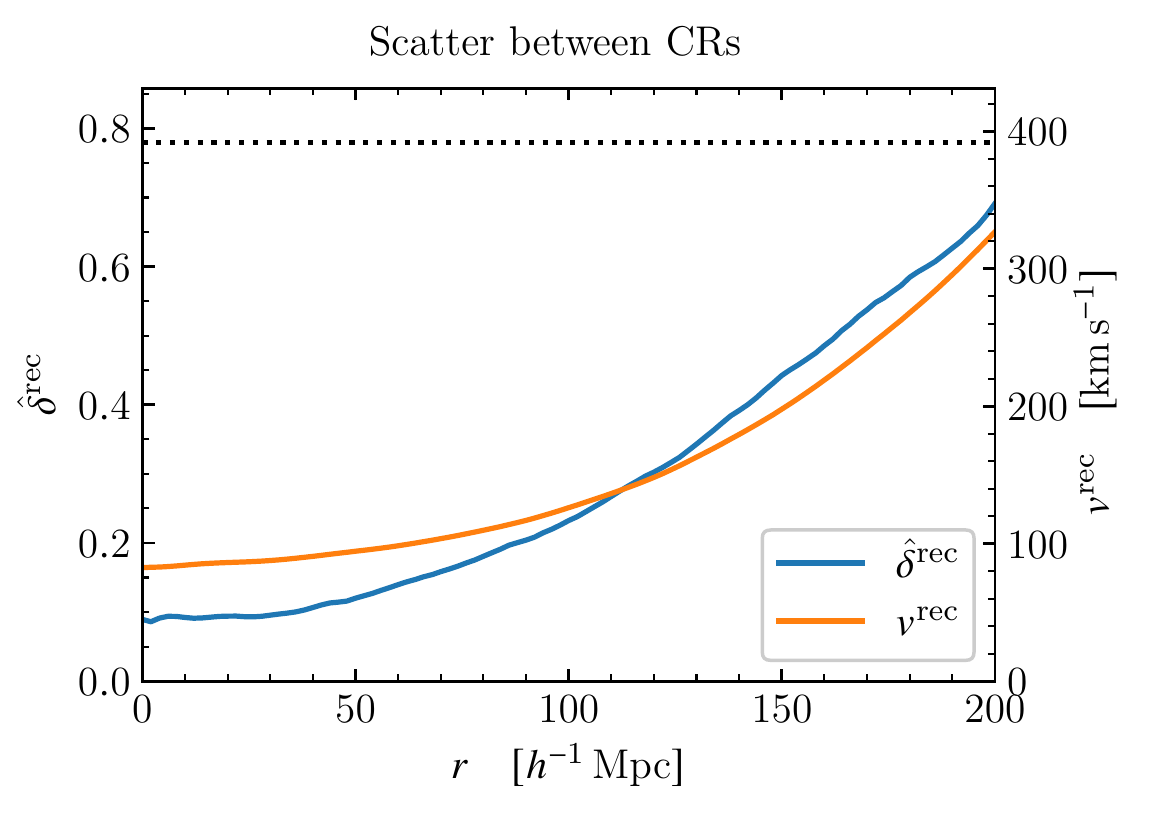}
	\caption{Standard deviation of the reconstructed normalized density contrast (left axis, blue line) and absolute peculiar velocity (right axis, orange line) over a set of 50 CRs, shown as a function of radius. The dotted horizontal line marks the square roots of both fields' cosmic variances, $\sigma^\text{cos}_{\norm{\delta}} = 0.78$ and $\sigma_v^\text{cos} = 392 \kms$.}
	\label{fig:solid_angle_average_of_reconstructed_fields}
\end{figure}

To quantify the uncertainty due to shot noise in the reconstructed fields given the adopted values of $\fsig$ and $\bext$, we compute the standard deviation between individual CRs around the WF estimate. Since the selection function of 2MRS is isotropic,\footnote{It is isotropic after repopulating the ZOA as described in \cref{sec:data:2MRS}.} the shot noise and thus the uncertainty can only depend on radius. Therefore, we additionally average the computed variance over the full solid angle. In \cref{fig:solid_angle_average_of_reconstructed_fields} we plot the resulting standard deviations of $\drec$ and $\vrec*$. For reference, we also mark the square root of their cosmic variances, $\sigma^\text{cos}_{\norm{\delta}} = 0.78$ and $\sigma_v^\text{cos} = 392 \kms$. The uncertainties in both fields show a similar radial increase from the origin, where the noise is weakest, towards the boundary $\rmax$ of the reconstruction volume: $\sim 0.1 - 0.7$ for $\drec$ ($\sim 0.1 - 0.9 \, \sigma^\text{cos}_{\norm{\delta}}$), and $\sim 80 - 320 \kms$ for $\vrec*$ ($\sim 0.2 - 0.85 \, \sigma^\text{cos}_v$).

Compared to the intrinsic error of our reconstruction method, quantified by the scatter between mocks in \cref{fig:solid_angle_averaged_residual_fields}, the shot noise  in $\drec$ is typically by a factor of 2 to 3 smaller at $r < 100 \hmpc$ (up to a factor of 6 at the (mock) Virgo distance), but grows more rapidly towards larger radii. At $\rmax = 200 \hmpc$, the shot noise error is slightly larger than the intrinsic error. For $\vrec*$ the shot noise and intrinsic errors relative to $\sigma^\text{cos}_v$ are comparable in size for all radii.

%%%%%%%%%%%%%%%%%%%%
\subsection{Cosmography}
\label{sec:reconstructed_fields_on_grid:cosmography}
\begin{figure*}
	\centering
	\includegraphics[width=0.94\textwidth]{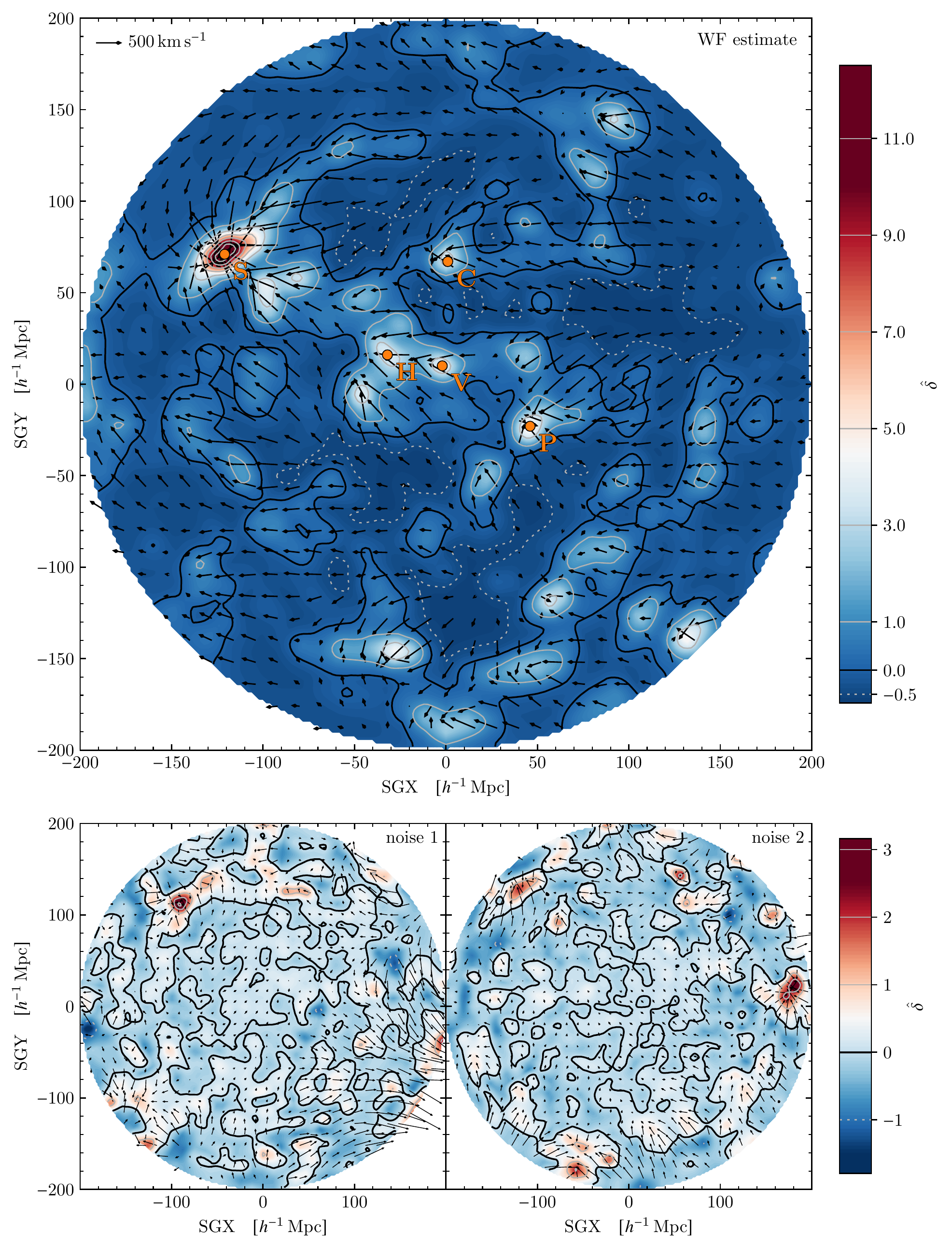}
	\caption{Top: WF estimate of the reconstructed normalized density contrast (heat map and contours) and peculiar velocity (arrows) in the supergalactic plane. Dominant structures are marked and labelled in orange: \textbf{S}hapley, \textbf{C}oma, \textbf{H}ydra-Centaurus, \textbf{V}irgo, \textbf{P}erseus-Pisces. Bottom: Two examples of residual noise realizations of the same fields. Adding these to the WF estimate in the top panel yields the corresponding full CRs. The contours show the values of $\hat{\delta}$ as marked in the colour bars. The arrow length scale is the same in all panels. An arrow corresponding to $500 \kms$ is given as reference.}
	\label{fig:SGP_reconstruction_and_realizations}
\end{figure*}

The upper panel of \cref{fig:SGP_reconstruction_and_realizations} plots the WF estimates of $\drec$ and $\vrec$ in a cut through the supergalactic plane (SGP). We marked some of the dominant visible overdensities, namely the Shapley Concentration at $(\text{SGX}, \text{SGY}) \approx (-120, 70) \hmpc$, the Coma Supercluster at $(0, 70) \hmpc$, the Hydra-Centaurus Supercluster at $(-30, 15) \hmpc$, the Perseus-Pisces Supercluster at $(45, -25) \hmpc$ and the Virgo Supercluster around $(0, 10) \hmpc$. Shapley is also clearly identifiable by the strong convergence of the peculiar velocity field at its position. In a less pronounced way, the same can be seen for for Perseus-Pisces and Coma. For the other marked overdensities, the velocity field at the shown smoothing scale $\rsmooth = 5 \hmpc$ shows only a weak convergence, since it is dominated by the flow towards Shapley.

To illustrate the variations between different CRs discussed in \cref{sec:reconstructed_fields_on_grid:uncertainty}, we also plot the residual noise obtained for two examples of individual realizations in the SGP in the lower panels of \cref{fig:SGP_reconstruction_and_realizations}. The radial increase of the noise amplitude towards the boundary of the reconstruction volume is clearly visible.

%%%%%%%%%%%%%%%%%%%%
\subsection{Bulk flow and LG motion}
\label{sec:reconstructed_fields_on_grid:bulk_flow}
\begin{figure}
	\centering
	\includegraphics[width=\linewidth]{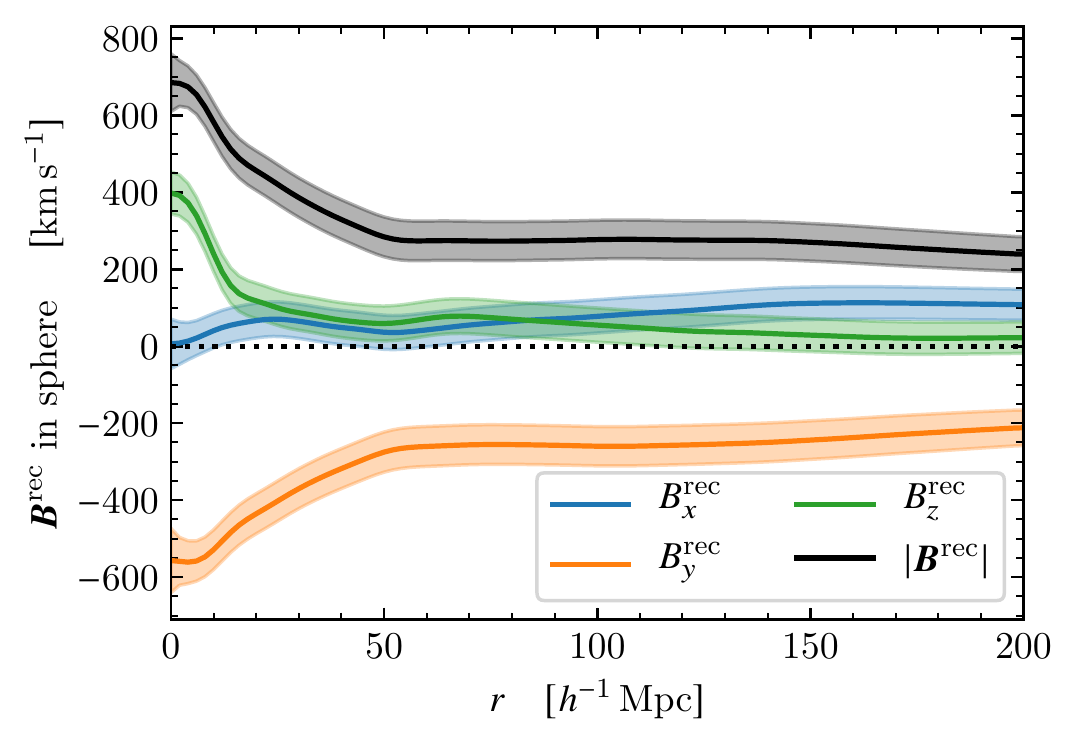}
	\caption{Reconstructed bulk velocity in spheres of radius $r$. The lines show the WF estimate of the Cartesian components in galactic coordinates and the total velocity as indicated in the figure. The shaded areas mark the $1 \sigma$ scatter between 50 CRs.}
	\label{fig:reconstructed_bulk_flow}
\end{figure}

\Cref{fig:reconstructed_bulk_flow} shows the WF estimate and standard deviation (over 50 CRs) of the total reconstructed bulk velocity $\brec$ in (top-hat) spheres of different radii, computed by volume-averaging $\vrec$ (including $\bext$). To resolve the flow on small distances, a Gaussian smoothing scale of only $\rsmooth = 1 \hmpc$ was used to compute $\vrec$ in this case. The bulk velocity amplitude $\brec*$ decreases from $\sim 690 \kms$ at $r = 0$ to $\sim 270 \kms$ at $r = 50$, then stays approximately constant up to $r = 150 \hmpc$, before it mildly decreases again for larger distances. The dominant component is in the galactic $y$-direction. When comparing to the bulk flows measured in \cite{nusser_cosmological_2011} up to a distance of $100 \hmpc$ using SFI++, we find agreement in the total velocity and the three individual components within $\sim 1 \sigma$ for nearly all distances. More significant deviations are only found for $\brec*_y$ at $r \lesssim 30 \hmpc$ (up to $1.7 \sigma$) and $\brec*_x$ at $r \gtrsim 60 \hmpc$ (up to $2.5 \sigma$). The reconstructed bulk velocity within the total reconstruction volume, $\rmax = 200 \hmpc$, is $\brec*_{200} = 239 \pm 45 \kms$ towards $l = 297 \pm 10^\circ$, $b = 5 \pm 10^\circ$, in good agreement with the result found in \cite{carrick_cosmological_2015}.

The value of $\brec$ at the origin, $\brec*_0 = 685 \pm 75 \kms$ towards $l = 270.6 \pm 6.6^\circ$, $b = 35.5 \pm 7.2^\circ$, describes the reconstructed LG velocity smoothed on the scale $\rsmooth = 1 \hmpc$. It agrees within $1 \sigma$ with the (unsmoothed) observed LG velocity $\vLG* = 620 \pm 15 \kms$ towards $l = 271.9 \pm 2.0^\circ$, $b = 29.6 \pm 1.4^\circ$ \citep{aghanim_planck_2020}. Considering that we explicitly excluded the local neighbourhood within $cz < 2000 \kms$ when inferring $\fsig$ and $\bext$ via the \vvcomp, this good agreement presents a strong consistency check of our results.

%%%%%%%%%%%%%%%%%%%%%%%%%%%%%%%%%%%%%%%%
\section{Summary and discussion}
\label{sec:conclusion}

%%%%%%%%%%%%%%%%%%%%
\subsection{CORAS framework}
\label{sec:conclusion:CORAS}
We have presented the CORAS framework for the reconstruction of density and peculiar velocity fields from all-sky redshift surveys. CORAS aims at maximal exploitation of the data in the Bayesian sense, in order to extract cosmological information on all observationally probed scales. For this purpose, we combine a variance-minimizing Wiener filter (WF) with the technique of constrained realizations (CRs) \citep{bertschinger_path_1987,hoffman_constrained_1991}. The CRs sample the distribution of realizations around the WF estimate compatible with the observed data, allowing for a faithful error estimation in the reconstructed fields and any inferred parameters. We account for the non-Gaussianity of the late small-scale structures by constructing the CRs from random log-normal density and Poisson-sampled galaxy realizations. We furthermore adopt a linear relation between (redshift galaxy) density and peculiar velocity fields. By additionally assuming statistical isotropy of the survey, we can apply the WF in spherical Fourier-Bessel (SFB) space \citep{fisher_wiener_1995} and thus achieve high computational efficiency and scalability.

A key ingredient in the CORAS framework is the test and calibration of the reconstruction pipeline using realistic mock data. This allows us to account for the intrinsic statistical and systematic errors introduced by our approximate modelling. For this purpose, we resorted to the \textsc{MultiDark} simulation run MDPL2 \citep{riebe_multidark_2013,klypin_multidark_2016}, identified 17 subvolumes that closely represent the LG environment, and extracted corresponding mock 2MRS and CF3 catalogs from the accompanying semi-analytic \textsc{sage} galaxy catalog \citep{knebe_multidark-galaxies_2018}. There are, however, limitations to the accuracy of these mocks. In particular, we still lack a simulation that provides a more detailed match to the coherent flows in the cosmological neighbourhood within $\sim$20 Mpc.
At present, even detailed constrained simulations of the local Universe, like those from the CLUES \citep{klypin_constrained_2003,gottloeber_constrained_2010,carlesi_constrained_2016,sorce_cosmicflows_2016}, ELUCID \citep{yang_elucid_2018} or HESTIA \citep{libeskind_hestia_2020} projects do not seem to reproduce the degree of coherence of the observed local flow. To constrain the simulations to the observed structures, CLUES and HESTIA use the velocity field obtained from Cosmicflows-2 \citep{tully_cosmicflows-2_2013}, while ELUCID uses the density field reconstructed from SDSS \citep{yang_evolution_2012,abazajian_seventh_2009}. Ideally, we would like to use both velocity and (redshift) density data simultaneously to generate new, more detailed constrained realizations. At the same time, these need to be large enough to cover the whole 2MRS reconstruction volume, while also providing a sufficiently accurate model of galaxy formation and a sufficiently high mass resolution.

%%%%%%%%%%%%%%%%%%%%
\subsection{Comparison to other methods}
\label{sec:conclusion:comparison_to_other_methods}
One of the main differences between CORAS and other reconstruction methods based on the linear theory assumption concerns the treatment of non-Gaussianities in the observed distribution of structures. Reconstructions based on peculiar velocity data often exploit that the linearity assumption for the velocity field is valid down to significantly smaller scales than for the density field. Because of this, WF/CR reconstructions based on Gaussian random field realizations \citep[e.\,g.][]{zaroubi_wiener_1999,courtois_three-dimensional_2011} or hierarchical Bayesian modelling assuming a Gaussian prior \citep[e,\,g.][]{lavaux_bayesian_2016,graziani_peculiar_2019} are viable. Linear reconstructions based on redshift data, on the other hand, typically perform a direct smoothing of the observed galaxy distribution \citep[e.\,g.][]{yahil_redshift_1991,carrick_cosmological_2015,boruah_cosmic_2020} or generate CRs by sampling from a high-dimensional log-normal Poissonian posterior distribution \citep[e.\,g.][]{kitaura_recovering_2010,jasche_bayesian_2010}. In CORAS, we combine the advantages of different approaches: we adopt the fast and conservative WF estimator, but account for non-Gaussianities by using the WF to generate CRs based on random log-normal density and Poisson-sampled galaxy realizations.

A different approach that goes beyond the linear theory assumption is ``physical forward modelling'' \citep[e.\,g.][]{wang_elucidexploring_2014,kitaura_cosmic_2020,jasche_physical_2019}. In this approach, CRs of the initial Gaussian density and velocity fields are generated, which are compatible with observations when evolved forward in time using a model of nonlinear structure growth. Usually, Lagrangian Perturbation Theory \citep[e.\,g.][]{moutarde_precollapse_1991}, generalizations thereof \citep{kitaura_cosmological_2013} or a particle-mesh model \citep[e.\,g.][]{klypin_three-dimensional_1983} are employed. This has the advantage of directly accounting for the dynamical development of non-Gaussianities rather than having to assume some specific functional form of the non-Gaussian probability distribution of the evolved fields. By combining this with a nonlinear galaxy bias model, these methods aim at a more accurate description of the small-scale statistics of the reconstructed signal than linear reconstruction methods.

On scales $\gtrsim 5 \hmpc$, however, it was found in \cite{keselman_performance_2017} that reconstructions based on nonlinear dynamics are compatible with linear theory when applied to realistic redshift space catalogs. The advantage of nonlinear dynamical descriptions appears to be counteracted by the increased sensitivity to selection effects, shot noise and incoherent motions on small scales leading to fingers-of-god and multi-flow regions in redshift space. Another source of information loss on small scales is related to the fact that galaxies today correspond to a much larger Lagrangian (initial) volume. For example, a galactic halo of mass $10^{12}M_\odot$ corresponds to a Lagrangian sphere of $2.3$ Mpc (comoving). Physical forward modelling furthermore comes at the price of high numerical complexity and computational cost, even when using highly efficient algorithms like Hamiltonian Monte Carlo sampling.

\begin{figure*}
	\centering
	\includegraphics[width=\textwidth]{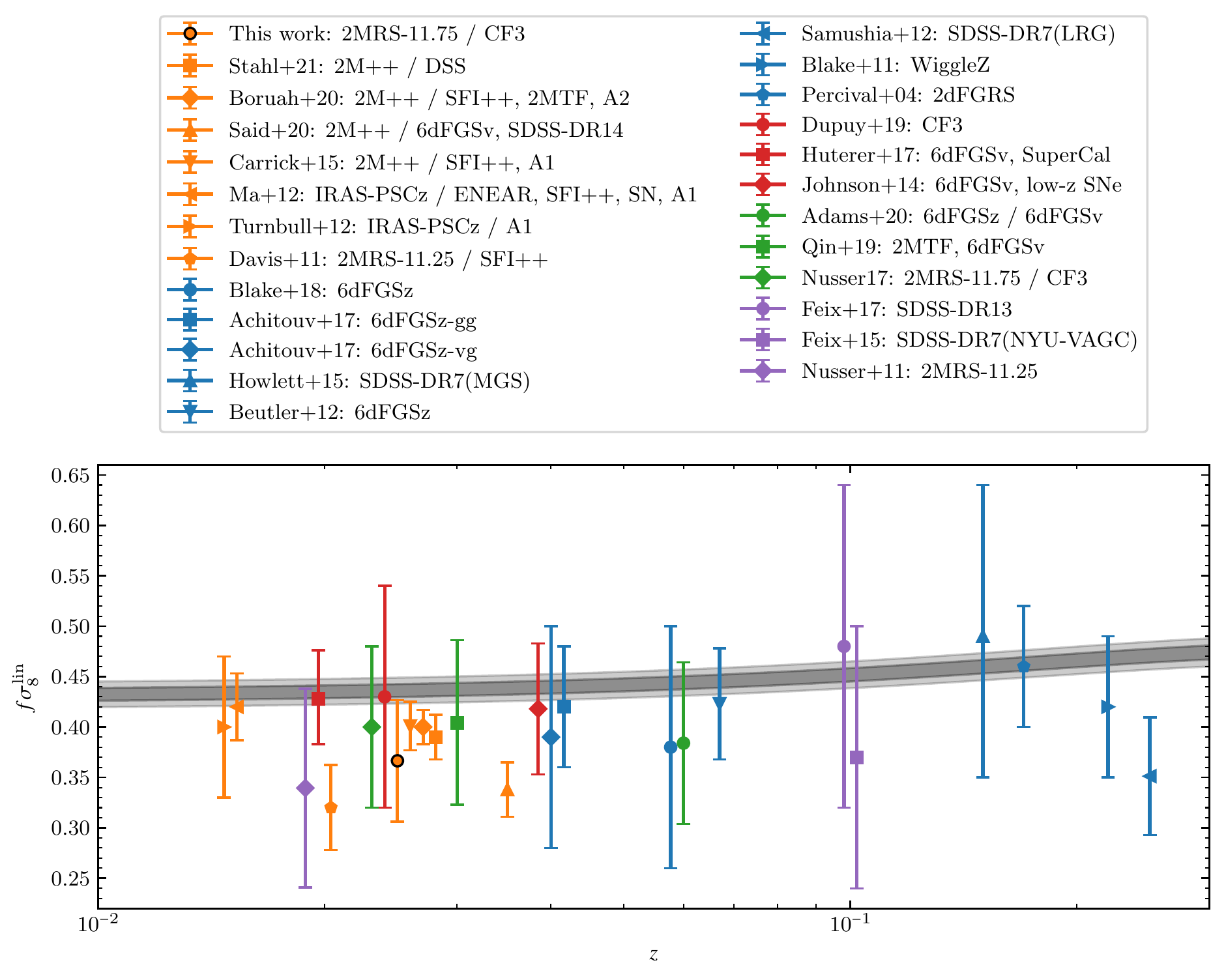}
	\caption{Comparison of our result for $\fsiglin$ (black-edged orange circle) with other low-redshift measurements using various methods:	\protect\vvcomp\ (orange) \protect\citep{stahl_peculiar-velocity_2021,boruah_cosmic_2020,said_joint_2020,carrick_cosmological_2015,ma_comparison_2012,turnbull_cosmic_2012,davis_local_2011}; correlations of the redshift-space density (blue) \protect\citep{blake_power_2018,achitouv_consistency_2017,howlett_clustering_2015,beutler_6df_2012,samushia_interpreting_2012,blake_wigglez_2011,percival_2df_2004}, peculiar velocities \protect\citep{dupuy_estimation_2019,huterer_testing_2017,johnson_6df_2014} or both (green) \protect\citep{adams_joint_2020,qin_redshift-space_2019,nusser_velocitydensity_2017}; and galaxy luminosity variations (purple) \protect\citep{feix_speed_2017,feix_growth_2015,nusser_new_2011}. If the effective $z$ of a measurement was not explicitly specified, we estimated it based on the employed data. To improve clarity, points sharing the same $z$ are slightly offset. The shaded areas mark the 1 and $2 \sigma$ confidence limits inferred from the
	Planck-18 results \protect\citep{aghanim_planck_2020-1}.}
	\label{fig:fsigma8_literature_comparison}
\end{figure*}

%%%%%%%%%%%%%%%%%%%%
\subsection{Application to 2MRS and Cosmicflows-3}
\label{sec:conclusion:application_to_2MRS_and_CF3}
We have focused on low redshift data and applied CORAS to reconstruct the density and peculiar velocity fields within $200 \hmpc$ from the Two-Micron All-Sky Redshift Survey (2MRS) \citep{huchra_2mass_2012,macri_2mass_2019}. The reconstructed velocity field was then compared to the observed velocities obtained from Cosmicflows-3 (CF3) \citep{tully_cosmicflows-3_2016}. Using a maximum-likelihood estimator, we constrained the normalized growth rate $\fsig$, the bulk flow contribution $\bext$ from sources beyond $200 \hmpc$, and, as a nuisance parameter, the dimensionless Hubble parameter $\hobs$ used in the computation of the observed galaxy distances. The parameters have further been calibrated using the results of applying the same estimator to the set of mocks. Assuming a Planck-18 cosmology \citep{aghanim_planck_2020-1}, we also translated $\fsig$ to its linear value $\fsiglin$.

All inferred parameter values together with their errors are listed in \cref{tab:parameter_values} for several choices of the smoothing scale $\rsmooth$ used in the reconstruction. The results obtained for $\rsmooth = 15 \hmpc$, above which linear theory becomes a viable approximation, are adopted as our conservative final parameter estimates. The result for $\hobs$ agrees well with the value found in the analysis of CF3 in \citep{tully_cosmicflows-3_2016}. The results for the other parameters are discussed in detail in \cref{sec:conclusion:comparison_of_fsigma8,sec:conclusion:comparison_of_bulk_flow}.

We furthermore confirmed the quality of the agreement between the reconstructed and observed velocities by comparing them on a point-by-point basis after tensor-smoothing both on the same scale. Additionally, we showed that the correlation function of the residual between reconstructed and observed velocities is strongly reduced compared to the individual correlation functions. The reconstructed velocities can, for example, be used to reduce the systematic and statistical errors in the determination of the Hubble constant \citep[e.\,g.][]{scolnic_complete_2018,mukherjee_velocity_2021,howlett_standard_2020,nicolaou_impact_2020,sedgwick_effects_2020,boruah_peculiar_2020}.

%%%%%%%%%%%%%%%%%%%%
\subsection[$\fsig$ versus $\beta$]{$\boldsymbol{\fsig}$ versus $\boldsymbol{\beta}$}
\label{sec:conclusion:fsigma8_vs_beta}
Traditionally, the quantity $\beta=f/b$ has been widely adopted as the parameter to be inferred from analysis of redshift space distortions and the comparison between the velocity recovered from the distribution of galaxies and the independently observed peculiar galaxy velocities. Indeed, assuming linear galaxy bias, linear theory directly yields this parameter.

Recently, however, the parameter $\beta$ has been abandoned in favour of $\fsig=\beta \siggal$. While $\fsig$ is desirable since it involves only the cosmological parameters independent of galaxy bias and distance, it is also associated with several shortcomings. First, the directly measured quantity is $\beta$, which is sensitive to the scales probed by the data and not necessarily $8\hmpc$. Therefore, strictly speaking, the relevant quantity is $\beta \siggal = \fsig b_8/b$, where $b$ is the galaxy bias on the scales probed by the data and $b_8=\siggal/\sig$ is the bias relevant to an $8\hmpc$ scale. Thus, there is an assumption that $b$ is independent of scale. Second, $\siggal$ should be computed from the galaxy distribution in real space, which is not easily obtained from redshift space surveys. Indeed, at $8\hmpc$ the clustering is affected by linear as well as nonlinear effects (including incoherent motions).

Thus, inferring $\siggal$ precisely requires careful calibration or an extrapolation of the measured power spectrum on larger scales. In our test on mock galaxy catalogs we found that the real-space result for $\siggal$ is smaller by a factor of $\sim 1.1$ than the redshift-space result, with a weak dependence on distance. We assumed that the same constant rescaling holds in the actual 2MRS, and accounted for it in the calibration of $\fsig$.

Although $\beta$ is the directly measured parameter, we follow the trend of using $\fsig$, keeping in mind the aforementioned caveats.

%%%%%%%%%%%%%%%%%%%%
\subsection[Comparison of $\fsig$]{Comparison of $\boldsymbol{\fsig}$}
\label{sec:conclusion:comparison_of_fsigma8}
In \cref{fig:fsigma8_literature_comparison} we compare our result for the linear normalized growth rate, $\fsiglin = 0.367 \pm 0.060$, to other values obtained using various methods at low redshifts. In addition, we plot the Planck-18 result \citep{aghanim_planck_2020-1}.
We find an agreement with all of these within at most $1.1 \sigma$. It is interesting that most points lie below the $\Lambda$CDM curve, but except for the result obtained by \cite{said_joint_2020} none of them individually show a statistically significant deviation. Of course, one could make an attempt at assessing the $\Lambda$CDM prediction using the combination of all measured $\fsig$ listed in the figure. However, this would have to be done with great care since many of these points are obtained from overlapping data and similar methods.

Our estimated errors are larger than those of some other results inferred via \vvcomp\ for approximately the same reconstruction volume, e.\,g.~\cite{carrick_cosmological_2015,boruah_cosmic_2020,stahl_peculiar-velocity_2021}. This is mostly explained by the intrinsic reconstruction errors that we account for when calibrating the value of $\fsig$ using realistic mocks. Partially, it is also due to excluding CF3 groups below $\czmin = 2000 \kms$, which typically have the smallest observed distance errors but would dominate our estimate for the external bulk flow contribution. Our \emph{uncalibrated} results for $\rsmooth \lesssim 10 \hmpc$ have an error of $\sim 0.03$ at $\czmin = 2000 \kms$ and even only $\sim 0.02$ at $\czmin = 0$, comparable to those estimated in \cite{carrick_cosmological_2015,boruah_cosmic_2020,stahl_peculiar-velocity_2021}.

However, these small scales and distances are affected by incoherent small-scale velocities that are not captured by linear theory. We use the scatter between $\fsig$ inferred for different mocks to estimate the resulting systematic error. This error contribution happens to be smallest for $\rsmooth \approx 10 \hmpc$ and $\czmin \approx 2000 \kms$, but is of comparable size for the range $5 \hmpc \lesssim \rsmooth \lesssim 15 \hmpc$ and $1000 \kms \lesssim \czmin \lesssim 4000 \kms$ (for $\czmin \lesssim 1000 \kms$ it strongly increases). It thus limits the potential gain in constraining power obtained from including more small-scale and -distance information -- at least as long as the linear theory assumption is kept. It is furthermore advisable to avoid relying on the accuracy of the calibration on small nonlinear scales. Hence, we prefer the slightly more conservative estimate of $\fsig$ (and the other parameters) found for $\rsmooth = 15 \hmpc$.

In our analysis of the mocks, the smoothing scale at which the raw $\fsig$ estimate approximately matches the true value was found to be $\sim 7.5 \pm 2 \hmpc$. It is thus notably larger and more uncertain than the scale of $4 - 5 \hmpc$ found in \cite{berlind_biased_2000,carrick_cosmological_2015,hollinger_assessing_2021} by directly comparing simulated galaxy or halo velocities with linear velocity predictions obtained in \rspace. This further demonstrates that it is crucial to use mocks which resemble the actual data as closely as possible and to perform exactly the same analysis for both, to avoid biasing the results.

We point out that our results for $\fsig$ and $\fsiglin$ would increase by $\sim 4 \%$ if in the calibration we had used the mock $\sig$ value 0.95 measured from the MDPL2 particle distribution in \cite{hollinger_assessing_2021} instead of the value 0.91 computed from the adopted Cosmic Emu power spectrum. Relative to our total error estimate, however, this would only correspond to a change by $\sim 0.3 \sigma$.

%%%%%%%%%%%%%%%%%%%%
\subsection{Comparison of bulk flows and LG motion}
\label{sec:conclusion:comparison_of_bulk_flow}
\begin{figure*}
	\centering
	\includegraphics[width=\textwidth]{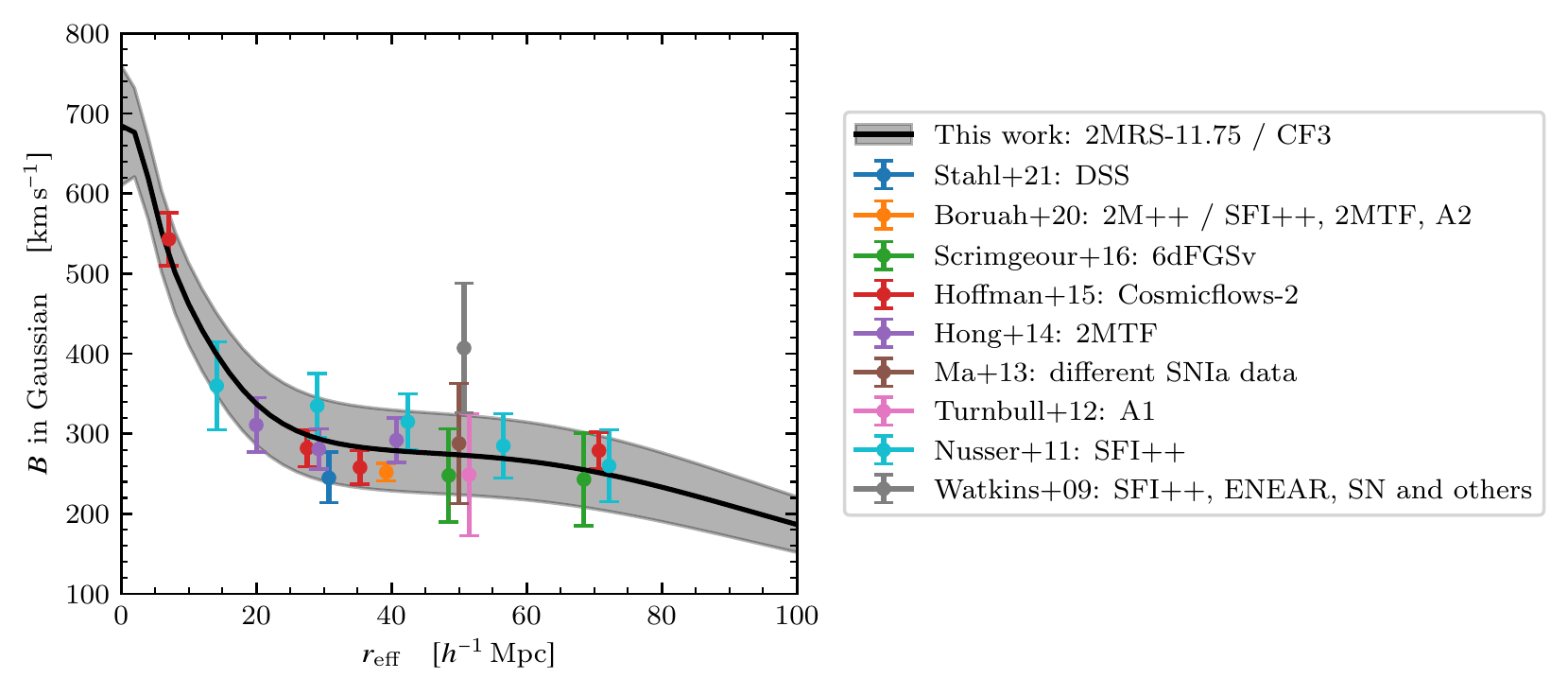}
	\caption{Comparison of our result for the bulk flow in Gaussian windows of effective radius $\reff$ (black line with $1 \sigma$ shaded area) with other measurements \protect\citep{stahl_peculiar-velocity_2021,boruah_cosmic_2020,scrimgeour_6df_2016,hoffman_cosmic_2015,hong_2mtf_2014,ma_cosmic_2013,turnbull_cosmic_2012,nusser_cosmological_2011,watkins_consistently_2009}. To improve clarity, points sharing the same $\reff$ are slightly offset. The bulk flows in \protect\cite{hoffman_cosmic_2015,nusser_cosmological_2011} (green and pink points) were measured in top-hat windows. We translate their values to approximately equivalent Gaussian windows via $\reff \approx r / \sqrt{2}$.}
	\label{fig:bulk_flow_literature_comparison}
\end{figure*}

In \cref{fig:bulk_flow_literature_comparison} we compare our results for the reconstructed bulk flow amplitude within Gaussian windows of different effective radii $\reff$ with those from a number of different measurements in the literature. We find agreement well within $1 \sigma$ with all literature results except for that by \cite{watkins_consistently_2009}, which still agrees within $1.4 \sigma$, however, and is a slight outlier among all data points. For the direction of the bulk flow, we also find agreement at the $1 \sigma$ level with most literature results. Only the directions found by \cite{turnbull_cosmic_2012} and \cite{scrimgeour_6df_2016} deviate more significantly, by 1.5 and $2 \sigma$, respectively. For reference, we quote our result at $\reff = 50 \hmpc$, $\brec* = 274 \pm 50 \kms$ towards $l = 287 \pm 9^\circ$, $b = 11 \pm 10^\circ$.

Our result for the external bulk flow contribution $\bext$ from sources beyond $200 \hmpc$, as specified in \cref{tab:parameter_values}, is in good agreement with the results found in \cite{carrick_cosmological_2015,boruah_cosmic_2020,stahl_peculiar-velocity_2021}. Furthermore, the total reconstructed motion of the LG with respect to the CMB, smoothed with a $\rsmooth = 1 \hmpc$ Gaussian, is $685 \pm 75 \kms$ towards $l = 270.6 \pm 6.6^\circ$, $b = 35.5 \pm 7.2^\circ$. It agrees within $1 \sigma$ with the observed CMB dipole \citep{aghanim_planck_2020}. The 2MRS-internal contribution is $541 \pm 71 \kms$ towards $l = 257 \pm 10^\circ$, $b = 43 \pm 11^\circ$, and accounts for $\sim 83 \%$ of the observed dipole. This is consistent at the $1 \sigma$ level with previous studies of the convergence of the clustering dipole \citep{bilicki_is_2011}.

%%%%%%%%%%%%%%%%%%%%
\subsection{Code and data products}
\label{sec:conclusion:code_and_data_products}
The code of CORAS is publicly available.\footnote{See footnote \ref{foot:coras_link}.} It is written in object-oriented C++ and is as such not only fast but also very modular. This allows the user to easily modify or exchange individual parts of the CR generation pipeline, e.\,g.~the selection function estimator, the RSD correction algorithm or the generator of random signal and data realizations. Hence, we strongly encourage anyone to use and adapt CORAS to their desired needs.

The reconstructed density and peculiar velocity fields on a grid are made directly available alongside the code of CORAS. Furthermore, the reconstructed velocities at the positions of the CF3 galaxies and groups are published in the Extragalactic Distance Database \citep{tully_extragalactic_2009}.\footnote{They are available at \href{http://edd.ifa.hawaii.edu}{http://edd.ifa.hawaii.edu}.}

%%%%%%%%%%%%%%%%%%%%
\subsection{Future prospects}
\label{sec:conclusion:future_prospects}
In future work, we plan to improve CORAS by directly including observed peculiar velocities as additional constraints into the CR generation, similar to \cite{zhu_reconstruction_2020}. This would render the separate step of constraining an external velocity contribution obsolete. Beyond that, we will investigate the use of more refined, galaxy bias and sampling models based on the halo occupation distribution \citep{peacock_halo_2000,berlind_halo_2002} or the related cumulative luminosity function \citep{yang_constraining_2003}.

Furthermore, we will prepare CORAS for a future application to the upcoming all-sky spectral survey SPHEREx, which is expected to measure redshifts of several hundred million galaxies up to $z > 1$ with an accuracy range of $\sigma_z/(1+z) \sim 0.3 - 10 \%$. For this purpose, we will extend our reconstruction method by explicitly accounting for redshift errors in the generation of CRs. This will also enable us to apply CORAS to photometric redshift surveys. The vastly larger number of observed galaxies compared to 2MRS furthermore requires the optimization of CORAS for highly parallelized computing architectures. The efficient and scalable SFB space method, which specifically exploits the statistical isotropy of all-sky surveys, will prove particularly beneficial in using this future wealth of data to further advance our understanding of cosmology.

\section*{Acknowledgements}
We thank Gustavo Yepes for providing the MDPL2 density and velocity fields used in the test of CORAS. For publishing our reconstructed velocities in the Extragalactic Distance Database, we would like to thank Brent Tully.  We are furthermore grateful to Martin Feix and Robert Reischke for many helpful discussions, and want to thank Thomas Jarrett, Ofer Lahav, Benjamin Wallisch and Steffen Hagstotz for their valuable input. We also thank the referee, Michael J.~Hudson, for his comments, which helped to greatly improve the paper. This work was supported in part by a Technion fellowship and by the Israel Science Foundation grant ISF 936/18.

CORAS uses the GSL \citep{galassi_gnu_2009} and FFTW3 \citep{frigo_design_2005} libraries.

The CosmoSim database used in this paper is a service by the Leibniz-Institute for Astrophysics Potsdam (AIP). The MultiDark database was developed in cooperation with the Spanish MultiDark Consolider Project CSD2009-00064.

The authors gratefully acknowledge the Gauss Centre for Supercomputing e.V. (\href{https://www.gauss-centre.eu/}{https://www.gauss-centre.eu/}) and the Partnership for Advanced Supercomputing in Europe (PRACE, \href{https://prace-ri.eu/}{https://prace-ri.eu/}) for funding the MultiDark simulation project by providing computing time on the GCS Supercomputer SuperMUC at Leibniz Supercomputing Centre (LRZ, \href{https://www.lrz.de/}{https://www.lrz.de/}).

%%%%%%%%%%%%%%%%%%%%%%%%%%%%%%%%%%%%%%%%%%%%%%%%%%
\section*{Data Availability}
Generated data: The code of CORAS and the reconstructed fields on a grid are publicly available at \href{https://github.com/rlilow/CORAS}{https://github.com/rlilow/CORAS}. The reconstructed velocities at the positions of CF3 galaxies and groups are publicly available in the Extragalactic Distance Database (EDD) \citep{tully_extragalactic_2009} at \href{http://edd.ifa.hawaii.edu}{http://edd.ifa.hawaii.edu}.

Used data: The 2MRS catalog is publicly available alongside its publications \cite{huchra_2mass_2012} (original data release) and \cite{macri_2mass_2019} (additional and corrected data). The employed 2MRS group catalog as well as the Cosmicflows-3 catalog are publicly available in the EDD. The power spectrum emulator Cosmic Emu \citep{heitmann_miratitan_2016} is publicly available at \href{https://github.com/lanl/CosmicEmu}{https://github.com/lanl/CosmicEmu}. The \textsc{sage} galaxy catalog of the MDPL2 simulation is publicly available in the CosmoSim database at \href{https://www.cosmosim.org}{https://www.cosmosim.org}. The MDPL2 density and velocity fields are not publicly available.

% The inclusion of a Data Availability Statement is a requirement for articles published in MNRAS. Data Availability Statements provide a standardised format for readers to understand the availability of data underlying the research results described in the article. The statement may refer to original data generated in the course of the study or to third-party data analysed in the article. The statement should describe and provide means of access, where possible, by linking to the data or providing the required accession numbers for the relevant databases or DOIs.

%%%%%%%%%%%%%%%%%%%% REFERENCES %%%%%%%%%%%%%%%%%%

% The best way to enter references is to use BibTeX:

\bibliographystyle{mnras}
\bibliography{references} % if your bibtex file is called example.bib

% Alternatively you could enter them by hand, like this:
% This method is tedious and prone to error if you have lots of references
%\begin{thebibliography}{99}
%\bibitem[\protect\citeauthoryear{Author}{2012}]{Author2012}
%Author A.~N., 2013, Journal of Improbable Astronomy, 1, 1
%\bibitem[\protect\citeauthoryear{Others}{2013}]{Others2013}
%Others S., 2012, Journal of Interesting Stuff, 17, 198
%\end{thebibliography}

%%%%%%%%%%%%%%%%%%%%%%%%%%%%%%%%%%%%%%%%%%%%%%%%%%

%%%%%%%%%%%%%%%%% APPENDICES %%%%%%%%%%%%%%%%%%%%%

\appendix

%%%%%%%%%%%%%%%%%%%%%%%%%%%%%%%%%%%%%%%%
\section[Distance dependence of $\siggal$]{Distance dependence of $\boldsymbol{\sig^\text{\MakeLowercase{g}}}$}
\label{sec:sigma_galaxy_estimator}
For a flux limited survey, only galaxies above a threshold luminosity that increases with distance are observable. Since galaxy biasing depends on luminosity, we expect a distance dependence of the galaxy density fluctuation amplitude, $\siggal(r)$. To estimate this directly from the galaxy survey, we create a volume-limited subsample containing only those galaxies with a redshift distance $s < r$ and an absolute magnitude large enough to be observable if placed at the distance $r$. We then convolve the discrete distribution of those galaxies with an $8 \hmpc$ scale top-hat filter, compute its volume-averaged RMS value and subtract the contribution from shot noise. As described in \cref{sec:data:2MRS}, we place the 2MRS galaxies at the mean redshift distances of their associated galaxy group \citep{tully_galaxy_2015} to collapse fingers-of-god.

\begin{figure}
	\centering
	\includegraphics[width=\linewidth]{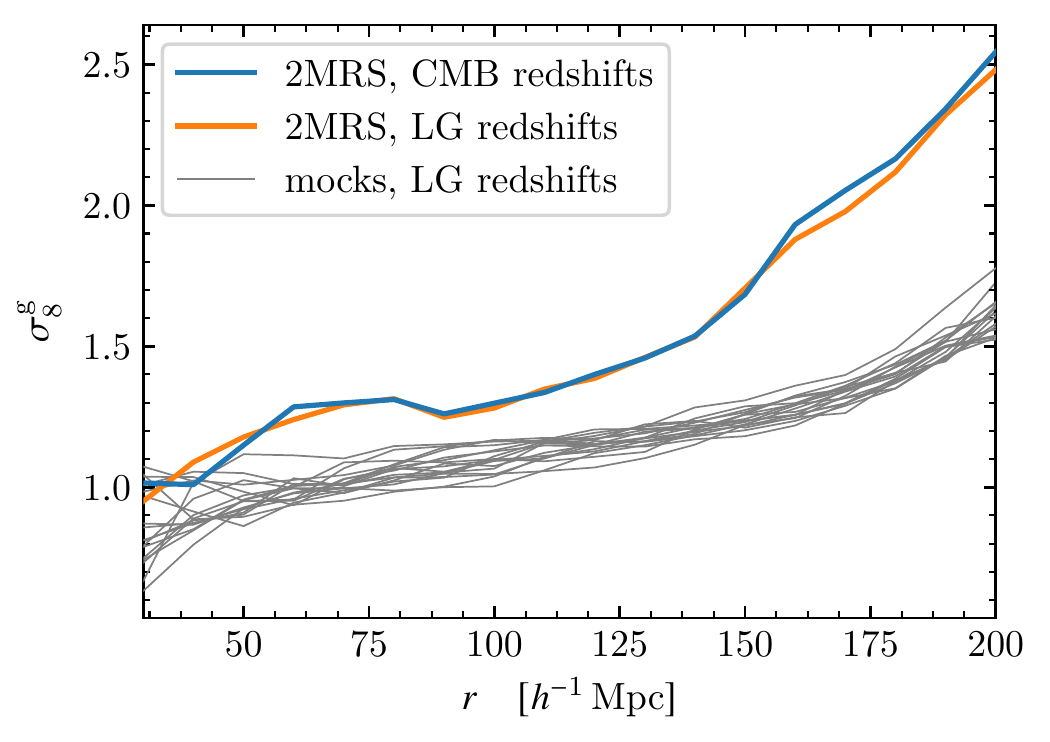}
	\caption{Dependence of $\siggal$ on distance, derived from volume-limited subsamples of 2MRS (thick lines) and the set of mock catalogs (thin grey lines). The two thick lines for 2MRS correspond to the CMB (blue) and LG (orange) frame redshifts, and are nearly identical. The same holds for the mocks, and only the LG redshift results are plotted for them.}
	\label{fig:radial_dependence_of_sigma8_galaxy}
\end{figure}

The results for $\siggal(r)$, using either CMB or LG frame redshifts, are shown in \cref{fig:radial_dependence_of_sigma8_galaxy}. The estimates for both reference frames are nearly identical. Up to a constant factor of $\sim 1.2$, they also agree well with the number-weighted result found for $\siggal(r) / \sig^\mathrm{g*}$ in \cite{carrick_cosmological_2015}, where $\sig^\mathrm{g*} = 0.99 \pm 0.04$ is the galaxy density fluctuation amplitude of $L_*$ galaxies. They combine the empirically determined bias-luminosity relation of \cite{westover_galaxy_2007} with the Schechter luminosity function model \citep{schechter_analytic_1976}. The relative factor can partially be explained by the uncertainty in $\sig^\mathrm{g*}$ and partially by the difference between computing $\siggal$ in real and redshift space.

The thin grey lines in \cref{fig:radial_dependence_of_sigma8_galaxy} show the results obtained for the set of mock 2MRS galaxy catalogs described in \cref{sec:test_on_mocks:mock_generation}. While there is some scatter between the individual mocks, they show consistently lower values than the results obtained from the actual 2MRS catalogs for distances above $50 \hmpc$. This implies that the galaxy bias of the more luminous galaxies is lower in the mocks than in the actual 2MRS. We suspect this to be a limitation of the semi-analytic galaxy formation model \textsc{sage}, as \cite{knebe_multidark-galaxies_2018} showed that there are deviations between the clustering of \textsc{sage} galaxies in MDPL2 and those of the SDSS DR7 sample \citep{strauss_spectroscopic_2002}. However, since we are explicitly normalizing the observed (mock) galaxy density field to the radially dependent $\siggal(r)$ estimated from the (mock) data itself, this mismatch should not affect the test and calibration of our method on the simulated mocks.

%%%%%%%%%%%%%%%%%%%%%%%%%%%%%%%%%%%%%%%%
\section{Details on SFB decomposition}
\label{sec:SFB_details}
For the SFB decomposition of the density and velocity fields, we employ boundary conditions that impose a vanishing density contrast for distances $r \geq \rmax$. As shown in \cite{fisher_wiener_1995}, the corresponding radial modes $k_{ln}$ are defined via
\begin{equation}
    j_{l-1}(k_{ln} \, \rmax) = 0
    \label{eq:app:radial_SFB_modes}
\end{equation}
and the SFB base functions satisfy the orthogonality relations
\begin{align}
	\int \d \Omega \, Y_{lm}(\vartheta, \varphi) \, Y^*_{l'm'}(\vartheta, \varphi) &= \kronecker_{ll'} \, \kronecker_{mm'} \,, \\
	\int\limits_0^{\mathclap{\rmax}} \d r \, r^2 \, j_l(k_{ln} r) \, j_l(k_{ln'} r) &= \kronecker_{nn'} \, C_{ln}^{-1} \,,
\end{align}
where $\d \Omega = \sin(\vartheta) \, \d \vartheta \, \d \varphi$ and the normalization coefficients are given by
\begin{equation}
	C_{ln}^{-1} = \frac{\rmax^3}{2} \, \bigl[j_l(k_{ln} \, \rmax)\bigr]^2 \,.
	\label{eq:app:SFB_normalization_coefficient_for_potential_boundary_conditions}
\end{equation}
With this orthogonality, the coefficients $\norm{\delta}^r_{lmn}$ of the SFB expansion of the normalized density contrast in \cref{eq:method:SFB_expansion_of_the_normalized_density_contrast} are obtained via \cref{eq:method:SFB_coefficients_of_density}.

Assuming a potential flow, $\vect{v} = \vect{\nabla} \psi$, \cref{eq:method:linear_relation_between_normalized_density_contrast_and_peculiar_velocity} yields $\vect{\nabla}^2 \psi = - \fsig H \, \norm{\delta}$. Since the SFB base functions are eigenfunctions of the Laplace operator, this relation can be inverted in to obtain the SFB coefficients of the velocity potential, $\psi_{lmn} = \fsig H \, \norm{\delta}_{lmn}^r / k_{ln}^2$. Taking the gradient of the SFB expansion of $\psi$ then yields,
\begin{align}
	v_r(\vect{r}) &= \fsig H \, \sum_{lmn} \, C_{ln} \, \norm{\delta}^r_{lmn} \, \frac{j'_l(k_{ln} r)}{k_{ln}} \, Y_{lm}(\vartheta, \varphi) \,,
	\label{eq:app:radial_component_of_peculiar_velocity_field_from_density_contrast_SFB_coefficients} \\
	v_\vartheta(\vect{r}) &= \fsig H \, \sum_{lmn} \, C_{ln} \, \norm{\delta}^r_{lmn} \, \frac{j_l(k_{ln} r)}{2 k_{ln}^2 r}
	\label{eq:app:theta_component_of_peculiar_velocity_field_from_density_contrast_SFB_coefficients} \\
	&\phantom{=} \times \Bigl[\N^+_{lm} \, Y_{l,m+1}(\vartheta, \varphi) \, \e^{-\ii \varphi} - \N^-_{lm} \, Y_{l,m-1}(\vartheta, \varphi) \, \e^{\ii \varphi}\Bigr] \,,
	\nonumber \\
	v_\varphi(\vect{r}) &= \fsig H \, \sum_{lmn} \, C_{ln} \, \norm{\delta}^r_{lmn} \, \frac{j_l(k_{ln} r)}{2 \ii k_{ln}^2 r}
	\label{eq:app:phi_component_of_peculiar_velocity_field_from_density_contrast_SFB_coefficients} \\
	&\phantom{=} \times \Bigl[\cos(\vartheta) \Bigl(\N^+_{lm} \, Y_{l,m+1}(\vartheta, \varphi) \, \e^{-\ii \varphi} + \N^-_{lm} \, Y_{l,m-1}(\vartheta, \varphi) \, \e^{\ii \varphi}\Bigr)
	\nonumber \\
	&\phantom{=} - 2 m \sin(\vartheta) \, Y_{lm}(\vartheta, \varphi)\Bigr]
	\nonumber
\end{align}
with the shorthand notation $\sum_{lmn}$ for the triple-sum in \cref{eq:method:SFB_expansion_of_the_normalized_density_contrast} and $\N^{\pm}_{lm} = \sqrt{(l \mp m) (l \pm m + 1)}$. Once we know the coefficients $\norm{\delta}^r_{lmn}$, we can thus directly evaluate both the density contrast and velocity fields. More details on the derivation of these expressions can be found in \cite{fisher_wiener_1995}.

Estimating $\norm{\delta}^r_{lmn}$ from the redshift survey involves several steps. First, we use \cref{eq:method:estimator_of_density_field} to estimate the observed density contrast in \sspace. Inserting this estimator into the \sspace\ equivalent of \cref{eq:method:SFB_coefficients_of_density} then yields the \sspace\ SFB coefficients $\norm{\delta}^{s,\mathrm{D}}_{lmn}$ given in \cref{eq:method:normalized_density_contrast_data}, which contains the \sspace\  monopole contribution

\begin{equation}
    \begin{split}
	    \norm{M}^s_n &= \int\limits_0^{\mathclap{\rmax}} \d s \, s^2 \int \d \Omega \, \phi(s) \, \norm{w}(s) \, j_0(k_{0n} s) \, Y^*_{00}(\vartheta, \varphi) \\
	    &= \sqrt{4 \pi} \int\limits_0^{\mathclap{\rmax}} \d s \, s^2 \, \phi(s) \, \norm{w}(s) \, j_0(k_{0n} s) \,.
	\end{split}
	\label{eq:app:SFB_density_contrast_monopole_contribution_redshift_space}
\end{equation}

Accounting for linear RSDs, the $s$- and \rspace\ coefficients of the density contrast are related by \cref{eq:method:relation_between_redshift_and_real_space_density_contrast_SFB_coefficients_via_coupling_matrix}, i.\,e.~a multiplication with the coupling matrix
\begin{equation}
	\begin{split}
		(\matr{Z}_l)_{nn'} &= \kronecker_{nn'} - \fsig \, C_{ln'} \, \int\limits_0^{\mathclap{\rmax}} \d r \, r^2 \, \phi(r) \, \norm{w}(r) \, j_l(k_{ln} r) \\
		&\phantom{=} \times \biggl[j''_l(k_{ln'} r) + \biggl(2 + \td{\ln{\phi(r)}}{\ln{r}}\biggr) \\
		&\phantom{=} \times \biggl(\frac{j'_l(k_{ln'} r)}{k_{ln'} r} - \alpha \, \kronecker_{l1} \, \frac{1 - j_0(k_{1n'} \rmax)}{3 k_{1n'} r}\biggr)\biggr] \,,
	\end{split}
	\label{eq:app:coupling_matrix_between_redshift_and_real_space_density_contrast_SFB_coefficients}
\end{equation}
derived in \cite{fisher_wiener_1995}.\footnote{\Cref{eq:app:coupling_matrix_between_redshift_and_real_space_density_contrast_SFB_coefficients} is true for any choice of boundary conditions. For our specific choice, it further simplifies, as \cref{eq:app:radial_SFB_modes} implies $j_0(k_{1n'} \rmax) = 0$.} If the observed redshifts are in LG frame, we need to set $\alpha = 1$ to subtract the observer's velocity. If they are in CMB frame, on the other hand, this is not necessary and $\alpha = 0$. The RSD correction can introduce a small spurious mean density contrast in the reconstruction volume, $\langle\norm{\delta}^{r,\mathrm{D}}\rangle$, which is incompatible with the imposed SFB boundary conditions. Therefore, we have to subtract the corresponding \rspace\ monopole contribution
\begin{equation}
	\begin{split}
	    \norm{M}^r_n &= \langle\norm{\delta}^{r,\mathrm{D}}\rangle \, \int\limits_0^{\mathclap{\rmax}} \d r \, r^2 \int \d \Omega \, j_0(k_{0n} r) \, Y^*_{00}(\vartheta, \varphi) \\
	    &= \langle\norm{\delta}^{r,\mathrm{D}}\rangle \, \sqrt{4 \pi} \, \rmax^3 \, \frac{j_1(k_{0n} \rmax)}{k_{0n} \rmax}
	\end{split}
	\label{eq:app:SFB_density_contrast_monopole_contribution_real_space}
\end{equation}
after multiplying $\norm{\delta}^{s,\mathrm{D}}_{lmn}$ with the inverse coupling matrix.

Finally, we obtain the estimate of the \rspace\ SFB coefficients of the normalized density contrast in \cref{eq:method:minimum_variance_estimate_in_SFB_space} by inverting the coupling relation and applying a WF. The latter contains the signal and noise correlation matrices
\begin{align}
	\begin{split}
		(\matr{S}_l)_{nn'} &= \frac{2}{\pi} \, \int\limits_0^\infty \d k \, k^2 \, P_{\norm{\delta}}(k) \, \int\limits_0^{\mathclap{\rmax}} \d r \, r^2 \, j_l(k_{ln} r) \, j_l(k r) \\
		&\phantom{=} \times \int\limits_0^{\mathclap{\rmax}} \d r' \, r'^2 \, j_l(k_{ln'} r') \, j_l(k r') \\
		&\approx \kronecker_{nn'} \, P_{\norm{\delta}}(k_{ln}) \, C_{ln}^{-1} \,,
	\end{split}
	\label{eq:app:signal_matrix} \\
	(\matr{N}_l)_{nn'} &= \int\limits_0^{\mathclap{\rmax}} \d r \, r^2 \, \frac{\norm{w}(r)}{\bar{n}^\gal} \, j_l(k_{ln} r) \, j_l(k_{ln'} r) \,,
	\label{eq:app:noise_matrix}
\end{align}
with the normalized density contrast power spectrum $P_{\norm{\delta}} = P_\delta / \sig^2$. To arrive at the second line of \cref{eq:app:signal_matrix}, we approximated the second radial integral over the two Bessel functions by $\dirac(k-k_{ln'}) \, \pi / (2 k^2)$ \citep{fisher_wiener_1995}.

%%%%%%%%%%%%%%%%%%%%%%%%%%%%%%%%%%%%%%%%
\section{Selection function estimator}
\label{sec:selection_function_estimator}
The selection function $\phi$ gives the fraction of observable galaxies at a certain distance. Assuming a spatially homogeneous distribution of galaxies, it can in principle be inferred directly from the comoving redshift distance histogram of the survey,
\begin{equation}
	\td{N(s)}{s} \propto - \phi(s) \, s^2 \,.
\end{equation}
In practice, though, the mean galaxy number density per spherical shell will experience some fluctuations around the mean over the total survey volume, especially at short distances. One way to avoid this problem, is to infer $\phi$ from an externally provided luminosity function. The caveat of this method is that it is susceptible to deviations of the true survey luminosity function from the model. 

Here, we adopt the $F/T$ estimator \citep{davis_survey_1982,branchini_linear_2012}, which yields the radially binned selection function from the survey itself without fitting local density variations. For each bin $s_j$, the number of galaxies at a distance $s \leq s_j$ which would also be visible if placed at a larger distance $s > s_j$ is denoted by $T_j$. The subset of those galaxies that would only be visible up to the next bin $s_{j+1}$ but not further out is $F_j$. Then,
\begin{equation}
	- \frac{F_j}{T_j} \approx \frac{\Delta \varphi(s_j)}{\phi(s_j)} \approx \td{\ln{\phi(s)}}{s} \,\biggr|_{s = s_j} \, \Delta s_j \,.
\end{equation}

To determine if a galaxy with apparent magnitude $m$ and redshift $z$ would be observable at a given distance, we compare its absolute magnitude
\begin{equation}
	M = m(z) - 25 - 5 \, \log_{10}\biggl(\frac{d_\lum(z)}{\mpc*}\biggr) - K\bigl(z^{\obs,\mathrm{hc}}\bigr) + Q(z)
	\label{eq:selection_function_estimator:absolute_apparent_magnitude_relation}
\end{equation}
to the absolute magnitude corresponding to the flux limit of the survey $m_\text{max}$ at that distance. Here, $K$ describes the effect of the so-called $k$-correction, related to the fact that we are observing only a certain bandwidth of the full luminosity. It was found in \cite{bell_optical_2003} that for 2MRS it is well-described by
\begin{equation}
	K\bigl(z^{\obs,\mathrm{hc}}\bigr) = - 2.1 \, z^{\obs,\mathrm{hc}} \,.
\end{equation}
Note that $K$ is a function of the directly observed redshift in heliocentric frame, $z^{\obs,\mathrm{hc}}$.
The term $Q$ describes a correction due to the luminosity evolution, and was determined for 2MRS as \citep{bell_optical_2003}
\begin{equation}
	Q(z) = 0.8 \, z \,.
\end{equation}

%%%%%%%%%%%%%%%%%%%%%%%%%%%%%%%%%%%%%%%%
\section{Generating random signal and data realizations}
\label{sec:generating_random_signal_and_data_realizations}
To construct CRs of the density contrast, as defined in \cref{eq:method:constrained_realizations_around_wiener_filter}, we first need to generate random realizations of both the normalized density contrast signal, $\norm{\delta}^\mathrm{R,S}$, and data, $\norm{\delta}^\mathrm{R,D}$, for which we adapt the approach of \cite{agrawal_generating_2017}. The log-normal distributed signal $\norm{\delta}^\mathrm{R,S}$ is generated from a realization of the associated zero-mean Gaussian log-density field $g = \ln(1 + \norm{\delta}^\mathrm{R,S}) - \langle \ln(1 + \norm{\delta}^\mathrm{R,S}) \rangle$ and transforming it according to
\begin{equation}
	\norm{\delta}^\mathrm{R,S}(\vect{r}) = \e^{g(\vect{r}) - \frac{\sigma_g^2}{2}} - 1 \,,
\end{equation}
where $\sigma_g^2 = \langle g^2\rangle$ is the variance of $g$. The log-density field itself is most easily generated in Fourier space,
\begin{equation}
	g\bigl(\vect{k}\bigr) = \sqrt{\frac{P_g(k) \, V}{2}} \, (q_1 + \ii q_2) \,,
\end{equation}
where $q_1$ and $q_2$ are two independent Gaussian random numbers with zero mean and unit variance, $V$ is the volume of the periodic box, and $P_g$ is the power spectrum of the log-density field. The latter is obtained by first inverse Fourier transforming the normalized density contrast power spectrum $P_{\norm{\delta}}$,\footnote{We first perform a Gaussian pre-smoothing on a scale large enough to be resolved in our FFT box. We found a pre-smoothing scale of $1 \hmpc$ to be suitable.} which yields the normalized density contrast correlation function $\xi_{\norm{\delta}}$. It is related to the log-density correlation function $\xi_g$ via \citep{coles_lognormal_1991}
\begin{equation}
	\xi_g(r) = \ln\bigl(1 + \xi_{\norm{\delta}}(r)\bigr) \,,
\end{equation}
from which we then get $P_g$ by another Fourier transform.\footnote{At very high wavenumbers $k$ the log-density power spectrum $P_g$ can acquire negative values, since a log-normal field can not perfectly reproduce any desired covariance \citep{xavier_improving_2016}. We regularize $P_g$ by setting those negative values to zero, which was found to only have a minor impact.} To ensure the field $g(\vect{r})$ to be real, we additionally impose the condition $g(-\vect{k}) = g^*(\vect{k})$ when drawing the random realizations. All Fourier transforms are performed using the FFTW library \citep{frigo_design_2005}.

To obtain the respective data realization $\norm{\delta}^\mathrm{R,D}$, we draw the number of associated galaxies in each Cartesian cell of volume $\Delta V$ from a Poisson distribution with mean
\begin{equation}
	\bar{N}_j = \bar{n}^\gal \, \phi(r_j) \, \bigl(1 + \norm{\delta}^\mathrm{R,S}(\vect{r}_j)\bigr) \, \Delta V \,,	
\end{equation}
where $\vect{r}_j$ denotes the position of the $j$th cell, $\bar{n}^\gal$ is the mean galaxy number density of the survey, and the selection function $\phi$ ensures that only the average observable fraction of galaxies is generated. The positions of those galaxies are then uniformly distributed within the cell. Afterwards, we apply the \rspace\ equivalent of \cref{eq:method:normalized_density_contrast_data} to compute the desired random data SFB modes $\norm{\delta}_{lmn}^{r,\mathrm{R,D}}$. Note that we need to use the weighting function $\norm{w}^\mathrm{R}(r) \coloneqq 1/\phi(r)$ instead of \cref{eq:method:weighting_function} for that, as the random galaxy positions have already been sampled from an underlying normalized density field, such that no further division by $\siggal(r)$ is needed. Since the SFB modes are directly computed in \rspace, we also skip the linear RSD correction step in the computation of the WF estimate $\norm{\delta}^\mathrm{R,W}$ in \cref{eq:method:minimum_variance_estimate_in_SFB_space}. The CR of the normalized density contrast is then given by $\norm{\delta}^\mathrm{C} = \norm{\delta}^\mathrm{R,S} - \norm{\delta}^\mathrm{R,W} + \norm{\delta}^\mathrm{W}$.

The CR of the velocity field $\vect{v}^\mathrm{C}$ corresponding to $\norm{\delta}^\mathrm{C}$ is obtained by inverting the linear relation in \cref{eq:method:linear_relation_between_normalized_density_contrast_and_peculiar_velocity} for $\norm{\delta}^\mathrm{R,S}$, $\norm{\delta}^\mathrm{R,W}$ and $\norm{\delta}^\mathrm{W}$ separately. For the random signal, this is achieved by evaluating the Cartesian Fourier modes of the velocity given the normalized density contrast,
\begin{equation}
	\vect{v}^\mathrm{R,S}\bigl(\vect{k}\bigr) = \fsig H \, \frac{\ii \vect{k}}{k^2} \, \norm{\delta}^\mathrm{R,S}\bigl(\vect{k}\bigr) \,,
	\label{eq:generating_random_signal_and_data_realizations:linear_relation_between_density_contrast_and_velocity_in_fourier_space}
\end{equation}
and transforming those back into real space. The two WF estimates, $\vect{v}^\mathrm{R,W}$ and $\vect{v}^\mathrm{W}$, on the other hand, are obtained from the SFB decomposition of $\norm{\delta}^\mathrm{R,W}$ and $\norm{\delta}^\mathrm{W}$, respectively, by means of \cref{eq:app:radial_component_of_peculiar_velocity_field_from_density_contrast_SFB_coefficients,eq:app:theta_component_of_peculiar_velocity_field_from_density_contrast_SFB_coefficients,eq:app:phi_component_of_peculiar_velocity_field_from_density_contrast_SFB_coefficients}. Finally, the three contributions to the constrained linear velocity field realization are combined as $\vect{v}^\mathrm{C} = \vect{v}^\mathrm{R,S} - \vect{v}^\mathrm{R,W} + \vect{v}^\mathrm{W}$.

%%%%%%%%%%%%%%%%%%%%%%%%%%%%%%%%%%%%%%%%
\section{Tensor-smoothing}
\label{sec:tensor_smoothing}
Following \cite{dekel_potential_1990}, we define the tensor-smoothed radial velocity field
\begin{equation}
    \tenssmooth{v}_r(\vect{s}) \coloneqq \sum_{j=1}^{N^\obs} \, \tenssmooth{W}(\vect{s},\vect{s}_j) \, v_r(\vect{s}_j) \,,
    \label{eq:method:tensor_smoothing}
\end{equation}
where the tensor-smoothing function $\tenssmooth{W}$ accounts for the correct averaging of radial velocities from different directions $\vect{s}_j / s_j$ for a given window function $W$,
\begin{align}
    \tenssmooth{W}(\vect{s},\vect{s}_j) &\coloneqq W(\vect{s},\vect{s}_j) \; \frac{\vect{s}^\top \, A^{-1}(\vect{s}) \, \vect{s}_j}{s \, s_j} \,, \\
    A(\vect{s}) &\coloneqq \sum_{j=1}^{N^\obs} \, W(\vect{s},\vect{s}_j) \; \frac{\vect{s}_j \, \vect{s}_j^\top}{s_j^2} \,.
\end{align}
We choose a weighted Gaussian window function,
\begin{equation}
    W(\vect{s},\vect{s}_j) = \frac{1}{\tenssmooth{\sigma}_j^2} \, \exp\biggl[-\frac{(\vect{s} - \vect{s}_j)^2}{2 \, \ssmooth*^2}\biggr] \,.
\end{equation}
To ensure that the matrix $A$ is invertible and well-conditioned, we adopt an adaptive smoothing scale,
\begin{equation}
    \ssmooth*(\vect{s}) = \text{min}\biggl(\ssmooth, \frac{s_8(\vect{s})}{2}\biggr) \,,
\end{equation}
where $\ssmooth$ denotes the minimal smoothing scale and $s_8(\vect{s})$ is the radius of a sphere centered around $\vect{s}$ containing 8 observed data points. We furthermore use the inverse variance of the radial peculiar velocity as a weighting, $\tenssmooth{\sigma}_j^2 = \bigl(\sigma_{v_r,j}^\obs\bigr)^2 + \bigl(\sigma_{v_r}^\text{cos}(\ssmooth)\bigr)^2$, accounting for both the error on the observed velocity as well as the cosmic variance of the underlying peculiar velocity field. The latter is computed from the same power spectrum adopted in the WF, using a Gaussian window of width $\ssmooth$. For the considered tensor smoothing scales, $\ssmooth = 10$ to $30 \hmpc$, it ranges from $\sigma_{v_r}^\text{cos}(\ssmooth) = 197$ to $134 \kms$. In practice, this contribution is only relevant for small distances $s \lesssim 10 \hmpc$, where it avoids an over-weighting of the most nearby points.

The observed radial velocities and their errors are obtained from the observed distance moduli via the linearized relation in \cref{eq:method:expected_distance_modulus_linearized_in_peculiar_velocity},
\begin{equation}
    v_{r,j}^\obs = v_r^\obs(\vect{s}_j) = \frac{\mu(z_j^\obs) - \mu_j^\obs}{\eta(z_j^\obs)} \,, \quad
    \sigma_{v_r,j}^\obs = \frac{\sigma_{\mu,j}^\obs}{\eta(z_j^\obs)} \,.
\end{equation}
To obtain the tensor-smoothed observed velocity $\tenssmooth{v}_r^\text{obs}$, we use the unsmoothed observed velocities $v_r^\obs(\vect{s}_j)$ as the input in \cref{eq:method:tensor_smoothing}. For the tensor-smoothed reconstructed velocity $\tenssmooth{v}_r^\text{rec}$, we use the reconstructed velocity smoothed on $5 \hmpc$ as the input because this is the smallest smoothing scale for which we estimate the parameters $\vect{\Theta} = (\fsig, \bext, \hobs)$.

We calculate the error in $\tenssmooth{v}_r$ using the relation
\begin{equation}
    \sigma_{\tenssmooth{v}_r}^2(\vect{s}) \approx \sum_{j=1}^{N^\obs} \, \tenssmooth{W}^2(\vect{s},\vect{s}_j) \, \bigl[\sigma_{v_r,j}^2 + \bigl(v_r(\vect{s}_j) - \tenssmooth{v}_r(\vect{s})^2\bigr)\bigr]
\end{equation}
derived in \cite{dekel_potential_1990}. The two contributions account for the error $\sigma_{v_r,j}$ in the unsmoothed velocities and the sample variance, respectively. Corrections from higher orders in $\sigma_{v_r,j}$ as well as the effect of these errors on the smoothing window position are neglected.

%%%%%%%%%%%%%%%%%%%%%%%%%%%%%%%%%%%%%%%%%%%%%%%%%%

% Don't change these lines
\bsp	% typesetting comment
\label{lastpage}
\end{document}